\documentclass[12pt,twoside, onecolumn,journal,draftclsnofoot]{IEEEtran}%
\usepackage{amssymb,amsmath,latexsym,graphicx,version,bm}%\usepackage{cuted},hfont,curves,epic,eepic,curves,finemath,
\newtheorem{problem}{Problem}
\newtheorem{remark}{Remark}

\newtheorem{definition}{Definition}
\newtheorem{lemma}{Lemma}
\newtheorem{theorem}{Theorem}
\newtheorem{proposition}{Proposition}

\begin{document}
%=-=-=-=-=-=-=-=-=-=-=-=-=-=-=-=-=-=-=-=-=-=-=-=-=-=-=-=-=-=-=-=-=-=-=-=-=-=-=-=-=-=-
%\title{Equivalence between Sum-Rate Optimal Multi-Code CDMA and Restricted FDMA Systems}
\title{$\;$\\ Sum-Rate Optimal Multi-Code CDMA Systems: An Equivalence Result}
%\title{Equivalence of Sum-Rate Optimal Restricted Multiple-Access Systems--Part I: FDMA and TDMA}
%\title{Equivalence of Sum-Rate Optimal Restricted Multiple-Access Systems--Part II: Multi-Code CDMA}
%=-=-=-=-=-=-=-=-=-=-=-=-=-=-=-=-=-=-=-=-=-=-=-=-=-=-=-=-=-=-=-=-=-=-=-=-=-=-=-=-=-=-
\author{$\;$\\Yeo Hun Yun and Joon Ho Cho, \emph{Member, IEEE}
\thanks{
%This work was supported in part by the Ministry of Knowledge Economy, Korea, under the grant NIPA-2011-C1090-1011-0011 for the BrOMA-ITRC@POSTECH supervised by the NIPA and in part by the Ministry of Education, Science and Technology under the NRF grants no. 2010-0000622 and ????.
The material in this paper was presented in part at the 2010 Information Theory and Applications workshops (ITA 2010), San Diego, CA, Jan. 2010.
}
\thanks{The authors are with the Department of Electrical Engineering, Pohang University of Science and Technology (POSTECH), Pohang, Gyeongbuk 790-784, Korea (e-mail: \{yhym205, jcho\}@postech.ac.kr).}}
%=-=-=-=-=-=-=-=-=-=-=-=-=-=-=-=-=-=-=-=-=-=-=-=-=-=-=-=-=-=-=-=-=-=-=-=-=-=-=-=-=-=-
\date{}
\markboth{Submitted to IEEE Transactions on Information Theory (ver 1.0)} {Yun and Cho: Sum-Rate Optimal Multi-Code CDMA Systems: An Equivalence Result}
%=-=-=-=-=-=-=-=-=-=-=-=-=-=-=-=-=-=-=-=-=-=-=-=-=-=-=-=-=-=-=-=-=-=-=-=-=-=-=-=-=-=-
\maketitle
%=-=-=-=-=-=-=-=-=-=-=-=-=-=-=-=-=-=-=-=-=-=-=-=-=-=-=-=-=-=-=-=-=-=-=-=-=-=-=-=-=-=-
\begin{abstract}
In this paper, the sum rate of a multi-code CDMA system with asymmetric-power users is maximized, given a processing gain and a power profile of users.
Unlike the sum-rate maximization for a single-code CDMA system, the optimization requires the joint optimal distribution of each user's  power to its multiple data streams as well as the optimal design of signature sequences.
The crucial step is to establish an equivalence of the multi-code CDMA system to restricted FDMA and TDMA systems.
The CDMA system has upper limits on the numbers of multi-codes of users, while the FDMA and the TDMA systems have upper limits on the bandwidths and the duty cycles of users, respectively, in addition to total bandwidth constraint.
The equivalence facilitates the complete characterization of the maximum sum rate of the multi-code CDMA system and also provides new insights into the single- and the multi-code CDMA systems in terms of the parameters of the equivalent FDMA and TDMA systems.
\end{abstract}
%=-=-=-=-=-=-=-=-=-=-=-=-=-=-=-=-=-=-=-=-=-=-=-=-=-=-=-=-=-=-=-=-=-=-=-=-=-=-=-=-=-=-
%=-=-=-=-=-=-=-=-=-=-=-=-=-=-=-=-=-=-=-=-=-=-=-=-=-=-=-=-=-=-=-=-=-=-=-=-=-=-=-=-=-=-
\section{Introduction}
%=-=-=-=-=-=-=-=-=-=-=-=-=-=-=-=-=-=-=-=-=-=-=-=-=-=-=-=-=-=-=-=-=-=-=-=-=-=-=-=-=-=-

\IEEEPARstart{T}{ime-division} multiple-access (TDMA) and code-division multiple-access (CDMA) have been intensively studied for more than last three decades, and have been serving as the major multiple-access schemes for the second and the third generation wireless cellular systems, respectively \cite{GSM}--\cite{WCDMA}.
Recently, the fourth generation wireless cellular system has started to be deployed worldwide, for which the frequency-division multiple-access (FDMA) serves in the form of discrete Fourier transform-spread orthogonal frequency-division multiplexing, a.k.a., single-carrier FDMA \cite{LTE}.
The main contribution of this paper is the complete characterization of the maximum sum rate of a multi-code CDMA system by establishing an equivalence among these three multiple-access schemes.

Since the primary performance limiting factor of a CDMA system is multiple-access interference (MAI), a lot of research has been conducted to mitigate the detrimental effect of the MAI through system parameter optimizations \cite{Viterbi: book}, \cite{Verdu : book}.
At the transmitter side, signature sequences have long been identified as important design parameters and hence optimized under various criteria \cite{Gold: 1}-\cite{Guess3}.
Particularly in \cite{Rupf}-\cite{Ulukus4}, the Shannon-theoretic sum rate is maximized through sequence optimization given a processing gain and a power profile of users in order to find the fundamental performance limit of the CDMA system that operates over the multiple-access channel (MAC) corrupted by additive white Gaussian noise (AWGN).

In \cite{Rupf}, the signature sequences are optimized to maximize the sum rate of a symbol-synchronous CDMA system with symmetric- or equal-power users.
For convenience, let the system be called underloaded, critically-loaded, and overloaded, respectively, if the number of users is less than, equal to, and greater than the processing gain.
Then, it is shown that orthogonal sequences are optimal for underloaded or critically-loaded systems, while Welch bound equality (WBE) sequences \cite{Welch} are optimal for overloaded systems.
It is also shown that the maximum sum rate is less than the sum capacity of the MAC for underloaded systems, while they are equal for critically-loaded or overloaded systems.

The sequence design problem considered in \cite{Rupf} is generalized in \cite{Viswanath : optimal0} to accommodate asymmetric- or unequal-power users.
Again for underloaded or critically-loaded systems, orthogonal sequences are optimal that completely remove the MAI.
For overloaded systems, it is shown that orthogonal sequences are optimal for oversized users and generalized WBE (GWBE) sequences \cite{Viswanath : optimal} are optimal for non-oversized users, where the oversized users are defined as the relatively strong-power users satisfying a certain condition in terms only of the processing gain and the power profile of users.
It is also shown that the maximum sum rate becomes equal to the sum capacity of the MAC if and only if the system is overloaded without any oversized user or critically-loaded only with symmetric-power users.
Thus, with asymmetric-power CDMA users, a strict loss in sum rate relative to unrestricted multiple-access achieving the sum capacity of the MAC is experienced not only by underloaded systems but also by critically-loaded and even by some overloaded systems.

The results in \cite{Rupf} and \cite{Viswanath : optimal0} are further extended in \cite{Ulukus4} to a symbol-asynchronous but chip-synchronous system.
Once again, orthogonal sequences are optimal for underloaded or critically-loaded systems.
For overloaded systems, it is shown that orthogonal sequences having only one non-zero element are optimal for oversized users and generalized asynchronous WBE (GAWBE) sequences are optimal for non-oversized users, where the users are classified in the same way as the users in the symbol-synchronous system are.
It is also shown that, if the asynchronous CDMA system has the same processing gain and the same power profile of users as the synchronous system does, the former has the same maximum sum rate and the same necessary and sufficient condition to achieve the sum capacity of the MAC as the latter does.
Thus, the asynchronous CDMA with asymmetric-power users becomes an optimal multiple-access scheme only when the system is overloaded without any oversized user.

In \cite{Guess3}, it is pointed out that the sub-optimality of these CDMA systems in achieving the sum capacity of the MAC is due to the restriction that allows the transmission of only a single data stream of scalar symbols per user and that, consequently, a multi-dimensional signaling such as multi-carrier CDMA can significantly expand the capacity region.
However, the sequence optimizations for the CDMA with multi-dimensional signaling have been limited so far to the sum-power minimization subject to the equal signal-to-interference-plus-noise ratio of users \cite{Wong: multicode} and the characterization of either the capacity region given a constraint on the total power of users or the power region given constraints on user rates \cite{Guess3}.
In \cite{Honig2}, the asymptotic sum rate is derived for the CDMA with multi-dimensional signaling.
However, isometric or random signature sequences are only employed without any sequence optimization.

In this paper, we consider a multi-code CDMA system as a representative CDMA system with multi-dimensional signaling, and maximize its sum rate for asymmetric-power users.
It is well known that the multi-code CDMA has an advantage over the single-code CDMA in that it can better exploit the system resources by naturally supporting various types of data with different rate requirements \cite{Wong: multicode}, \cite{Guess3}.
The crucial step is to establish an equivalence of the multi-code CDMA system to restricted FDMA and TDMA systems.
The CDMA system has upper limits on the numbers of multi-codes of users, while the FDMA and the TDMA systems have upper limits on the bandwidths and the duty cycles of users, respectively, in addition to total bandwidth constraint.
We classify the multi-code users into oversized, critically-sized, and undersized users.
The equivalence then provides a physical meaning to this classification rule by using the power spectral densities (PSDs) of users in the equivalent FDMA and TDMA systems.
Unlike the definition of an oversized user in \cite{Viswanath : optimal0} and \cite{Ulukus4} for the single-code CDMA system, this classification rule for the multi-code system is applicable to any combination of the processing gain and the power profile of users.

Note that, unlike the sum-rate maximization in \cite{Rupf}--\cite{Ulukus4} for the single-code CDMA system, the sum-rate maximization for the multi-code system requires the optimal distribution of each user's  power to its multiple data streams as well as the optimal design of the signature sequences.
We introduce the notions of FDMA-equivalent bandwidth and TDMA-equivalent duty cycle, and use them to find the optimal power distribution and to obtain an insight into how to allocate the signal dimension or the system bandwidth to the users.

We derive the maximum sum rate of the multi-code CDMA system and show that it is the same as that of the equivalent FDMA and TDMA systems.
We also derive the necessary and sufficient condition for the sum-rate optimal multi-code CDMA system to achieve the sum capacity of the MAC.
The equivalence says that this condition is nothing but the condition for the equivalent FDMA and TDMA systems to achieve the sum capacity of the MAC by having a flat PSD for each user and equal PSDs for all users.
It turns out that, unlike the single-code CDMA system, the multi-code system can achieve the sum capacity of the MAC even for underloaded systems if and only if there is no oversized multi-code user.

We show that, similar to the single-code CDMA system, all optimal signature sequences of the oversized multi-code users are orthogonal sequences.
Interestingly, we also show that even some signature sequences of the non-oversized users can be orthogonal sequences, though in general they are GWBE sequences.

With the system complexity taken into account, we derive both the maximum number of orthogonal sequences and the minimum number of signature sequences that still retain the maximum sum rate given the upper limits on the numbers of multi-codes.
We examine whether these apparently contradicting objectives of the maximization of the number of orthogonal sequences and the minimization of the number of signature sequences can be achieved at the same time or not.
In addition, we find the minimal upper limit profile on the numbers of multi-codes to achieve the sum capacity of the MAC.
The finiteness of this minimal upper limit profile shows that the CDMA as a multiple-access scheme is not inherently sub-optimal but just becomes sub-optimal when the excessive restriction is imposed on the numbers of multi-codes.

We briefly discuss how to extend the results in this paper to the symbol-asynchronous but chip-synchronous system, which completes the generalization of the theory developed in \cite{Rupf}--\cite{Ulukus4} for the single-code CDMA systems to the multi-code systems.
We also compare the spectral efficiency of the sum-rate optimal CDMA systems with that of the CDMA systems having random sequences \cite{Honig2}--\cite{Verdu : fading}, both in no fading and Rayleigh flat-fading channels.

This way, the equivalence facilitates the complete characterization of the maximum sum rate of the multi-code CDMA system and also provides new insights into the single- and the multi-code CDMA systems in terms of the parameters of the equivalent FDMA and TDMA systems.
The rest of this paper is organized as follows.
In Section~\ref{sec: sig sys}, the signal and system models are described and the optimization problems are formulated for the multi-code CDMA system.
In Section~\ref{sec: FDMA TDMA}, a detour is made to the optimal bandwidth allocation problem for the restricted FDMA system and the optimal duty cycle allocation problem for the restricted TDMA system, and their maximum sum rates are characterized.
In Section~\ref{sec: CDMA}, the optimal solutions to the optimization problems formulated in Section~\ref{sec: sig sys} are derived.
Numerical results and discussions are provided in Section~\ref{sec: discussion}, and concluding remarks are offered in Section~\ref{sec: Con}.

%=-=-=-=-=-=-=-=-=-=-=-=-=-=-=-=-=-=-=-=-=-=-=-=-=-=-=-=-=-=-=-=-=-=-=-=-=-=-=-=-=-=-
\section{Signal Model and Problem Formulation}\label{sec: sig sys}
%=-=-=-=-=-=-=-=-=-=-=-=-=-=-=-=-=-=-=-=-=-=-=-=-=-=-=-=-=-=-=-=-=-=-=-=-=-=-=-=-=-=-
Suppose that there are $K$ active users transmitting to a single receiver over a MAC.
Similar to \cite{Viswanath : optimal0}, it is assumed that the users are symbol synchronous, possibly have an asymmetric-power profile, and employ CDMA as a multiple-access scheme.
Later in Section~\ref{sec: CDMA-asynch}, we will consider the CDMA with symbol-asynchronous but chip-synchronous users like that in \cite{Ulukus4}.
The major difference from \cite{Rupf}--\cite{Ulukus4} is that the users in this paper are multi-code users that may transmit more than one data streams by using multiple signature sequences or codes.

% 정반합: Go, Tense일치: Go, 정확한 단어사용: Go

We model the received signal $\underline{\bm{y}}$ of this multi-code CDMA system by using an $N$-by-$1$ real-valued random vector
\begin{equation}\label{eq: sig model}
\underline{\bm{y}}\triangleq \sum_{k=1}^{K}\underline{\bm{x}}_k+\underline{\bm{w}},
\end{equation}
where $N$ is called the processing gain, $\underline{\bm{x}}_k$ is the signal component from the $k$th multi-code user, and $\underline{\bm{w}}$ is the ambient AWGN component.
The signal component $\underline{\bm{x}}_k$ is given by
\begin{equation}
\underline{\bm{x}}_k \triangleq \sum_{l=1}^{n_k}\bm{d}_{k,l}\underline{s}_{k,l},
\end{equation}
where $n_k\geq 1$ is the number of active data streams of the $k$th multi-code user, and $\bm{d}_{k,l}$ and $\underline{s}_{k,l}$ are the data symbol and the associated signature sequence of the $l$th data stream of the $k$th user, respectively.
The AWGN component
\begin{equation}
\underline{\bm{w}}\sim \mathcal{N}(\underline{0}_N,\sigma^2 I_N)
\end{equation}
is assumed to have mean zero and variance $\sigma^2>0$ per dimension, where $\underline{0}_N$ is the $N$-by-$1$ all-zero vector and $I_N$ is the $N$-by-$N$ identity matrix.

% 정반합: Go, Tense일치: Go, 정확한 단어사용: Go

With the signal model in (\ref{eq: sig model}), we solve the following problems in this paper.
For convenience in comparing the results with those in \cite{Rupf}--\cite{Ulukus4} for the single-code CDMA system, the components of the received signal and the signature sequences are modeled to be real, though the extension to a complex-valued baseband model is straightforward for real-valued passband signaling.
For consistency to \cite{Guess3}, the single- or the multi-code CDMA system is called overloaded if the number of active users is greater than the processing gain.
In this paper, we further classify the non-overloaded systems to have the following definition.

\begin{definition}
A multi-code CDMA system with number of users $K$ and processing gain $N$ is called
\begin{equation}\label{def: loading}
\begin{array}{rll}
\text{(a)}&\!\!\text{overloaded}, & \text{if } K>N,\\
\text{(b)}&\!\!\text{critically-loaded}, & \text{if } K=N, \text{ and}\qquad\\
\text{(c)}&\!\!\text{underloaded}, & \text{if } K<N.
\end{array}
\end{equation}
\end{definition}

The first problem is to maximize the sum rate of the system, given the average signal power $p_k>0$ and the upper limit $\bar{n}_k\in \mathbb{N}$ on the number of active data streams or, equivalently, active signature sequences of the $k$th multi-code user, i.e.,
\begin{IEEEeqnarray}{C}
\frac{1}{N}\mathbb{E}\left\{\|\bm{x}_k\|^2\right\}=p_k,\IEEEyessubnumber\label{eq: power}\\%
\noalign{\noindent{\text{and}}\vspace{\jot}}%
n_k\leq \bar{n}_k,\IEEEyessubnumber
\end{IEEEeqnarray}
for $k=1,2,\cdots,K$, where $\mathbb{N}$ is the set of all natural numbers.
As the definition of the sum rate, we adopt the total information rates of users per chip at which the users can transmit reliably.
This system with upper limits on $(n_k)_k$ will be called a restricted multi-code CDMA system or, just simply, a multi-code CDMA system.
Then, the system with no upper limit on some $(n_k)_k$ and that with no upper limit on all $(n_k)_k$, respectively, may be called a partially-restricted and an unrestricted multi-code CDMA system.
We are mostly interested in the restricted and the unrestricted systems, but the extension of the results in this paper to the partially-restricted systems is straightforward.
These upper limits may be imposed to prevent a small group of multi-code users from occupying most of the signal dimension or the system bandwidth.
This will become clear once the notions of the FDMA-equivalent bandwidth and the TDMA-equivalent duty cycle are introduced and the necessary and sufficient condition for the multi-code CDMA system to achieve the sum capacity of the MAC is derived in Section~\ref{sec: CDMA}.
Unlike the sum-rate maximization in \cite{Rupf}--\cite{Ulukus4} for the single-code CDMA system, that for the multi-code system requires to find the optimal power distribution as well as the optimal sequence design, which will be shown shortly.

The second problem is to find the multi-code CDMA systems that maximize the number of orthogonal signature sequences for each multi-code user, among the sum-rate optimal multi-code CDMA systems found in solving the first problem.
To maximize the sum rate of a multiple-access scheme with non-orthogonal user signals, joint encoding and decoding are required that perform, e.g., time sharing or rate splitting combined with superposition encoding and successive interference cancelation decoding \cite{NIT}, \cite{Urbanke: RS}.
Since orthogonal channelization among users significantly simplifies the encoding and the decoding procedures, this maximization of the number of orthogonal sequences may contribute a lot to reducing the overall system complexity.
Recall that, in the sum-rate maximization for the single-code CDMA system \cite{Viswanath : optimal0}, a group of users called oversized users are allocated orthogonal signature sequences and non-oversized users are allocated GWBE sequence.
In this paper, we show that all optimal signature sequences of the oversized or the critically-sized multi-code users are orthogonal sequences, and that some optimal signature sequences of the undersized multi-code users can also be orthogonal sequences.

The third problem is to find the multi-code CDMA systems that minimize the number of active signature sequences for each multi-code user, among the sum-rate optimal multi-code CDMA systems found in solving the first problem.
Since the constraint on the number of active signature sequences may be imposed not only to prevent a small group of multi-code users from occupying most of the signal dimension or the system bandwidth but also to limit the complexity of the transmitters and the receiver, this minimization may contribute to further reducing the overall system complexity.
Interestingly, it turns out that the objectives of the second and the third problems can be achieved simultaneously.

As mentioned in the Introduction, the major differences from \cite{Guess3}, where the CDMA with multi-dimensional signaling is also considered, are that all the above problems are related to the maximum sum rate and that the upper limits on the numbers of multi-codes of users are given as the constraints.
We also briefly discuss in Section~\ref{sec: optimal solution2} what is the minimal upper limit profile on the numbers of multi-codes to achieve the sum capacity of the MAC.

The decision parameters of these optimization problems are the active signature sequences and their numbers, and the powers distributed to the multiple data streams of each user.
For notational simplicity in formulating the problems, we define the signature-sequence matrix $S_k$, the data-symbol vector $\underline{\bm{d}}_k$, and the data-correlation matrix $P_k$ of the $k$th multi-code user as
\begin{IEEEeqnarray}{rCl}
S_k &\triangleq&  [\underline{s}_{k,1},\underline{s}_{k,2},\cdots,\underline{s}_{k,\bar{n}_k}],\IEEEyessubnumber\\%
\;\;\underline{\bm{d}}_k &\triangleq& [\bm{d}_{k,1},\bm{d}_{k,2},\cdots,\bm{d}_{k,\bar{n}_k}]^T,\text{ and}\IEEEyessubnumber\\%
P_k &\triangleq&  \mathbb{E}\{\underline{\bm{d}}_k \underline{\bm{d}}_k^{T}\},\IEEEyessubnumber\label{eq: P_k}
\end{IEEEeqnarray}
respectively, where the superscript $^{T}$ denotes the transposition.
Then, since $\underline{\bm{x}}_k=S_k\underline{\bm{d}}_k$, the power constraint (\ref{eq: power}) can be rewritten as
\begin{equation}
\frac{1}{N}\text{tr}\left(S_k P_k S_k^{T}\right)=p_k,\forall k,
\end{equation}
where $\text{tr}(\cdot)$ denotes the trace.
Now, whenever $n_k<\bar{n}_k$ is needed, the power of the data symbol $\bm{d}_{k,l}$ can simply be set to zero for $l=n_k+1,n_k+2,\cdots,\bar{n}_k$, without loss of generality.

% 댓구: Go , Conciseness: Go, 첨언은 Note that: Go, Gap을 만들고 바로 메우고: Go

It is well known \cite{Cover} that a zero-mean Gaussian input distribution maximizes the mutual information between the input and the output of a Gaussian MAC.
Since $\underline{\bm{x}}_k=S_k\underline{\bm{d}}_k$, this requires zero-mean Gaussianity of the data symbols.
The data symbols from different multi-code users are assumed statistically independent in this MAC.
Hence, we can obtain the maximum sum rate of this symbol-synchronous multi-code CDMA system by solving
\begin{problem}\label{prob: CDMA 1}
\begin{IEEEeqnarray}{ll}
\underset{P,S}{\text{maximize}}&\frac{1}{2N}\log\det\left(I_N+\frac{1}{\sigma^2}S P S^{T}\right)\IEEEyessubnumber\label{eq: obj}\IEEEeqnarraynumspace\\
\text{subject to}\;\;&\frac{1}{N}\text{tr}\left(S_k P_k S_k^{T}\right)=p_k,\forall k,\IEEEyessubnumber
\end{IEEEeqnarray}
where the first decision parameter
\begin{subequations}
\begin{equation}
P \triangleq \text{diag}(P_1,P_2,\cdots,P_K)\label{eq: P}
\end{equation}
is a $(\sum_{k=1}^{K}\bar{n}_k)$-by-$(\sum_{k=1}^{K}\bar{n}_k)$ data-correlation matrix and the second parameter
\begin{equation}
S \triangleq [S_1,S_2,\cdots,S_K]\label{eq: S}
\end{equation}
is an $N$-by-$(\sum_{k=1}^{K}\bar{n}_k)$ signature-sequence matrix.
Throughout this paper, logarithmic functions have base $2$, unless otherwise specified.
Thus, the unit of the sum rate here is [bits/chip] or, equivalently, [bits/dimension].
\end{subequations}
\end{problem}
% first and second: Go, 물흐르듯하는 로직: Go, 독립된 problem formulation: Go,

Note in (\ref{eq: obj}) that, unlike the single-code CDMA system in \cite{Rupf}--\cite{Ulukus4}, the multi-code system requires the joint optimization of the data-correlation matrix $P$ and the signature-sequence matrix $S$ in order to maximize the sum rate.
Note also that, for each $k$, the $\bar{n}_k$-by-$\bar{n}_k$ matrix $P_k$ is positive semi-definite but not necessarily diagonal at this point because the data symbols of the $k$th multi-code user may be correlated.

\begin{lemma}\label{lemma: diagonal}
Without loss of generality, we can restrict the decision parameters of {Problem~\ref{prob: CDMA 1}}, respectively, to a diagonal matrix $P$ with non-negative entries and to a rectangular matrix $S$ with column vectors of norm $N$.
\end{lemma}

%{\bf Move to appendix?}
\begin{IEEEproof}
See Appendix~A.
\end{IEEEproof}

In this paper, we completely characterize the maximum sum rate achievable by the multi-code CDMA system, but only derive the optimal solutions in the form described in {Lemma~\ref{lemma: diagonal}} without loss of generality.
Thus, the problem of designing $P_k$ for each $k$ becomes that of distributing $p_k$ to the $\bar{n}_k$ multiple data streams of the $k$th user.
Just remember that, once such an optimal pair of ${P_k^*}$ and ${S_k^*}$ is obtained for all $k$, any orthogonal matrix $V_k$ preserves the optimality of $V_k{P_k^*}V_k^T$ and ${S_k^*}V_k^T$ as the optimal data-correlation and signature-sequence matrices of the $k$th multi-code user, respectively, because $\sum_{k=1}^{K}{S_k^*} {P_k^*} {S_k^*}^{T}=\sum_{k=1}^{K}({S_k^*}V_k^T)(V_k{P_k^*}V_k^T)({S_k^*}V_k^T)^T$.\footnote{In this paper, the superscript $^*$ denotes the optimality of the quantity not the complex conjugate of it.}
Thus, our solutions accommodate in effect a multi-code user that transmits a correlated vector symbol as well as that transmits a uncorrelated scalar symbol.

Once the data-correlation matrix $P_k$ defined as (\ref{eq: P_k}) is chosen to be diagonal, it can be rewritten as
\begin{equation}
P_k=\text{diag}(p_{k,1},p_{k,2},\cdots,p_{k,\bar{n}_k}),
\end{equation}
in terms of the power $p_{k,l}\triangleq\frac{1}{N}\mathbb{E}\{\|\bm{d}_{k,l}\underline{s}_{k,l}\|^2\}$ of the $l$th data stream of the $k$th multi-code user for $l=1,2,\cdots,\bar{n}_k$.
So, {Problem~\ref{prob: CDMA 1}} can be converted to an equivalent problem

\begin{problem}\label{prob: CDMA 2}
\begin{IEEEeqnarray}{ll}
\underset{P,S}{\text{maximize}}&\frac{1}{2N}\log\det\left(I_N+\frac{1}{\sigma^2}SPS^{T}\right)\IEEEyessubnumber\IEEEeqnarraynumspace\\
\text{subject to}\;\;&\sum_{l=1}^{\bar{n}_k}{p_{k,l}}=p_k,\forall k, \text{ and}\IEEEyessubnumber\\%
&p_{k,l}\geq 0, \forall k, \forall l,\IEEEyessubnumber
\end{IEEEeqnarray}
where the power matrix $P$ is diagonal with $p_{k,l}$ being the $(\sum_{k'=1}^{k-1}\bar{n}_{k'}+l)$th diagonal entry, and the signature-sequence matrix $S$ has $\underline{s}_{k,l}$ of norm $N$ as the $(\sum_{k'=1}^{k-1}\bar{n}_{k'}+l)$th column vector.
Here and henceforth, we adopt the convention that the sum $\sum (\cdot)$ is zero if the lower limit of the sum is greater than its upper limit.
\end{problem}

As mentioned just after {Lemma~\ref{lemma: diagonal}}, we consider only a diagonal matrix as a feasible power matrix $P$.
So, the zero-mean Gaussian data symbols $(\bm{d}_{k,l})_l$ of the $k$th multi-code user become always statistically independent.
Thus, the data streams of a multi-code user are not distinguishable from independent data streams of single-code users.
This observation leads to the following definition.

\begin{definition}
Given a feasible power matrix $P$, the $k$th multi-code user is effectively the collection of $\bar{n}_k$ single-code users with power $(p_{k,l})_l$ for each $k$.
These users are called the virtual single-code users of the multi-code CDMA system.
\end{definition}

The notion of virtual users for CDMA with multi-dimensional signaling is originally introduced in \cite{Guess3}.
This view of a multi-code user as virtual single-code users motivates us to rewrite {Problem~\ref{prob: CDMA 2}} in the form of a double maximization given by
\begin{problem}\label{problem: CDMA outer inner}
\begin{IEEEeqnarray}{ll}
\underset{P}{\text{maximize}}& \left\{\underset{S}{\text{max}} \displaystyle\frac{1}{2N}\log\det\left(I_N+\displaystyle\frac{1}{\sigma^2}SPS^{T}\right)\right\}\IEEEyessubnumber\IEEEeqnarraynumspace\label{eq: inner}\\%
\text{subject to}\;&\;\sum_{l=1}^{\bar{n}_k}p_{k,l}=p_k,\forall k, \text{ and}\IEEEyessubnumber\label{eq: sumnk3}\\%
&\;p_{k,l}\geq 0, \forall k, \forall l.\IEEEyessubnumber\label{eq: sumnk33}
\end{IEEEeqnarray}
\end{problem}

Now, the inner optimization problem over $S$ given $P$ is nothing but the sequence optimization problem considered in \cite{Viswanath : optimal0} for the single-code CDMA system, whose solution can be identified and constructed by using the results therein.
For $P$ such that the virtual single-code CDMA system is not overloaded, i.e., the processing gain is no less than the number of active virtual single-code users, orthogonal sequences are optimal and the inner optimization leads to the sum rate $1/(2N)\sum_{k=1}^{K}\sum_{l=1}^{n_k}\log(1+p_{k,l}/\sigma^2)$ \cite{Viswanath : optimal0}.
For $P$ such that the virtual single-code CDMA system is overloaded, however, a glimpse at the optimal solution in \cite{Viswanath : optimal0} reveals that the maximum sum rate expressed in terms of $(p_{k,l})_{k,l}$ turns out not to be mathematically tractable at all for the outer optimization over the diagonal entries of $P$.
This is because the maximum sum rate involves, in effect, the determination of the largest virtual single-code user index of so-called oversized users after numbering the virtual single-code users in a non-increasing order of the signal power, which is a highly non-linear operation of $P$.
Thus, this double-maximization approach cannot be taken as is.
%and this is maximized by water filling applied to each user, which results in the equal-power distribution of $p_k$ to $(p_{k,l})_{l}$ and the maximum sum rate $1/(2N)\sum_{k=1}^{K}n_k \log(1+p_k/(\sigma^2n_k))$.

Instead, we make a detour to the sum-rate maximization for FDMA and TDMA systems in the next section.
It will be shown later in Section~\ref{sec: CDMA} that this detour is well worth making not only to solve {Problem~\ref{problem: CDMA outer inner}} by non-trivially extending the results in \cite{Rupf}--\cite{Ulukus4} but also to obtain insights into the sum-rate optimal single- and multi-code CDMA systems.
Both the maximum number of orthogonal sequences and the minimum number of signature sequences will also be found in Section~\ref{sec: CDMA} for the multi-code CDMA system to still satisfy the constraint on $n_k$ for all $k$ and, at the same time, retain the maximum sum rate.

%=-=-=-=-=-=-=-=-=-=-=-=-=-=-=-=-=-=-=-=-=-=-=-=-=-=-=-=-=-=-=-=-=-=-=-=-=-=-=-=-=-=-
\section{Maximum Sum Rates of Restricted FDMA and TDMA Systems}\label{sec: FDMA TDMA}
%=-=-=-=-=-=-=-=-=-=-=-=-=-=-=-=-=-=-=-=-=-=-=-=-=-=-=-=-=-=-=-=-=-=-=-=-=-=-=-=-=-=-
In this section, we characterize the maximum sum rate achievable by a restricted FDMA system, where upper limits are imposed on the bandwidths of users, and that achievable by a restricted TDMA system, where upper limits are imposed on the duty cycles of users, both given a power profile of users and total available system bandwidth.
It turns out that these two systems are equivalent in the sense that the sum-rate maximization problem for the FDMA system can be converted to that for the TDMA system, and vice versa, by the proper substitutions of system parameters.
It will be shown in the next section that these impositions of the upper limits are analogous to limiting the numbers of multi-codes of users for the multi-code CDMA system considered in the previous section.

%=================================================================
\subsection{Problem Formulation for FDMA System}
%=================================================================
Suppose that there is a $K$-user FDMA system with total available system bandwidth $w_\text{tot}>0$ [Hz] in real passband, two-sided PSD $N_0/2>0$ of the AWGN that corrupts the channel, and average power $p_k >0$ and upper limit $0<\bar{w}_k<\infty$ on the bandwidth $w_k$ of the $k$th user for $k=1,2,\cdots,K$.
In what follows,
\begin{IEEEeqnarray}{rCl}
\underline{p} &\triangleq& [p_1, p_2, \cdots, p_K]\IEEEyessubnumber\\%
\noalign{\noindent{\text{and}}\vspace{\jot}}%
\underline{\bar{w}} &\triangleq& [\bar{w}_1, \bar{w}_2, \cdots, \bar{w}_K]\IEEEyessubnumber
\end{IEEEeqnarray}
respectively, denote the power profile of users and the bandwidth-constraint profile of users that consists of the bandwidth upper limits of users.
Then, the maximum sum rate
\begin{equation}
\mathcal{C}_\text{FDMA}(\underline{p},\underline{\bar{w}},w_\text{tot},N_0)
\end{equation}
of this restricted FDMA system can be found by solving
\begin{problem}\label{prob: FDMA}
\begin{IEEEeqnarray}{rll}\label{eq: FDMA prob}
&\underset{(w_k)_k}{\text{maximize}}\;\;&\sum_{k=1}^{K}w_k\log\left(1+\frac{p_k}{N_0w_k}\right)\IEEEyessubnumber\label{obj}\\
&\text{subject to}\;\;& 0\leq w_k\leq\bar{w}_k, \forall k,\text{ and }\IEEEyessubnumber\label{bandub}\\
&&\sum_{k=1}^{K}w_k\leq w_{\text{tot}},\IEEEyessubnumber\label{tbandconst}
\end{IEEEeqnarray}
\end{problem}
where the decision parameter $(w_k)_k$ consists of the bandwidth $w_k$ to be allocated to the $k$th user for $k=1,2,\cdots,K$.
The objective function is the sum of the AWGN channel capacities of the users, so that the unit of the sum rate is [bits/second].
% Problem 1과 댓구.

It is straightforward to verify that the constraint set is convex.
Throughout this paper, we follow the convention
\begin{equation}\label{convention}
w_k \log\left(1+\frac{p_k}{N_0w_k}\right)=0, \text{ for $w_k=0 $},
\end{equation}
which makes the left side of (\ref{convention}) continuous at any $p_k\geq 0$ and $w_k\geq 0$.
This also makes sense physically because if no bandwidth is allocated to a user then the user cannot transmit any signal and, hence, cannot increase the sum rate.
By this convention, the objective function in (\ref{obj}) becomes a strictly concave function of $(w_k)_k$ well defined on the convex constraint set.
Thus, {Problem~\ref{prob: FDMA}} is a standard convex optimization problem \cite{Boyd}, which allows us to use general-purpose convex programming routines to find its solution at least numerically.
Instead, we provide in this section its unique optimal solution in a closed algorithmic expression, which will be used in the next section in characterizing the maximum sum rate of the multi-code CDMA system.

It is well known \cite[Ch. 15]{Cover} that, if an unrestricted FDMA system is considered instead, i.e., if the constraint (\ref{bandub}) of {Problem~\ref{prob: FDMA}} is replaced by $0\leq w_k,\forall k$, the unique optimal solution allocates
\begin{equation}
w_{\text{tot}}\cdot\left(\frac{p_k}{\sum_{k'=1}^{K}p_{k'}}\right)\label{proportional}
\end{equation}
to the $k$th user as its bandwidth.
Throughout this paper, we use the following term for this solution, which uses up all available system bandwidth $w_{\text{tot}}$ and allocates the $k$th user the optimal bandwidth that is proportional to its signal power $p_k$.
\begin{definition}
The optimal solution to the unrestricted FDMA problem is called the proportional-share bandwidth allocation scheme.
\end{definition}

It will be shown that this proportional-share bandwidth allocation scheme plays a crucial role in constructing the optimal solution also to {Problem~\ref{prob: FDMA}} having the bandwidth upper limits of users.

Note that the FDMA system under the proportional-share bandwidth allocation has the sum rate equal to the sum capacity
\begin{equation}\label{eq: MAC}
\mathcal{C}_{\text{MAC}}(\underline{p},w_\text{tot},N_0)\triangleq w_{\text{tot}}\log\left(1+\frac{\sum_{k=1}^{K}p_k}{N_0w_{\text{tot}}}\right)
\end{equation}
of the MAC and becomes one of the optimal multiple-access schemes.
Note also that, under the proportional-share bandwidth allocation, the optimal PSD of the $k$th user becomes flat and given by
\begin{equation}\label{eq: PSD}
\frac{\sum_{k'=1}^{K}p_{k'}}{w_{\text{tot}}},
\end{equation}
which is not a function of $k$.
Thus, it immediately follows that the sum capacity of the MAC is achieved by the FDMA system if and only if the PSD of the overall received signal is flat, i.e., all the frequency components in the system frequency band are loaded uniformly.
In Section~\ref{sec: CDMA}, this will be used to explain why the presence of an oversized user incurs a strict loss in the sum rate of the CDMA system.

%=================================================================
\subsection{Problem Formulation for TDMA System}
%=================================================================
Suppose that there is a $K$-user TDMA system with total available system bandwidth $w_\text{tot}>0$ [Hz] in real passband, two-sided PSD $N_0/2>0$ of the AWGN that corrupts the channel, and average power $p_k >0$ and upper limit $0<\bar{t}_k<\infty$ on the duty cycle $t_k$ of the $k$th user, for $k=1,2,\cdots,K$.
In particular, we consider the TDMA system with power control \cite[Sec. 4.6.2]{NIT}, where the transmission power of the $k$th user in its transmission period is $p_k/t_k$ because the average power is $p_k$ and the duty cycle is $t_k$.
If we define
\begin{equation}
\underline{\bar{t}} \triangleq [\bar{t}_1, \bar{t}_2, \cdots, \bar{t}_K]
\end{equation}
as the upper-limit profile on the duty cycles of users, then the maximum sum rate
\begin{equation}
\mathcal{C}_\text{TDMA}(\underline{p},\underline{\bar{t}},w_{\text{tot}},N_0)
\end{equation}
of this restricted TDMA system can be found by solving
\begin{problem}\label{prob: TDMA}
\begin{IEEEeqnarray}{rll}\label{eq: TDMA prob}
&\underset{(t_k)_k}{\text{maximize}}\;\;&\sum_{k=1}^{K}t_kw_\text{tot}\log\left(1+\frac{p_k}{N_0t_kw_\text{tot}}\right)\IEEEyessubnumber\label{objT}\\
&\text{subject to}\;\;& 0\leq t_k\leq\bar{t}_k, \forall k,\text{ and }\IEEEyessubnumber\label{bandubT}\\
&&\sum_{k=1}^{K}t_k\leq 1,\IEEEyessubnumber\label{tbandconstT}
\end{IEEEeqnarray}
\end{problem}
where the decision parameter $(t_k)_k$ consists of the duty cycle $0\leq t_k \leq \min(1,\bar{t}_k)$ to be allocated to the $k$th user for $k=1,2,\cdots,K$, and their sum cannot exceed the unity.
The objective function is the sum of the AWGN channel capacities of the users scaled by the unitless duty cycles, so that the unit of the sum rate is again [bits/second].
% Problem 1과 댓구.

It is straightforward to see that {Problem~\ref{prob: TDMA}} can be converted to {Problem~\ref{prob: FDMA}} and vice versa, by substituting $t_kw_{\text{tot}}$ and $\bar{t}_kw_{\text{tot}}$ with $w_k$ and $\bar{w}_k$, respectively, for all $k$, which implies
\begin{equation}\label{eq: TDMA FDMA}
\mathcal{C}_\text{TDMA}(\underline{p},\underline{\bar{t}},w_{\text{tot}},N_0)=\mathcal{C}_\text{FDMA}(\underline{p},\underline{\bar{t}}w_\text{tot},w_\text{tot},N_0).
\end{equation}
Thus, in the rest of this section, we focus mostly on characterizing the maximum sum rate of the restricted FDMA system because the solution to {Problem~\ref{prob: TDMA}} can then be readily obtained from that to {Problem~\ref{prob: FDMA}} by the simple substitutions.

Similar to the FDMA case in the previous subsection, if an unrestricted TDMA system is considered instead, i.e., if the constraint (\ref{bandubT}) of {Problem~\ref{prob: TDMA}} is replaced by $0\leq t_k,\forall k$, the unique optimal solution allocates
\begin{equation}
\frac{p_k}{\sum_{k'=1}^{K}p_{k'}}\label{proportionalT}
\end{equation}
to the $k$th user as its duty cycle.

\begin{definition}
The optimal solution to the unrestricted TDMA problem is called the proportional-share duty-cycle allocation scheme.
\end{definition}

By this equivalence, the proportional-share duty-cycle allocation scheme uses up all available transmission time and allocates the $k$th user the optimal duty cycle that is proportional to its signal power $p_k$.
Thus, the maximum sum rate of the unrestricted TDMA system becomes exactly the same as (\ref{eq: MAC}).
Note that, under this proportional-share duty-cycle allocation, the TDMA becomes one of the optimal multiple-access schemes that achieve the sum capacity of the MAC.
Note also that the optimal PSD of the $k$th user in its transmission period becomes flat and is the same as (\ref{eq: PSD}).
Thus, it immediately follows that the sum capacity of the MAC is achieved by the TDMA system if and only if the PSD of the overall received signal is flat over the entire transmission periods of users.\footnote{Since the system is assumed band-limited, this argument \emph{approximately} holds if the transmission period of each user is chosen sufficiently large.}

%=================================================================
\subsection{Optimal Solution to Restricted FDMA System}
%=================================================================

In this subsection, we present the optimal solution to {Problem~\ref{prob: FDMA}} in a closed algorithmic expression with three steps.
The derivation of the optimal solution to {Problem~\ref{prob: TDMA}} is then straightforward by the simple substitutions of $w_k$ and $\bar{w}_k$ with $t_kw_{\text{tot}}$ and $\bar{t}_kw_{\text{tot}}$, respectively, for all $k$.
Consider the PSD $p_k/w_k$ of the $k$th user's signal in the FDMA system, along with the following definition.
\begin{definition}
The quantity $p_k/\bar{w}_k$ is called the minimal PSD of the $k$th user.
\end{definition}

The first step of the three-step algorithm to construct the optimal solution is to reorder the users in a non-increasing order of their minimal PSDs $(p_k/\bar{w}_k)_k$.
So, in what follows, we assume that the users are already numbered as
\begin{equation}
\frac{p_1}{\bar{w}_1}\geq \frac{p_2}{\bar{w}_2} \geq \cdots \geq \frac{p_K}{\bar{w}_K}.\label{powerordered}
\end{equation}
Note that, due to the bandwidth constraint (\ref{bandub}), the PSD of the $k$th user is lower bounded by its minimal PSD.

The second step begins with classifying the users as follows.
The physical meaning of this classification rule will be provided in the next subsection.

\begin{definition}\label{def: classification}
The $k$th user is tested by the rule
\begin{IEEEeqnarray}{C}
\hat{w}_k\triangleq\left(w_\text{tot}-\sum_{k'=1}^{k-1}\bar{w}_{k'}\right)\frac{p_k}{\sum_{k'=k}^{K}p_{k'}}\gtreqqless \bar{w}_k,\label{eq: classification rule}\\
\noalign{\noindent{\text{and classified as}}\vspace{\jot}}%
%\left\{
\begin{array}{rll}
\text{(a)}&\!\!\text{an oversized FDMA user}, & \text{if } \hat{w}_k>\bar{w}_k,\\
\text{(b)}&\!\!\text{a critically-sized FDMA user}, & \text{if } \hat{w}_k=\bar{w}_k, \text{ and}\qquad\\
\text{(c)}&\!\!\text{an undersized FDMA user}, & \text{if } \hat{w}_k<\bar{w}_k.
\end{array}\label{eq: classification}%\right.
\end{IEEEeqnarray}
\end{definition}

Note that this user classification is not affected by the noise level $N_0$, similar to the fact that the user classification for the overloaded single-code CDMA system in \cite{Viswanath : optimal0} is not affected by the noise variance but only by the processing gain and the power profile of users.

In \cite{Viswanath : optimal0} and \cite{Viswanath : optimal}, the term oversized user is introduced for the single-code CDMA system, but no further classification is made for non-oversized users.
It will be shown that the above definition of the oversized FDMA user naturally extends that in \cite{Viswanath : optimal0} and \cite{Viswanath : optimal} to the multi-code CDMA system.
It will be also shown that the further classification in (\ref{eq: classification}) of the non-oversized users into critically-sized and undersized users turns out to be very useful in finding the maximum number of orthogonal signature sequences and the minimum number of signature sequences in the sum-rate optimal multi-code CDMA system.

% 의미가 무엇인지에 대한 호기심 자극, 기존의 것과의 공통점 차이점 예고.

Here follows a simple consequence of {Definition~\ref{def: classification}}.

\begin{lemma}\label{lemma: kappa}
There exists a unique integer ${K_1}\in\{0,1,\cdots,K\}$ such that every user with index $k\leq {K_1}$ is oversized, while every user with index $k>{K_1}$ is non-oversized, i.e.,
\begin{IEEEeqnarray}{rCl}\label{eq: kapp}
\hat{w}_k>\bar{w}_k, & \text{ for } & 1\leq k\leq{K_1},\IEEEyessubnumber\label{eq: kappa}\\
\noalign{\noindent{\text{and}}\vspace{\jot}}
\hat{w}_k\leq\bar{w}_k,& \text{ for } &{K_1}<k\leq K.\IEEEyessubnumber\label{eq: kappa2}
\end{IEEEeqnarray}
\end{lemma}

\begin{IEEEproof}
See Appendix~B.
\end{IEEEproof}

Thus, ${K_1}$ is the number of oversized users in the system.
The final third step to construct the optimal solution is to allocate the system bandwidth to the users.

\begin{proposition}\label{proposition: FDMAsol}
Given a power profile $\underline{p}$, a bandwidth-constraint profile $\underline{\bar{w}}$, and a total system bandwidth $w_\text{tot}$, the optimal solution $({w_k^*})_k$ to {Problem~\ref{prob: FDMA}} is given by
\begin{equation}\label{eq: optband}
{w_k^*}=\left\{
\begin{array}{ll}
\!\!\bar{w}_k,& \text{for }1\leq k\leq {K_1},\\
\!\!\!\!\left(w_\text{tot}-\!\!\displaystyle\sum_{k'=1}^{{K_1}}\bar{w}_{k'}\right)\!\displaystyle\frac{ p_{k}}{\sum_{k'={K_1}+1}^{K}p_{k'}},& \text{for }{K_1}<k\leq K.
\end{array}
\right.
\end{equation}
Consequently, the maximum sum rate of the restricted FDMA system becomes
\begin{IEEEeqnarray}{ll}\label{eq: FDMAmaxsumrate}
\mathcal{C}_\text{FDMA}(\underline{p},\underline{\bar{w}},w_\text{tot},N_0)= \sum_{k=1}^{{K_1}}\bar{w}_k\log\left(1+\frac{p_k}{N_0\bar{w}_k}\right)\nonumber\\
\;\;\,+\left(w_\text{tot}-\sum_{k=1}^{{K_1}}\bar{w}_k\right)\log\left(1+\frac{\sum_{k={K_1}+1}^{K}p_k}{N_0(w_\text{tot}-\sum_{k=1}^{{K_1}}\bar{w}_k)}\right).\IEEEeqnarraynumspace
\end{IEEEeqnarray}
\end{proposition}

\begin{IEEEproof}
See Appendix~C.
\end{IEEEproof}

The proof in the appendix just verifies that the presented three-step algorithm indeed generates an optimal solution, without showing how it is derived.
A detailed derivation can be found in \cite{Yun: thesis}, which expands that in \cite{Cho: ITA10} of the sum-rate optimal restricted FDMA system with an equal upper limit on $w_k,\forall k$.

%=================================================================
\subsection{Remarks on Optimal Solution}\label{sec: properties}
%=================================================================

In this subsection, we make some remarks on the optimal solution (\ref{eq: optband}) and the maximum sum rate (\ref{eq: FDMAmaxsumrate}).
In particular, an iterative algorithm is presented, which alternatively constructs the optimal solution by repeatedly utilizing the proportional-share bandwidth allocation scheme.
The properties of the user PSDs of the optimal solution are also examined, which turn out to be useful in finding the necessary and sufficient condition to achieve the sum capacity of the MAC and later in obtaining insights into the optimal solutions to the multi-code CDMA problem.
We also extend the restricted FDMA problem to allow zero-power inactive users, which is needed in the next section.
Again, the results in this subsection can also be applied to the restricted TDMA problem after the simple substitutions.

\begin{remark}\label{remark: 1a}
The optimal solution to {Problem~\ref{prob: FDMA}} can also be constructed by using the iterative algorithm presented in TABLE~\ref{table: 1}.
\end{remark}

\begin{table}%[tbp]
\caption{An Iterative Algorithm to Construct the Optimal Solution to the Restricted FDMA Problem}\label{table: 1}\vspace*{-0.1in}
\hrulefill\\
\begin{itemize}
\item[1:] \textbf{REPEAT}
\item[2:] Renumber the users in the user list in a non-increasing order of the minimal PSDs
\item[3:] Compute the users' due shares of the bandwidths under the proportional-share allocation \item[4:] \begin{itemize} \item[]\textbf{IF} the first user has the due share greater than its bandwidth upper limit\end{itemize}
\item[5:] \begin{itemize} \item[]\textbf{THEN} allocate the bandwidth upper limit to the first user, subtract the upper limit from the total system bandwidth, and remove the first user from the user list\end{itemize}
\item[6:] \begin{itemize} \item[]\textbf{ELSE} allocate the due shares to the users, and remove all the users from the user list\end{itemize}
\item[7:] \textbf{UNTIL} no user is left in the user list
\end{itemize}
\hrulefill
\end{table}

\begin{IEEEproof}
The following is a case-by-case verification that the iterative algorithm in TABLE~\ref{table: 1} indeed constructs the optimal solution in (\ref{eq: optband}).
There are three cases to consider.

If ${K_1}=0$, then $\hat{w}_1 \leq\bar{w}_1$ by {Definition~\ref{def: classification}}.
Since $\hat{w}_1$ is equal to the due share\footnote{In what follows, `due share' and `due rate' are those computed under the proportional-share allocation.} $w_\text{tot}p_1/\sum_{k=1}^{K}p_{k}$ to the first user, the iterative algorithm allocates the due shares to all the users and removes them from the user list by the line $6$ of TABLE~\ref{table: 1}.
Then, it terminates because the condition in the line $7$ is satisfied.
It can be verified that the resultant allocation becomes identical to the optimal allocation described by {Proposition~\ref{proposition: FDMAsol}} for ${K_1}=0$.

If $1\leq {K_1} <K$, then $\hat{w}_k>\bar{w}_k, \forall k\leq{K_1}$.
Since $\hat{w}_1$ is equal to the due share to the first user, the iterative algorithm allocates the bandwidth upper limit to the first user and removes it from the user list by the line $5$ of TABLE~\ref{table: 1}.
Then, it goes back to the line $2$ because the condition in the line $7$ is not yet satisfied.
Now, suppose that the above procedure is repeated $k(<{K_1})$ times.
Similarly, it can be shown that the iterative algorithm allocates the bandwidth upper limit to the renumbered first user, removes this first user from the user list, and goes back to the line $2$.
Suppose that the above procedure is repeated ${K_1}$ times.
Since $\hat{w}_{{K_1}+1}$ of the original $({K_1}+1)$th user is not greater than the user's bandwidth upper limit by assumption, the algorithm performs the proportional-share bandwidth allocation to all the users remaining in the user list, removes them from the user list, and terminates.
It can be verified again that the resultant allocation becomes identical to the optimal allocation for $1\leq {K_1} <K$.

If ${K_1}=K$, then $\hat{w}_k>\bar{w}_k,\forall k$.
Similar iterations are conducted to the case with $1\leq {K_1} <K$, but now until the $K$th user is eventually allocated its bandwidth upper limit.
Then, it terminates because there is no user left.
Once again, it can be easily verified that the resultant allocation becomes identical to the optimal allocation for ${K_1} =K$.
Therefore, the conclusion follows.
\end{IEEEproof}

By the above equivalence between the closed algorithmic solution presented in the previous subsection and the iterative algorithm in Table~\ref{table: 1}, both of which lead to (\ref{eq: optband}), we can freely choose whichever convenient in what follows.

The proof of {Remark~\ref{remark: 1a}} shows that if the iterative algorithm terminates in $k<K$ iterations then ${K_1}=k-1$.
If it terminates in $K$ iterations and the only one user in the user list has the due share computed in the line 4 of the $K$th iteration no greater than its bandwidth upper limit, then ${K_1}=K-1$, otherwise ${K_1}=K$.
Thus, we can find the number ${K_1}$ of oversized users from the iterative algorithm that iterates $\min({K_1}+1,K)$ times to construct the optimal bandwidths of the users.

As shown in the proof, for $k\leq {K_1}+1$, the $k$th user as the renumbered first user has the due share computed in the line $3$ of the $k$th iteration equal to $\hat{w}_k$, which is used in the classification rule (\ref{eq: classification}).
Thus, by {Definition~\ref{def: classification}}, it turns out that the line $4$ of TABLE~\ref{table: 1} actually tests whether the renumbered first user is oversized or not, i.e., whether $\hat{w}_k>\bar{w}_k$ or not.
If so, then by the line $5$ the user is allocated its bandwidth upper limit $\bar{w}_k$ that is less than its due share $\hat{w}_k$.
This justifies why we name the oversized users as such by precisely explaining what the physical meaning of $\hat{w}_k$ in (\ref{eq: classification rule}) is for the oversized users.

By the line $6$ of TABLE~\ref{table: 1}, non-oversized users are allocated the bandwidths that are equal to their due shares computed in the line $3$ of the $({K_1}+1)$th iteration.
Despite this commonality of the non-oversized users, we further classify them into the critically-sized and the undersized users as done in {Definition~\ref{def: classification}}.
To see why, we alter {Lemma~\ref{lemma: kappa}} and introduce the number ${K_2}$ of non-undersized users in the system.

\begin{lemma}\label{lemma: kappa'}
There exists a unique integer ${K_2}\in\{0,1,\cdots,K\}$ such that all the users with index $k\leq {K_2}$ are non-undersized, while all the other users with index $k>{K_2}$ are undersized, i.e.,
\begin{IEEEeqnarray}{rCl}\label{eq: kapp'}
\hat{w}_k\geq \bar{w}_k, & \text{ for } & 1\leq k\leq{K_2},\IEEEyessubnumber\label{eq: kappa'}\\
\noalign{\noindent{\text{and}}\vspace{\jot}}
\hat{w}_k<\bar{w}_k,& \text{ for } &{K_2}<k\leq K.\IEEEyessubnumber\label{eq: kappa'2}
\end{IEEEeqnarray}
\end{lemma}

\begin{IEEEproof}
Omitted. It can be proved in almost the same way as {Lemma~\ref{lemma: kappa}} is proved.
\end{IEEEproof}

\begin{remark}\label{remark: 2a}
The $k$th user is
\begin{IEEEeqnarray}{rll}
\text{(a) }& \text{oversized},& \text{ for $1\leq k \leq {K_1}$,} \nonumber\\
\text{(b) }& \text{critically-sized},& \text{ for ${K_1} < k \leq {K_2}$, and} \nonumber\\
\text{(c) }& \text{undersized},& \text{ for ${K_2} < k \leq K$.} \nonumber
\end{IEEEeqnarray}
\end{remark}

\begin{IEEEproof}
{Lemmas~\ref{lemma: kappa}} and {\ref{lemma: kappa'}} show that the unique boundary indexes ${K_1}$ and ${K_2}$ can be identified, respectively, by which the oversized and the non-oversized users are separated and by which the non-undersized and the undersized users are separated.
Since ${K_1}\leq {K_2}$, the conclusion follows immediately.
\end{IEEEproof}

The following remark justifies why we name the critically-sized and the undersized users as such.

\begin{remark}\label{remark: 3a}
The optimal bandwidths of the non-oversized users satisfy the following properties.
\begin{itemize}
\item[(a)] The optimal bandwidth of a critically-sized user, if such a user exists, equals the user's bandwidth upper limit.
\item[(b)] The optimal bandwidth of an undersized user, if such a user exists, is less than the user's bandwidth upper limit.
\end{itemize}
\end{remark}

\begin{IEEEproof}
%See Appendix~E.
To prove {Remark~\ref{remark: 3a}}-(a), assume that at least one critically-sized user exists.
Then, by {Definition~\ref{def: classification}} and {Remark~\ref{remark: 2a}}, the $({K_1}+1)$th user becomes a critically-sized user that satisfies $\hat{w}_{{K_1}+1}=\bar{w}_{{K_1}+1}$.
Also, by {Definition~\ref{def: classification}} and (\ref{eq: optband}), we have $w_{{K_1}+1}^*=\hat{w}_{{K_1}+1}$.
Thus, $w_{{K_1}+1}^*=\bar{w}_{{K_1}+1}$.
For the situations with more than one critically-sized users, since the optimal bandwidths of the non-oversized users follow the proportional-share allocation rule, it can be shown that {Remark~\ref{remark: 3a}}-(a) is still true by applying a form of the algebraic method of componendo and dividendo\footnote{In this paper, we use only the form $a/b=c/d \Rightarrow (a+c)/(b+d)=(a-c)/(b-d)=a/b$, for $b\ne 0, d\ne 0, b+d\ne 0$, and $b-d\ne 0$.} to {Definition~\ref{def: classification}} and (\ref{eq: optband}) for ${K_1} < k\leq {K_2}$.
To prove {Remark~\ref{remark: 3a}}-(b), assume that at least one undersized user exists.
Then, by {Definition~\ref{def: classification}} and {Remark~\ref{remark: 2a}}, the $({K_2}+1)$th user becomes an undersized user that satisfies $\hat{w}_{{K_2}+1}<\bar{w}_{{K_2}+1}$.
Also, by {Definition~\ref{def: classification}} and (\ref{eq: optband}), we have $w_{{K_2}+1}^*=\hat{w}_{{K_2}+1}$.
Thus, $w_{{K_2}+1}^*<\bar{w}_{{K_2}+1}$.
For the situations with more than one undersized users, let $s$ be the common PSD of the non-oversized users.
Then, by (\ref{powerordered}), (\ref{eq: optband}), and the result $w_{{K_2}+1}^*<\bar{w}_{{K_2}+1}$ just derived, we have ${w_k^*}=p_k/s=p_k/(p_{{K_2}+1}/w_{{K_2}+1}^*)<p_k/(p_{{K_2}+1}/\bar{w}_{{K_2}+1})\leq p_k/(p_k/\bar{w}_k)=\bar{w}_k, \forall k\geq {K_2}$.
Therefore, the conclusion follows.
\end{IEEEproof}

{Remark~\ref{remark: 3a}}-(a) says that not only an oversized user but also a critically-sized user has the optimal bandwidth equal to its bandwidth upper limit.
Unlike an oversized user, however, the upper limit is now equal to the critically-sized user's due share computed in the line $3$ of the $({K_1}+1)$th iteration after removing all the oversized users from the user list.
This justifies why we name the critically-size users as such.

{Remark~\ref{remark: 3a}}-(b) says that, though the non-oversized users are all allocated the due shares computed in the line $3$ of the $({K_1}+1)$th iteration, an undersized user is different from a critically-sized user in that its optimal bandwidth is less than the upper limit.
This justifies why we name the undersized users as such.

The proof of {Remark~\ref{remark: 3a}}-(b) shows that the test applied to the first non-oversized user performs nothing but the comparison of the upper limit $\bar{w}_{{K_1}+1}$ with the non-oversized user's due share $\hat{w}_{{K_1}+1}$ computed in the line $3$ of the $({K_1}+1)$th iteration.
Moreover, by {Lemma~\ref{lemma: kappa}}, we do not need to check whether a user is non-oversized or not any more once the first non-oversized user is found.
Thus, the physical meaning of $\hat{w}_k$ in (\ref{eq: classification rule}) is also explained for the non-oversized users and shown to be consistent with that for the oversized users.

The proof of {Remark~\ref{remark: 3a}}-(b) also shows that the proportional-share allocation scheme applied to the non-oversized users results in equal PSDs for these users.
The following remark details the properties of the user PSDs of the optimal restricted FDMA system.

\begin{remark}\label{remark: 4a}
The user PSDs of the optimal solution have the following properties.
\begin{itemize}
\item[(a)] The optimal PSDs are equal among the non-oversized users, if they exist.
\item[(b)] The optimal PSDs are non-increasing in the user index.
\item[(c)] The optimal PSD of an oversized user is always greater than those of non-oversized users, if they exist.
\end{itemize}
\end{remark}

\begin{IEEEproof}
%See Appendix~F.
{Remark~\ref{remark: 4a}}-(a) can be immediately seen from {Proposition~\ref{proposition: FDMAsol}}.
If ${K_1}=0$, then the optimal solution (\ref{eq: optband}) reduces to the proportional-share bandwidth allocation scheme.
Since the PSD of the $k$th user satisfies $p_k/{w_k^*}=\sum_{k'=1}^{K}p_{k'}/w_\text{tot}, \forall k$, the PSDs are equal, so that non-increasing.

If $1\leq {K_1} <K$, then the PSDs of the oversized users are given by $p_k/\bar{w}_k,\forall k\leq{K_1}$.
Thus, by the assumption (\ref{powerordered}), the PSDs of the oversized users are non-increasing.
Since the proportional-share bandwidth allocation is performed to the non-oversized users, the PSDs of the non-oversized users are all equal to that of the $({K_1}+1)$th user, i.e.,
$p_k/{w_k^*}=\sum_{k'={K_1}+1}^{K}p_{k'}/(w_\text{tot}-\sum_{k'=1}^{{K_1}}\bar{w}_{k'}), \forall k>{K_1}$.
Moreover, the PSDs of the ${K_1}$th and the $({K_1}+1)$th users satisfy
\begin{IEEEeqnarray}{ll}\label{eq: less}
\frac{p_{{K_1}}}{{w_{{K_1}}^*}}=\frac{p_{K_1}}{\bar{w}_{K_1}}&>\frac{\sum_{k'={K_1}}^{K}p_{k'}}{w_\text{tot}-\sum_{k'=1}^{{K_1}-1}\bar{w}_{k'}}\IEEEyessubnumber\label{eq: 56a}\\
&>\frac{\sum_{k'={K_1}+1}^{K}p_{k'}}{w_\text{tot}-\sum_{k'=1}^{{K_1}}\bar{w}_{k'}}=\frac{p_{{K_1}+1}}{w_{{K_1}+1}^*}\IEEEyessubnumber\label{eq: 56b}
\end{IEEEeqnarray}
where (\ref{eq: 56a}) comes from {Definition~\ref{def: classification}} and the fact that the ${K_1}$th user is oversized, and (\ref{eq: 56b}) comes from the facts that an oversized user is allocated its bandwidth upper limit and that $b/a>d/c$ implies $d/c>(d-b)/(c-a)$ for all real constants $c>a>0,b>0,$ and $d>0$.
Hence, the PSDs are non-increasing.
Also, (\ref{eq: less}) proves {Remark~\ref{remark: 4a}}-(c).

If ${K_1}=K$, then the PSDs of the optimal solution are given by $p_k/\bar{w}_k,\forall k$.
Thus, by the assumption (\ref{powerordered}), the PSDs are non-increasing.
Therefore, {Remark~\ref{remark: 4a}}-(b) is true in all cases.
\end{IEEEproof}

The above remark implies that, if $1\leq{K_1}<K$, then we have
\begin{equation}
\frac{p_1}{w_1^*} \geq \cdots \geq \frac{p_{K_1}}{w_{K_1}^*}> \frac{p_{{K_1}+1}}{w_{{K_1}+1}^*}= \cdots =\frac{p_K}{w_K^*}.
\end{equation}
By {Remarks~\ref{remark: 3a}} and {\ref{remark: 4a}}, we have the following alternative expressions of the optimal bandwidth allocation and the maximum sum rate.
Recall that ${K_2}$, defined in {Lemma~\ref{lemma: kappa'}}, is the number of non-undersized users in the system.

\begin{remark}\label{remark: 5}
The optimal solution $({w_k^*})_k$ in {Proposition~\ref{proposition: FDMAsol}} to {Problem~\ref{prob: FDMA}} can be rewritten as
\begin{equation}\label{optband2}
{w_k^*}=\left\{
\begin{array}{ll}
\!\bar{w}_k,&\!\text{for }1\leq k\leq {K_2},\\
\!\!\!\left(w_\text{tot}-\!\!\displaystyle\sum_{k'=1}^{{K_2}}\bar{w}_{k'}\right)\!\displaystyle\frac{ p_{k}}{\sum_{k'={K_2}+1}^{K}p_{k'}},&\! \text{for }{K_2}<k\leq K.
\end{array}
\right.
\end{equation}
Consequently, the maximum sum rate in (\ref{eq: FDMAmaxsumrate}) can be rewritten as
\begin{IEEEeqnarray}{ll}\label{eq: FDMAmaxsumrate2}
\mathcal{C}_\text{FDMA}(\underline{p},\underline{\bar{w}},w_\text{tot},N_0)= \sum_{k=1}^{{K_2}}\bar{w}_k\log\left(1+\frac{p_k}{N_0\bar{w}_k}\right)\nonumber\\
\;\;\,+\left(w_\text{tot}-\sum_{k=1}^{{K_2}}\bar{w}_k\right)\log\left(1+\frac{\sum_{k={K_2}+1}^{K}p_k}{N_0(w_\text{tot}-\sum_{k=1}^{{K_2}}\bar{w}_k)}\right).\IEEEeqnarraynumspace
\end{IEEEeqnarray}
\end{remark}

\begin{IEEEproof}
By {Remark~\ref{remark: 3a}}-(a), the optimal bandwidth of a critically-sized user, if exists, equals its bandwidth upper limit.
Moreover, by {Remark~\ref{remark: 4a}}-(a), the PSD of a critically-sized user, if exists, equals that of the other non-oversized users.
Thus, again by the method of componendo and dividendo, it can be shown that $(w_\text{tot}-\sum_{k=1}^{{K_1}}\bar{w}_k)/\sum_{k={K_1}+1}^{K}p_k=(w_\text{tot}-\sum_{k=1}^{{K_2}}\bar{w}_k)/\sum_{k={K_2}+1}^{K}p_k$.
Therefore, the conclusion follows.
\end{IEEEproof}

The following remark shows how the imposition of the upper limits affects the bandwidth allocation.

\begin{remark}\label{remark: 5.5}
If there exists a non-oversized user in the optimal restricted FDMA system, then the bandwidth allocated to the non-oversized user is always greater than or equal to its due share of the bandwidth, where the equality holds if and only if there is no oversized user.
\end{remark}
\begin{IEEEproof}
A sketch of the proof is as follows.
Since there exists at least one non-oversized user, we have $0\leq {K_1} <K$, which is divided into two cases of $K_1=0$ and $1\leq {K_1} < K$.

If ${K_1}=0$, then the algorithm in TABLE~\ref{table: 1} terminates in one iteration and the proportional-share allocation becomes optimal.
Thus, the statement is trivially true.

If $1\leq {K_1} < K$, then the first user is oversized and allocated a smaller bandwidth than its due share, i.e., $w_1^*=\bar{w}_1<\hat{w}_1$, by the line $5$ of the first iteration.
Since $\sum_{k=1}^{K}w_k^*=w_\text{tot}$, it leaves the remaining users in the user list a larger bandwidth $w_{\text{tot}}-w_1^*$ than the sum of their due shares $w_\text{tot}-\hat{w}_1$.
So, in the line $3$ of the second iteration, the computed due share of the bandwidth of each user in the user list becomes larger than the due share computed in the line $3$ of the first iteration.
There are two sub-cases to consider.
For ${K_1}=1$, the algorithm terminates after assigning the due shares to the users.
Thus, the statement is true.
For ${K_1}>1$, the first user in the user list is allocated a smaller bandwidth than the due share computed in the line $3$ of the second iteration and leaves the remaining users in the user list a larger bandwidth than the due share computed in the line $3$ of the second iteration.
In this way, the bandwidth left to the non-oversized users is always increasing as the iterations go on.
We keep continuing this iteration until the algorithm terminates.
Therefore, the conclusion follows.
\end{IEEEproof}

The user PSDs also help us visually and, hence, more straightforwardly understand the above remark.
A related numerical example can be found in Section~\ref{sec: discussion}.

From {Remark~\ref{remark: 5.5}}, we have the following consequences, which will be used later to explain the properties of the sum-rate optimal multi-code CDMA system.
\begin{remark}\label{remark: 5.6}
$\;$
\begin{itemize}
\item[(a)] If there exists a critically-sized user in the optimal restricted FDMA system, then its rate is always greater than or equal to its due rate, where the equality holds if and only if there is no oversized user.
\item[(b)] If there exist undersized users in the optimal restricted FDMA system, then the undersized users' sum rate is always greater than or equal to the sum of their due rates, where the equality holds if and only if there is no oversized user.
\item[(c)] If there exists an oversized user in the optimal restricted FDMA system, then the oversized users' sum rate is always less than the sum of their due rates.
\item[(d)] If there exists an oversized user in the optimal restricted FDMA system, then the first oversized user's rate is always less than its due rate. However, it is not necessarily true for other oversized users.
\end{itemize}
\end{remark}

\begin{IEEEproof}
Note that $w_k\log(1+p_k/(N_0w_k))$ is a monotone increasing function of $w_k$ and that a critically-size user is also a non-oversized user.
Thus, by {Remark~\ref{remark: 5.5}}, we have {Remark~\ref{remark: 5.6}}-(a).
Similarly, we can say the same as {Remark~\ref{remark: 5.6}}-(a) for each undersized user, which directly implies that the sum of the optimal bandwidths allocated to the undersized users is always greater than or equal to the sum of their due shares, where the equality holds if and only if there is no oversized user.
Note that the sum rate of the undersized users is the second term in the right side of (\ref{eq: FDMAmaxsumrate2}).
Thus, we have {Remark~\ref{remark: 5.6}}-(b) because $(\sum_{k={K_2}+1}^{K}w_k)\log(1+(\sum_{k={K_2}+1}^{K}p_k)/(N_0\sum_{k={K_2}+1}^{K}w_k))$ is a  monotone increasing function of $(\sum_{k={K_2}+1}^{K}w_k)$.
When there exists an oversized user, the sum of the optimal bandwidths allocated to the oversized users is always less than the sum of their due shares because the sum of the optimal bandwidths allocated to the non-oversized users is always greater than the sum of their due shares.
Thus, we have {Remark~\ref{remark: 5.6}}-(c) because a smaller bandwidth implies a lower rate even if the proportional-share allocation scheme is applied among the oversized users.
The first sentence of {Remark~\ref{remark: 5.6}}-(d) is straightforward from the algorithm in TABLE~\ref{table: 1}.
To show that this does not hold for the other oversized users, a counter example is enough, which can be found in Section~\ref{sec: discussion}.
Thus, we have the second sentence of {Remark~\ref{remark: 5.6}}-(d).
\end{IEEEproof}

{Remarks~\ref{remark: 5.5} and \ref{remark: 5.6}}-(c) show that the imposition of, for example, equal upper limits mitigates the unfairness inherent in the sum-rate maximization by reducing the sum rate of the strong-power oversized users, while increasing the sum rate of the non-oversized users.
However, {Remark~\ref{remark: 5.6}}-(d) emphasizes that an oversized $k (>1)$th user is not always allocated a smaller bandwidth than its due share computed in the line $3$ of the first iteration.
Rather, it is allocated a smaller bandwidth than its due share computed in the line $3$ of the $k$th iteration.

The next remark is on the necessary and sufficient condition for the optimal restricted FDMA system to achieve the sum capacity of the MAC.

\begin{remark}\label{remark: 5a}
The maximum sum rate (\ref{eq: FDMAmaxsumrate}) of the restricted FDMA system is upper bounded by the sum capacity of the MAC, i.e.,
\begin{subequations}
\begin{equation}\label{eq: CFDMA CMAC}
\mathcal{C}_\text{FDMA}(\underline{p},\underline{\bar{w}},w_\text{tot},N_0)\leq \mathcal{C}_{\text{MAC}}(\underline{p},w_\text{tot},N_0),
\end{equation}
where equality holds if and only if there is no oversized user, i.e., $K_1=0$ or, equivalently,
\begin{equation}\label{eq: nece suffi}
w_{\text{tot}}\cdot\left(\frac{p_1}{\sum_{k=1}^Kp_k}\right)\leq \bar{w}_1.
\end{equation}
\end{subequations}
\end{remark}

\begin{IEEEproof}
The constraint set of the restricted FDMA problem is a subset of that of the unrestricted FDMA problem.
In addition, the unrestricted FDMA as one of the various multiple-access schemes has its sum rate upper bounded by the sum capacity of the MAC.
Thus, we have (\ref{eq: CFDMA CMAC}).
It is well known \cite{Cover} that the FDMA system achieves the sum capacity if and only if the bandwidths are allocated to all users by applying the proportional-share allocation scheme.
Moreover, TABLE~\ref{table: 1} shows that this occurs if and only if the first user is non-oversized, i.e., (\ref{eq: nece suffi}) holds by {Definition~\ref{def: classification}}.
By {Lemma~\ref{lemma: kappa}}, there is no oversized user because the first user is not oversized.
Therefore, the conclusion follows.
\end{IEEEproof}

As mentioned earlier in this section, the proportional-share bandwidth allocation is equivalent to assigning a flat PSD for each user and equal PSDs for all users, which evenly loads all the frequency components in the system frequency band.
Thus, (\ref{eq: nece suffi}) is also the necessary and sufficient condition for the equal PSDs of users to result in not only a feasible but also the optimal solution to the sum-rate maximization problem for the restricted FDMA system.

Although no direct link has been yet established between the definitions of the oversized users for  the FDMA and for the single-code CDMA systems, a similar argument can be found in \cite{Viswanath : optimal0}, which says that the optimal overloaded single-code CDMA system achieves the sum capacity of the MAC if and only if there is no oversized user.
In the next section, it will be shown that the above remark not only verbally parallels the argument in \cite{Viswanath : optimal0} but also successfully  generalizes it to the multi-code CDMA system.

Until now, we have been assuming that all users are active and have positive powers, i.e., $p_k>0,\forall k$.
The following rather obvious remark, which is presented for the completeness of the argument, shows how to extend the results in this section to allow zero-power inactive users.
This result along with the convention $w_k \log(1+p_k/(N_0w_k))=0$ at $w_k=0$, for all $p_k\geq 0$, in (\ref{convention}) will be used in the next section, where we return to the sum-rate maximization problem for the multi-code CDMA system.
To avoid a vacuous situation, it is assumed that at least one user has a positive power.

\begin{remark}\label{remark: 7a}
Suppose that there are $K'$ users but only $K <K'$ users are active.
Then, ${w_k^*}$ is exactly the same as (\ref{eq: optband}) or, equivalently, (\ref{optband2}) for $k\leq K$, while it can be any value satisfying
\begin{equation}\label{eq: zero power}
\sum_{k=K+1}^{K'}{w_k^*}\leq w_{\text{tot}}-\sum_{k=1}^{K}{w_k^*}, \text{ and } 0\leq {w_k^*}\leq \bar{w}_k,
\end{equation}
for $k>K$.
\end{remark}

\begin{IEEEproof}
Since the users are numbered in a non-increasing order of the minimal PSDs, we have $p_k>0, \forall k\leq K$, and $p_k=0, \forall k>K$.
Note that a zero-power inactive user does not increase the sum rate even when a positive bandwidth is allocated to the user.
Thus, by simply ignoring the inactive users, we obtain ${w_k^*}$ as (\ref{eq: optband}) or, equivalently, (\ref{optband2}) for $k\leq K$, which results in the maximum sum rate given by (\ref{eq: FDMAmaxsumrate}).
If $\sum_{k=1}^{K}{w_k^*}<w_{\text{tot}}$, then the inactive users have the remaining bandwidth $w_{\text{tot}}-\sum_{k=1}^{K}{w_k^*}>0$ at their disposal.
Otherwise, no inactive user can be allocated a positive bandwidth at all.
Therefore, we have (\ref{eq: zero power}).
\end{IEEEproof}

By the simple substitutions of $w_k$ and $\bar{w}_k$ with  $t_kw_{\text{tot}}$ and $\bar{t}_kw_{\text{tot}}$, respectively, for all $k$, all the frequency-domain arguments in this subsection can be straightforwardly converted to the time-domain arguments for the restricted TDMA system.
It turns out in the next section that this equivalence between the optimal bandwidth allocation problem for the restricted FDMA system and the optimal duty-cycle allocation problem for the restricted TDMA system also extends to the optimal power distribution problem for the multi-code CDMA system.

%=-=-=-=-=-=-=-=-=-=-=-=-=-=-=-=-=-=-=-=-=-=-=-=-=-=-=-=-=-=-=-=-=-=-=-=-=-=-=-=-=-=-
\section{Maximum Sum Rate of Multi-Code CDMA Systems}\label{sec: CDMA}
%=-=-=-=-=-=-=-=-=-=-=-=-=-=-=-=-=-=-=-=-=-=-=-=-=-=-=-=-=-=-=-=-=-=-=-=-=-=-=-=-=-=-
In this section, we completely characterize the maximum sum rate achievable by the multi-code CDMA system considered in Section~\ref{sec: sig sys}.
Given an upper limit on $n_k$ for each $k$, we not only derive the maximum sum rate along with the jointly optimal power distribution and sequence design, but also find the maximum number of orthogonal signature sequences and the minimum number of signature sequences both with the maximum sum rate being still retained.
To do so, first, a rather unconventional approach of the replace-and-switch method is applied to {Problem~\ref{problem: CDMA outer inner}}.
Then, the notions of FDMA-equivalent bandwidth and TDMA-equivalent duty-cycle are introduced to establish the equivalence of the optimal multi-code CDMA system to the optimal restricted FDMA and TDMA systems.
We also extend the results to the symbol-asynchronous but chip-synchronous multi-code CDMA system.
Throughout this section, we assume that the users are numbered in a non-increasing order of $(p_k/\bar{n}_k)_k$, i.e.,
\begin{equation}
\frac{p_1}{\bar{n}_1}\geq \frac{p_2}{\bar{n}_2} \geq \cdots \geq \frac{p_K}{\bar{n}_K}.\label{CDMApowerordered}
\end{equation}

%----------------------------------------------------------------------
\subsection{Replace-and-Switch Method}\label{sec: CDMA-1}
%----------------------------------------------------------------------

Recall {Problem~\ref{problem: CDMA outer inner}} having the outer optimization over $P$ and the inner optimization over $S$, which is one of two double-maximization forms of {Problem~\ref{prob: CDMA 2}}.
It is already discussed at the end of Section~\ref{sec: sig sys} that, though this form has a well-known inner optimization problem, the outer optimization problem is not mathematically tractable.
The alternative double-maximization form having the outer optimization over $S$ and the inner one over $P$ may also be considered.
However, it can be easily seen that this form does not even result in a simple or a known optimization problem for the inner optimization.

Instead of taking such direct double-maximization approaches, we take an indirect approach of replacing the inner optimization of {Problem~\ref{problem: CDMA outer inner}} with an equivalent one and then switching the order of the inner and the outer optimizations in the hope that it may convert the problem into a solvable one.
This use of the replace-and-switch method is motivated by its success in solving the total power minimization problem in \cite{Cho: IT4}, where the jointly optimal power allocation and signature waveform design are derived for a continuous-time band-limited single-code CDMA system that meets the asymmetric SINR requirements of users at the output of the linear minimum mean-squared error (LMMSE) receivers.

To proceed, we define the signum function as

\begin{equation}
\text{sgn}(x)\triangleq \left\{
\begin{array}{rl}
1, &\text{ for } x>0,\\%
0, &\text{ for } x=0, \text{ and}\\%
-1,&\text{ elsewhere}.
\end{array}
\right.
\end{equation}
As seen below, this function is used just to count the positive numbers among non-negative numbers, though defined also for a negative argument.

\begin{lemma}\label{lemma: replace}
Given a feasible power matrix $P$, the inner optimization problem of {Problem~\ref{problem: CDMA outer inner}} over $S$ results in the same sum rate as {Problem~\ref{prob: FDMA}} does for the restricted FDMA system, where the FDMA system has the total number $\sum_{k=1}^K{\bar{n}_k}$ of positive-power active and zero-power inactive FDMA users, the power $p_{k,l}$ of the $(\sum_{k'=1}^{k-1}\bar{n}_k+l)$th user, one-sided PSD $2\sigma^2$ of the AWGN, the equal bandwidth upper limit $1/(2N)$, and the total system bandwidth $1/2$.
\end{lemma}

\begin{IEEEproof}
Given $P=\text{diag}(p_{1,1},p_{1,2},\cdots,p_{K,\bar{n}_K})$ satisfying $p_{k,l}\geq 0, \forall k, \forall l,$ and $\sum_{l=1}^{\bar{n}_k}p_{k,l}=p_k, \forall k$, the inner optimization problem in (\ref{eq: inner}) can be viewed as the sequence design problem for a single-code CDMA system considered in \cite{Viswanath : optimal0}, now having the number of active users equal to $\sum_{k=1}^K \sum_{l=1}^{\bar{n}_k}\text{sgn}(p_{k,l})$.
Suppose that this number is not greater than the processing gain, i.e., $\sum_{k=1}^K \sum_{l=1}^{\bar{n}_k}\text{sgn}(p_{k,l})\leq N$.
As shown in \cite{Viswanath : optimal0}, a complete orthogonalization of the signature sequences is possible in this case of a non-overloaded CDMA system with virtual single-code users and, consequently, the maximum sum rate is given by $(1/2N)\sum_{k=1}^{K}\sum_{l=1}^{\bar{n}_k}\log(1+Np_{k,l}/\sigma^2)$.
Note that the corresponding restricted FDMA system with parameters specified in this lemma has the sum of the bandwidth upper limits of the active FDMA users no greater than the total system bandwidth, i.e., $\sum_{k=1}^K \sum_{l=1}^{\bar{n}_k}\text{sgn}(p_{k,l})(1/2N)\leq 1/2$.
Thus, every active FDMA user is non-undersized by {Definition~\ref{def: classification}} and the maximum sum rate becomes equal to that of this single-code CDMA system.

Now, suppose that $\sum_{k=1}^K \sum_{l=1}^{\bar{n}_k}\text{sgn}(p_{k,l})> N$, i.e., the CDMA system with virtual single-code users is overloaded.
As shown in \cite{Viswanath : optimal0}, we first need to identify the oversized single-code users in this case to compute the maximum sum rate.
Note that the direct comparison of {Definition~\ref{def: classification}} with the definition in \cite[Eq. (6)]{Viswanath : optimal0} reveals that the corresponding restricted FDMA system with parameters specified in this lemma has the oversized user classification rule in common with the single-code CDMA system.
In addition, the direct comparison of (\ref{eq: FDMAmaxsumrate}) with the maximum sum rate in \cite[Eq. (7)]{Viswanath : optimal0} shows that the corresponding restricted FDMA system has the maximum sum rate in common with the single-code CDMA system.
Therefore, the conclusion follows.
\end{IEEEproof}

{Lemma~\ref{lemma: replace}} can be straightforwardly converted to the equivalence of the single-code CDMA system to the restricted TDMA system having equal upper limits $1/N$ on the duty cycles of $\sum_{k=1}^K \sum_{l=1}^{\bar{n}_k}\text{sgn}(p_{k,l})$ users.
This equivalence of the single-code CDMA system to the restricted FDMA and TDMA systems will be further extended in the next subsections to the equivalence of the multi-code CDMA system to the restricted FDMA and TDMA systems possibly having unequal upper limits on the bandwidths and the duty cycles of users, respectively.
Note that, since some $p_{k,l}$ of a feasible $P$ of the inner optimization in (\ref{eq: inner}) may be zero, it is now clear why the sum-rate maximization for the restricted FDMA and TDMA systems is extended at the end of Section~\ref{sec: FDMA TDMA} to include both active and inactive users.

In the proof of {Lemma~\ref{lemma: replace}}, we considered the CDMA system with number of active virtual single-code users equal to $\sum_{k=1}^K \sum_{l=1}^{\bar{n}_k}\text{sgn}(p_{k,l})$.
These virtual single-code users can be classified as follows.

\begin{definition}\label{def: virtual single-code}
The active virtual single-code users of a multi-code CDMA system are classified into oversized, critically-sized, or undersized virtual single-code users, if the corresponding users of the equivalent restricted FDMA system with parameters specified in {Lemma~\ref{lemma: replace}} are classified as such by {Definition~\ref{def: classification}}.
The non-oversized and the non-undersized virtual single-code users are also similarly defined among the active virtual single-code users.
\end{definition}

Note, as mentioned in the proof of {Lemma~\ref{lemma: replace}}, that this definition of the oversized virtual single-code user coincides with that of the oversized single-code user in \cite{Viswanath : optimal0}.
Now, {Lemma~\ref{lemma: replace}} allows us to replace the inner optimization problem of {Problem~\ref{problem: CDMA outer inner}} with the sum-rate maximization problem for the restricted FDMA system having the upper limit $1/(2N)$ on the bandwidth $w_{k,l}$ of the $(\sum_{k'=1}^{k-1}\bar{n}_k+l)$th user as

\begin{problem}\label{problem: replaced}
\begin{IEEEeqnarray}{ll}
\underset{(p_{k,l})_{k,l}}{\text{maximize}}&\left\{
\begin{array}{ll}
\underset{(w_{k,l})_{k,l}}{\text{max}}&\!\!\!\!\!\!\displaystyle\sum_{k=1}^{K}\sum_{l=1}^{\bar{n}_k}w_{k,l}\log\left(1+\frac{p_{k,l}}{2\sigma^2w_{k,l}}\right)\\
\text{subject to}\;\;&\!\!\!\!\!\!0\leq w_{k,l}\leq \displaystyle\frac{1}{2N},\forall k,\forall l, \text{ and}\\
&\!\!\!\!\!\!\displaystyle\sum_{k=1}^{K}\sum_{l=1}^{\bar{n}_k}w_{k,l}\leq \frac{1}{2},
\end{array}
\right.\nonumber\vspace{-10pt}\\
\IEEEyessubnumber\\
\text{subject to}\;\;&\;\sum_{l=1}^{\bar{n}_k}p_{k,l}=p_k,\forall k,\text{ and}\IEEEyessubnumber\label{eq: sumpkl}\\
&\;p_{k,l}\geq 0,\forall k,\forall l.\IEEEyessubnumber\label{eq: pkl}
\end{IEEEeqnarray}
\end{problem}

Recall that, once at least the optimal solution $({p_{k,l}^*})_{k,l}$ to the outer optimization problem of {Problem~\ref{problem: replaced}} is found somehow, the jointly optimal signature sequences can be straightforwardly identified and constructed by applying the results in \cite{Viswanath : optimal0} to solve the inner optimization problem of {Problem~\ref{problem: CDMA outer inner}}.
Notice, however, that this replacement of the inner optimization problem itself does not help at all yet in solving the outer optimization problem of {Problem~\ref{problem: replaced}} over $P$, because the inner optimization still returns the same objective function that is a highly non-linear function of $(p_{k,l})_{k,l}$.

\begin{lemma}\label{lemma: switch}
{Problem~\ref{problem: replaced}} has the maximum sum rate and the optimal solution in common with
\begin{problem}\label{problem: CDMA outer inner switched}
\begin{IEEEeqnarray}{ll}
\underset{(w_{k,l})_{k,l}}{\text{maximize}}&\left\{
\begin{array}{ll}
\underset{(p_{k,l})_{k,l}}{\text{max}}&\!\!\!\!\!\!\displaystyle\sum_{k=1}^{K}\sum_{l=1}^{\bar{n}_k}w_{k,l}\log\left(1+\frac{p_{k,l}}{2\sigma^2w_{k,l}}\right)\\
\text{subject to}\;\;&\!\!\!\!\!\!\displaystyle\sum_{l=1}^{\bar{n}_k}p_{k,l}=p_k,\forall k,\text{ and}\\
&\!\!\!\!\!\!p_{k,l}\geq 0,\forall k,\forall l,
\end{array}
\right.\nonumber\vspace{-10pt}\\
\IEEEyessubnumber\\
\text{subject to}\;\;&\;0\leq w_{k,l}\leq \frac{1}{2N},\forall k,\forall l,\text{ and}\IEEEyessubnumber\\
&\;\displaystyle\sum_{k=1}^{K}\sum_{l=1}^{\bar{n}_k}w_{k,l}\leq\frac{1}{2}.\IEEEyessubnumber
\end{IEEEeqnarray}
\end{problem}
\end{lemma}

\begin{IEEEproof}
The constraints of the inner optimization problem of {Problem~\ref{problem: replaced}} are not affected by the choice of $(p_{k,l})_{k,l}$.
In addition, those of the outer optimization problem are not affected by the choice of $(w_{k,l})_{k,l}$.
Thus, the optimal solution to {Problem~\ref{problem: replaced}} does not change even if we switch the order of the inner and the outer optimization problems.
Therefore, the conclusion follows.
\end{IEEEproof}

As shown in the next subsections, {Problem~\ref{problem: CDMA outer inner switched}} obtained by applying the simple trick of switching the order of the inner and the outer optimization problems of {Problem~\ref{problem: replaced}} turns out to be solvable.
Of course, {Lemma~\ref{lemma: switch}} and {Problem~\ref{problem: CDMA outer inner switched}} as well as {Definition~\ref{def: virtual single-code}} and {Problem~\ref{problem: replaced}} can be re-stated in terms of the parameters of the restricted TDMA system having the upper limit $1/N$ on the duty cycle
\begin{equation}\label{eq: wkl tkl}
t_{k,l}=2w_{k,l}
\end{equation}
of the $(\sum_{k'=1}^{k-1}\bar{n}_k+l)$th user.

%----------------------------------------------------------------------
\subsection{FDMA-Equivalent Bandwidth and TDMA-Equivalent Duty Cycle}\label{sec: CDMA-2}
%----------------------------------------------------------------------

In the previous subsection, the equivalence of the optimal single-code CDMA system was established to the optimal restricted FDMA and TDMA systems, respectively, with equal upper limits on the bandwidths and the duty cycles of users.
By using this equivalence, the replace-and-switch method successfully formulated {Problem~\ref{problem: CDMA outer inner switched}} that has the maximum sum rate and the optimal power distribution in common with {Problem~\ref{problem: CDMA outer inner}}.
In this subsection, we introduce the notions of FDMA-equivalent bandwidth and TDMA-equivalent duty cycle, and derive the optimal $(w_{k,l}^*)_{k,l}$ and $(t_{k,l}^*)_{k,l}$ for {Problems~\ref{problem: replaced}} and {\ref{problem: CDMA outer inner switched}}.
To proceed, we examine the inner optimization of {Problem~\ref{problem: CDMA outer inner switched}}.
% 앞서 subsection summary: Go, 이번 section이 왜 나누어져 있나: Go, 호기심 Gap: Go

\begin{lemma}\label{lemma: pk,l}
Given a feasible solution $(w_{k,l})_{k,l}$ to the outer optimization problem of {Problem~\ref{problem: CDMA outer inner switched}}, the inner optimization problem has the optimal solution given by
\begin{equation}\label{eq: pkltenta}
p_{k,l}=\left\{
\begin{array}{ll}
\text{arbitrary non-negative value satisfying}\\%
\;\;\;\;\qquad\qquad\sum_{l=1}^{\bar{n}_k}p_{k,l}=p_k,\text{ for }\sum_{l'=1}^{\bar{n}_k}w_{k,l'}=0,\\%
\!\!\left(\displaystyle\frac{w_{k,l}}{\sum_{l'=1}^{\bar{n}_k}w_{k,l'}}\right) p_k, \qquad\qquad\text{ for } \sum_{l'=1}^{\bar{n}_k}w_{k,l'}>0,
\end{array}
\right.
\end{equation}
for all $k$ and $l$.
\end{lemma}
\begin{IEEEproof}
Straightforward by applying the water-filling argument \cite{Cover} to $\bar{n}_k$ parallel AWGN channels with total power $p_k$ and channel bandwidths $(w_{k,l})_{l=1}^{\bar{n}_k}$, for each $k=1,2,\cdots,K$.
\end{IEEEproof}

Recall the convention $w_k \log(1+p_k/(N_0w_k))=0$ at $w_k=0$, for all $p_k\geq 0$, in (\ref{convention}).
Then, (\ref{eq: pkltenta}) allows us to write the outer problem of {Problem~\ref{problem: CDMA outer inner switched}} as
\begin{problem}\label{problem: CDMA outer inner switched plugged}
\begin{IEEEeqnarray}{ll}
\underset{(w_{k,l})_{k,l}}{\text{maximize}}&\sum_{k=1}^{K}\sum_{l=1}^{\bar{n}_k}w_{k,l}\log\left(1+\frac{p_{k}}{2\sigma^2\sum_{l'=1}^{\bar{n}_k}w_{k,l'}}\right)\IEEEyessubnumber\IEEEeqnarraynumspace\label{eq: obj FDMA2}\\
\text{subject to}\;\;&\;0\leq w_{k,l}\leq \frac{1}{2N},\forall k,\forall l,\text{ and}\IEEEyessubnumber\label{eq: pklconst}\\
&\;\displaystyle\sum_{k=1}^{K}\sum_{l=1}^{\bar{n}_k}w_{k,l}\leq\frac{1}{2},\IEEEyessubnumber
\end{IEEEeqnarray}
\end{problem}
which needs to be solved only for $(w_{k,l})_{l=1}^{\bar{n}_k}$.

Of course, {Lemma~\ref{lemma: pk,l}} and {Problem~\ref{problem: CDMA outer inner switched plugged}} can be re-stated in terms of the duty cycles $(t_{k,l})_{k,l}$ of the equivalent restricted TDMA system with $\sum_{k=1}^{K}\bar{n}_k$ users having the equal upper limit $1/N$ on their duty cycles.
To solve {Problem~\ref{problem: CDMA outer inner switched plugged}}, we introduce the following definition.
%This definition is motivated by the fact that the objective function in (\ref{eq: obj FDMA2}) depends on the decision parameter $(w_{k,l})_{k,l}$ only through $(\sum_{l=1}^{\bar{n}_k}w_{k,l})_k$.

\begin{definition}
Given a feasible solution to {Problem~\ref{problem: CDMA outer inner switched plugged}}, the FDMA-equivalent bandwidth $w_k$ and the TDMA-equivalent duty cycle $t_k$ of the $k$th multi-code user are defined as
\begin{IEEEeqnarray}{rCl}
w_k&\triangleq& \sum_{l=1}^{\bar{n}_k}w_{k,l},\IEEEyessubnumber\label{eq: w_k}\\%
\noalign{\noindent{\text{and}}\vspace{\jot}}%
t_k&\triangleq& \sum_{l=1}^{\bar{n}_k}t_{k,l}=2w_k,\IEEEyessubnumber\label{eq: t_k}
\end{IEEEeqnarray}
respectively, for $k=1,2,\cdots,K$.
\end{definition}

These definitions are motivated by the fact that the objective function in (\ref{eq: obj FDMA2}) depends on $(w_{k,l})_{k,l}$ only through $(w_k)_k$ or, equivalently, $(t_k)_k$.

\begin{proposition}\label{prop: eFDMA}
A profile $(w_{k,l})_{k,l}$ is an optimal solution to {Problem~\ref{problem: CDMA outer inner switched plugged}} if and only if it is a member of the set
\begin{equation}\label{eq: conditions for wkl}
\!\left\{\!(w_{k,l})_{k,l}: 0\leq w_{k,l}\leq \frac{1}{2N}, \forall k, \forall l, \!\text{ and }\!\sum_{l=1}^{\bar{n}_k}{w_{k,l}}={w_k^*}, \forall k\right\},
\end{equation}
where $({w_k^*})_k$ is given by
\begin{subequations}\label{eq: wk*}
\begin{equation}\label{eq: FDMA equi sol}
w_k^*\!=\!\left\{
\begin{array}{ll}
\!\!\!\displaystyle\frac{\bar{n}_k}{2N}, &\!\!\!\!\text{for } 1\leq k\leq {K_1},  \\%
\\%
\!\!\!\displaystyle\frac{1}{2N}\!\!\left(N-\!\!\displaystyle\sum_{k'=1}^{{K_1}}\bar{n}_{k'}\right)\!\displaystyle\frac{ p_{k}}{\sum_{k'={K_1}+1}^{K}p_{k'}}, &\!\!\!\!\text{for } {K_1}<k\leq K,\;\;
\end{array}
\right.
\end{equation}
or, equivalently,
\begin{equation}
w_k^*\!=\!\left\{
\begin{array}{ll}
\!\!\!\displaystyle\frac{\bar{n}_k}{2N}, &\!\!\!\!\text{for } 1\leq k\leq {K_2},  \\%
\\%
\!\!\!\displaystyle\frac{1}{2N}\!\Bigg(N-\!\!\sum_{k'=1}^{{K_2}}\bar{n}_{k'}\Bigg)\!\displaystyle\frac{ p_{k}}{\sum_{k'={K_2}+1}^{K}p_{k'}}, &\!\!\!\!\text{for } {K_2}<k\leq K,\;\;
\end{array}
\right.\label{eq: FDMA equi sol2}
\end{equation}
where ${K_1}$ and ${K_2}$ are, respectively, the numbers of oversized and non-undersized users in the restricted FDMA system with power $p_k$, bandwidth upper limit $\bar{n}_k/(2N)$, and total system bandwidth $1/2$.
\end{subequations}
\end{proposition}

\begin{IEEEproof}
Since the objective function in (\ref{eq: obj FDMA2}) is a function only of the FDMA-equivalent bandwidths, the constraint set can be partitioned for local searches into the subsets of feasible solutions having the common profile of $(w_k)_k$.
Moreover, the constraint (\ref{eq: pklconst}) implies
\begin{equation}\label{eq: addi}
0\leq w_k\leq \frac{\bar{n}_k}{2N},\forall k.
\end{equation}
Thus, (\ref{eq: w_k}) and (\ref{eq: addi}) can be imposed on {Problem~\ref{problem: CDMA outer inner switched plugged}} as additional constraints without altering the maximum sum rate and the set of the optimal solutions.
%Thus, the optimal FDMA-equivalent bandwidths $({w_k^*})_k$ of {Problem~\ref{problem: CDMA outer inner switched plugged}} can be found by solving
Consequently, {Problem~\ref{problem: CDMA outer inner switched plugged}} can be rewritten in a double maximization form given by
\begin{problem}\label{problem: feasibility}
\begin{IEEEeqnarray}{ll}
\underset{(w_{k,l})_{k,l}}{\text{maximize}}&\left\{
\begin{array}{ll}
\underset{(w_k)_k}{\text{max}}&\!\!\!\!\!\!\displaystyle\sum_{k=1}^{K}w_k\log\left(1+\frac{p_{k}}{2\sigma^2w_{k}}\right)\\
\text{subject to}\;\;&\!\!\!\!\!\!0\leq w_{k}\leq \displaystyle\frac{\bar{n}_k}{2N},\forall k, \text{ and }\\
&\!\!\!\!\!\!\displaystyle\sum_{k=1}^{K}w_{k}\leq \frac{1}{2},
\end{array}
\right.\nonumber\vspace{-10pt}\\
\IEEEyessubnumber\\
\text{subject to}\;\;&\;0\leq w_{k,l}\leq \frac{1}{2N},\forall k,\forall l, \text{ and }\IEEEyessubnumber\\
&\;\sum_{l=1}^{\bar{n}_k}w_{k,l}=w_{k}, \forall k.\IEEEyessubnumber
\end{IEEEeqnarray}
\end{problem}
By comparing the inner optimization problem of {Problem~\ref{problem: feasibility}} with {Problem~\ref{prob: FDMA}}, we straightforwardly see that the results in {Proposition~\ref{proposition: FDMAsol}} or, equivalently, those in {Remark~\ref{remark: 5}} can be immediately used to obtain $({w_k^*})_k$ as (\ref{eq: wk*}) by replacing the parameters $\bar{w}_k$ and $w_{\text{tot}}$ with $\bar{n}_k/(2N)$ and $1/2$, respectively.
Once ${w_k^*}$ is found for all $k$, the optimal solution $({w_{k,l}^*})_{k,l}$ can be obtained by solving the outer optimization problem of {Problem~\ref{problem: feasibility}}, which is just a simple feasibility test to check the membership to the set defined in (\ref{eq: conditions for wkl}).
Therefore, the conclusion follows.
\end{IEEEproof}

In {Proposition~\ref{prop: eFDMA}}, we found the optimal FDMA-equivalent bandwidths of the multi-code users by solving the corresponding restricted FDMA problem.
Henceforth, the multi-code users will be classified as follows in a similar way to the virtual single-code users.

\begin{definition}\label{def: over}
The users of the multi-code CDMA system are classified into oversized, critically-sized, or undersized multi-code users, if the corresponding users of the restricted FDMA system with power $p_k$, bandwidth upper limit $\bar{n}_k/(2N)$, and total system bandwidth $1/2$ are classified as such by {Definition~\ref{def: classification}}, i.e.,
the $k$th user is tested by the rule
\begin{subequations}
\begin{equation}\label{eq: classification rule2}
\hat{n}_k\triangleq\left(N-\sum_{k'=1}^{k-1}\bar{n}_{k'}\right)\frac{p_k}{\sum_{k'=k}^{K}p_{k'}}\gtreqqless \bar{n}_{k},
\end{equation}
and classified as
\begin{equation}\label{eq: classification2}
\begin{array}{rll}
\text{(a)}&\!\!\text{an oversized multi-code user}, & \text{if } \hat{n}_k>\bar{n}_{k},\\
\text{(b)}&\!\!\text{a critically-sized multi-code user}, & \text{if } \hat{n}_k=\bar{n}_{k}, \text{ and}\qquad\\
\text{(c)}&\!\!\text{an undersized multi-code user}, & \text{if } \hat{n}_k<\bar{n}_{k}.
\end{array}
\end{equation}
\end{subequations}
\end{definition}

Recall that the oversized users of an overloaded single-code CDMA system are defined as the relatively strong-power users satisfying the conditions in \cite[Eq. (5)]{Viswanath : optimal0} and, equivalently, those in \cite[Eq. (21)]{Ulukus4}.
It can be easily seen that these conditions are the same as the condition $\hat{n}_k>\bar{n}_{k}$ in {Definition~\ref{def: over}} if $\bar{n}_k$ is replaced by $1$ for all $k$.
Moreover, {Definition~\ref{def: over}} works for all system loading condition, while the conditions \cite[Eq. (5)]{Viswanath : optimal0} and \cite[Eq. (21)]{Ulukus4} for the single-code CDMA system do not always work due to division by zero.
Also recall that the corresponding oversized users of the optimal restricted FDMA system are allocated their bandwidth upper limits and have greater PSDs than non-oversized users.
This is because, if the $k$th user is oversized, its due share of the bandwidth computed in the line $3$ of the $k$th iteration in {TABLE~\ref{table: 1}} is greater than its bandwidth upper limit.
In this way, {Definition~\ref{def: over}} successfully generalizes the definition of the oversized user for the single-code system to the user classification rule for the multi-code CDMA system and provides it a physical meaning in terms of the parameters of the equivalent FDMA system.

Note that (\ref{eq: conditions for wkl}) allows freedom in choosing $(w_{k,l}^*)_l$ only for the non-undersized multi-code users with $\bar{n}_k>1$ because $\bar{n}_k/(2N) > w_k^*$ for such users.
Once a feasible optimal solution is found by using {Proposition~\ref{prop: eFDMA}}, the optimal solution $({p_{k,l}^*})_{k,l}$ to the inner and, equivalently, the outer optimization problems of {Problems~\ref{problem: replaced}} and {\ref{problem: CDMA outer inner switched}}, respectively, can be found by {Lemma~\ref{lemma: pk,l}}.
Thus, combined with the results in \cite{Viswanath : optimal0}, {Proposition~\ref{prop: eFDMA}} eventually leads to the optimal solution $(p_{k,l}^*,{\underline{s}_{k,l}^*})_{k,l}$ to {Problem~\ref{problem: CDMA outer inner}} as presented in the next subsection.
Of course, all the results in this subsection can be re-stated in terms of the parameters of the equivalent TDMA system.

%----------------------------------------------------------------------
\subsection{Optimal System and Its Properties}\label{sec: optimal solution2}
%----------------------------------------------------------------------

In this subsection, we present the sum-rate optimal multi-code CDMA system and examine its properties.
First, we establish the equivalence of the optimal multi-code CDMA system to the optimal restricted FDMA system and, then, present the necessary and sufficient condition for an optimal multi-code system to achieve the sum capacity of the MAC.
Throughout this subsection, the equivalence to the optimal restricted FDMA system is solely utilized just for simplicity, though (\ref{eq: TDMA FDMA}) says that the same arguments can be made by using the equivalence to the optimal restricted TDMA system.
In what follows,
\begin{equation}
\underline{\bar{n}}\triangleq [\bar{n}_1, \bar{n}_2, \cdots, \bar{n}_K]^T
\end{equation}
denotes the upper-limit profile on the numbers of multi-codes of users.

\begin{theorem}\label{theorem: CDMA max sum rate}
The maximum sum rate $\mathcal{C}_\text{CDMA}(\underline{p},\underline{\bar{n}},N,\sigma^2)$ of the multi-code CDMA system with power profile $\underline{p}$, upper-limit profile $\underline{\bar{n}}$, processing gain $N$, and variance $\sigma^2$ per dimension of the AWGN is equal to that of the restricted FDMA system with power profile $\underline{p}$, bandwidth-constraint profile $\underline{\bar{n}}/(2N)$, total system bandwidth $1/2$, and one-sided PSD $2\sigma^2$ of the AWGN, i.e.,
\begin{IEEEeqnarray}{lCl}\label{eq: CDMAmaxsumrate}
&&\mathcal{C}_\text{CDMA}(\underline{p},\underline{\bar{n}},N,\sigma^2)=
\mathcal{C}_\text{FDMA}\left(\underline{p},\frac{\underline{\bar{n}}}{2N},\frac{1}{2},2\sigma^2\right)\IEEEyessubnumber\label{eq: CDMAmaxsumrate0}\\%
&=&\sum_{k=1}^{{K_1}}\frac{\bar{n}_k}{2N}\log\left(1+\frac{Np_k}{\sigma^2\bar{n}_k}\right)\nonumber\\
&&+\frac{N-\sum_{k=1}^{{K_1}}\bar{n}_k}{2N}\log\left(1+\frac{N\sum_{k={K_1}+1}^{K}p_k}{\sigma^2(N-\sum_{k=1}^{{K_1}}\bar{n}_k)}\right)\IEEEyessubnumber\label{eq: CDMAmaxsumrate1}\IEEEeqnarraynumspace\\%
&=&\sum_{k=1}^{{K_2}}\frac{\bar{n}_k}{2N}\log\left(1+\frac{Np_k}{\sigma^2\bar{n}_k}\right)\nonumber\\
&&+\frac{N-\sum_{k=1}^{{K_2}}\bar{n}_k}{2N}\log\left(1+\frac{N\sum_{k={K_2}+1}^{K}p_k}{\sigma^2(N-\sum_{k=1}^{{K_2}}\bar{n}_k)}\right),\IEEEyessubnumber\label{eq: CDMAmaxsumrate2}\IEEEeqnarraynumspace
\end{IEEEeqnarray}
where ${K_1}$ and ${K_2}$ are the numbers of oversized and non-undersized users in the system, respectively.
\end{theorem}

\begin{IEEEproof}
Since the outer optimization problem of {Problem~\ref{problem: feasibility}} is a feasibility test, the maximum sum rate can be found just by solving the inner optimization problem of {Problem~\ref{problem: feasibility}}.
Thus, by {Propositions~\ref{proposition: FDMAsol}} and \ref{prop: eFDMA}, we have (\ref{eq: CDMAmaxsumrate0}) and (\ref{eq: CDMAmaxsumrate1}).
Moreover, by {Remark~\ref{remark: 5}}, we have (\ref{eq: CDMAmaxsumrate2}).
Therefore, the conclusion follows.
\end{IEEEproof}

{Theorem~\ref{theorem: CDMA max sum rate}} combined with the inner optimization problem of {Problem~\ref{problem: feasibility}} shows that limiting the numbers of multi-codes of users in CDMA corresponds to imposing the upper limits on the bandwidths of users in FDMA.
Recall that the unrestricted FDMA having only the total bandwidth constraint maximizes the sum rate by assigning each user the bandwidth that is proportional to its signal power, which implies a more bandwidth to a stronger-power user \cite{Cover}.
Thus, the above equivalence shows that the CDMA is a multiple-access scheme that mitigates the unfairness inherent in the sum-rate maximization by imposing upper limits on the FDMA-equivalent bandwidths of users.

\begin{theorem}\label{theorem: nece suffi}
The maximum sum rate (\ref{eq: CDMAmaxsumrate}) of the multi-code CDMA system is upper bounded by the sum capacity of the MAC, i.e.,
\begin{subequations}
\begin{equation}
\mathcal{C}_\text{CDMA}(\underline{p},\underline{\bar{n}},N,\sigma^2)\leq \mathcal{C}_{\text{MAC}}\left(\underline{p},\frac{1}{2},2\sigma^2\right),
\end{equation}
where equality holds if and only if there is no oversized multi-code user, i.e., $K_1=0$ or, equivalently,
\begin{equation}
N\cdot\left(\frac{p_1}{\sum_{k=1}^{K}p_{k}}\right)\leq \bar{n}_1.\IEEEyessubnumber\label{eq: iff}
\end{equation}
\end{subequations}
\end{theorem}

\begin{IEEEproof}
Straightforward by {Remark~\ref{remark: 5a}} and {Theorem~\ref{theorem: CDMA max sum rate}}.
\end{IEEEproof}

Especially for the single-code CDMA system, i.e., for $\bar{n}_k=1,\forall k$, the condition (\ref{eq: iff}) simplifies to
\begin{equation}\label{eq: nece suffi single}
N{p_1} \leq {\sum_{k=1}^{K}p_{k}},
\end{equation}
which is identical to the necessary and sufficient condition derived in \cite{Viswanath : optimal0} for the optimal single-code CDMA system to achieve the sum capacity of the MAC by having no oversized user.
Note that (\ref{eq: nece suffi single}) implies $Np_1\leq \sum_{k=1}^Kp_k\leq Kp_1$.
Thus, for (\ref{eq: nece suffi single}) to hold, it is necessary that the system is not underloaded, i.e., the number of active users must be no less than the processing gain.
On the contrary, note that (\ref{eq: iff}) implies $Np_1\leq\bar{n}_1\sum_{k=1}^{K}p_k\leq p_1\sum_{k=1}^{K}\bar{n}_k$ by the assumption (\ref{CDMApowerordered}).
Thus, for (\ref{eq: iff}) to hold, it is necessary that the total sum of the upper limits on the number of multi-codes of the active users is just no less than the processing gain, which includes the cases of non-underloaded systems.
So, the necessary and sufficient condition is successfully extended to the multi-code CDMA system by {Theorem~\ref{theorem: nece suffi}}, saying that the sum capacity of the MAC is achieved, regardless of the system loading, if and only if the first user and, consequently by {Lemma~\ref{lemma: kappa}}, every user is a non-oversized multi-code user.
This implies that the multi-code CDMA is better suited than the single-code CDMA in that it may achieve a higher sum rate and even be an optimal multiple-access scheme for underloaded systems by properly choosing the upper limits on the numbers of multi-codes of users.

Divided by $1/(2N)$, the left side of (\ref{eq: iff}) is nothing but the due share of the bandwidth to the first user of the equivalent FDMA system, and the right side is the first user's bandwidth upper limit.
If the inequality in (\ref{eq: iff}) holds, then the equivalent restricted FDMA system achieves the sum capacity of the MAC by having a flat PSD for each user and equal PSDs for all users.
Otherwise, the presence of an oversized user in the restricted FDMA system makes the PSD no longer flat, which leads to an inefficient utilization of the system bandwidth compared to the proportional-share allocation scheme that achieves the sum capacity of the MAC.
Thus, {Theorem~\ref{theorem: nece suffi}} implies that, if the upper limits $(\bar{n}_k)_k$ are chosen relatively large then, though the sum capacity may be achievable, it becomes possible for a small group of strong-power multi-code users to occupy most of the total FDMA-equivalent bandwidth due to the nature of the proportional-share bandwidth allocation.
If the upper limits are chosen relatively small, which is the case with the single-code system, then fairness among users may be increased at the cost of the maximum sum rate falling short of the sum capacity of the MAC.

Second, we present the optimal power distribution that achieves the maximum sum rate of the multi-code CDMA system jointly with the optimal signature sequences that will be presented next.
% 잘 나눠져 보이도록 First, Second, ..., Finally,...

\begin{theorem}\label{theorem: opt power}
The optimal power distribution for the $k$th multi-code user is given by
\begin{subequations}
\begin{equation}\label{eq: opt power}
{p_{k,l}^*}=\left\{
\begin{array}{ll}
\displaystyle\frac{p_k}{\bar{n}_k},& \text{for } 1\leq k\leq {K_2},\\%
\;\\%
\displaystyle\left(\frac{{w_{k,l}^*}}{ w_{k}^*}\right)p_k,& \text{for } {K_2} < k \leq K,
\end{array}
\right.
\end{equation}
for each $l\in\{1,2,\cdots,\bar{n}_k\}$, so that the optimal number $n_k^*$ of active data streams is given by
\begin{equation}
n_k^*=\left\{
\begin{array}{ll}
\bar{n}_k,& \text{for } 1\leq k\leq {K_2},\\%
\;\\%
\displaystyle\sum_{l=1}^{\bar{n}_k}\text{sgn}({p_{k,l}^*}),& \text{for } {K_2} < k \leq K,
\end{array}
\right.
\end{equation}
\end{subequations}
where ${w_{k,l}^*}$ and ${w_{k}^*}$, for $k>{K_2}$, are found as described in {Proposition~\ref{prop: eFDMA}}.
\end{theorem}

\begin{IEEEproof}
From the iterative algorithm in {TABLE~\ref{table: 1}}, we can easily see that the $k$th multi-code user with $p_k>0$ always has $w_k^*>0, \forall k$.
By substituting $w_{k,l}^*$ and $w_k^*, \forall k, \forall l,$ into (\ref{eq: pkltenta}) in {Lemma~\ref{lemma: pk,l}}, we have $p_{k,l}^*=(w_{k,l}^*/w_{k}^*)p_k, \forall k,\forall l$, which leads to $n_k^*=\sum_{l=1}^{\bar{n}_k}\text{sgn}({p_{k,l}^*}), \forall k$.
In particular, (\ref{eq: conditions for wkl}) and (\ref{eq: FDMA equi sol2}) in {Proposition~\ref{prop: eFDMA}} imply that, for $k\leq K_2$,
\begin{IEEEeqnarray}{rl}
&w_{k,l}^*=\frac{1}{2N},  \forall l, \IEEEyessubnumber\label{eq: wkl*K2}\\
\noalign{\noindent{\text{and}}\vspace{\jot}}%
&n_k^*=\bar{n}_k.\IEEEyessubnumber\label{eq: nk*K2}%
\end{IEEEeqnarray}
Therefore, we have the conclusion.
\end{IEEEproof}

Again by the method of componendo and dividendo, it can be shown that {Theorem~\ref{theorem: opt power}} holds with ${K_2}$ being replaced by ${K_1}$.
This theorem shows that, if the $k$th user is oversized or critically-sized, an optimal multi-code CDMA system equally distributes $p_k$ to its all $\bar{n}_k$ data streams, i.e., ${p_{k,l}^*}=p_k/\bar{n}_k$, for $l=1,2,\cdots,\bar{n}_k$.
However, if the $k$th user is undersized with $\bar{n}_k>1$, an optimal multi-code CDMA system has no unique power distribution and, consequently, no unique number of active data streams in general.
So, it becomes of interest to find a simple form among all optimal power distributions that can maximize the sum rate jointly with the optimal sequences.
The following remark shows that the equal distribution of each user's power to its multiple data streams is not only the unique optimal distribution for the non-undersized multi-code users but also an optimal distribution for the undersized multi-code users.

\begin{remark}\label{remark: equal power}
The equal distribution of the $k$th user's  power $p_k$ to its all $\bar{n}_k$ data streams for all $k$, i.e.,
\begin{IEEEeqnarray}{rCl}
p_{k,l}^*&=&\displaystyle\frac{p_k}{\bar{n}_k}, \forall k, \forall l,\IEEEyessubnumber\label{eq: equal power}\\%
\noalign{\noindent{\text{and}}\vspace{\jot}}%
n_k^*&=&\bar{n}_k, \forall k,\IEEEyessubnumber
\end{IEEEeqnarray}
can achieve the maximum sum rate of the multi-code CDMA system.
\end{remark}

\begin{IEEEproof}
Given the optimal FDMA-equivalent bandwidths $(w_k^*)_k$, consider the equal-bandwidth allocation of $w_k^*$ to $(w_{k,l})_{l}, \forall k$, given by $w_{k,l}={w_k^*}/\bar{n}_k, \forall k,\forall l$.
Then, it can be easily verified that this allocation satisfies the conditions in (\ref{eq: conditions for wkl}).
Thus, it is an optimal allocation by {Proposition~\ref{prop: eFDMA}}.
By (\ref{eq: opt power}), this equal-bandwidth allocation leads to the equal-power distribution (\ref{eq: equal power}).
Therefore, the conclusion follows.
\end{IEEEproof}

Third, we present the optimal signature sequences that achieve the maximum sum rate of the multi-code CDMA system jointly with the optimal power distribution that is just presented.
As mentioned earlier, once an optimal power distribution is found, the inner optimization problem of {Problem~\ref{problem: CDMA outer inner}} can be viewed as the sequence design problem for a CDMA system with virtual single-code users.
Thus, the results in \cite{Viswanath : optimal0} can be directly used to identify and construct the optimal signature sequences, which will yield orthogonal sequences for oversized virtual single-code users and GWBE sequences for non-oversized virtual single-code users.
However, unlike \cite{Viswanath : optimal0}, we have introduced the notions of critically-sized users in this paper.
This separation of the critically-sized users and the undersized users enables us to better design the system as shown below.

% 대비 시키는 말로 시작.
% 잘 나눠져 보이도록 First, Second, ..., Finally,...
% \mathcal{K}_2을 쓰는 우리 결과를 theorem으로 제시하는 것이 좋겠음.

\begin{theorem}\label{theorem: opt seq}
The optimal signature sequences are given by
\begin{equation}\label{eq: opt seq}
{\underline{s}_{k,l}^*}=\left\{
\begin{array}{r}
\!\text{arbitrary orthogonal vector of norm $N$ and}\;\;\;\;\;\;\\%
\!\!\!\text{length $N$,   for }(k,l)\in\mathcal{K}_2,\\
\!\!\!E\underline{\tilde{s}}_{k,l}^*, \qquad\qquad\qquad\qquad\;\;\;\;\;\,\;\;\;\;\text{for }(k,l)\notin\mathcal{K}_2,
\end{array}
\right.
\end{equation}
where $\mathcal{K}_2$ is the set of all virtual single-code user indexes $(k,l)$ such that the $(\sum_{k'=1}^{k-1}\bar{n}_k+l)$th virtual single-code user with power $p_{k,l}^*>0$ is non-undersized, $E$ is the $N$-by-$(N-|\mathcal{K}_2|)$ matrix consisting of the norm-$N$ and length-$N$ orthogonal basis vectors of the orthogonal complement of the span of $\{{\underline{s}_{k,l}^*}:(k,l)\in\mathcal{K}_2\}$, the sequences $(\underline{\tilde{s}}_{k,l}^*)_{k,l}$ satisfying $(k,l)\notin\mathcal{K}_2$ and $p_{k,l}^*>0$ are arbitrary length-$(N-|\mathcal{K}_2|)$ sequences, and the sequences $(\underline{\tilde{s}}_{k,l}^*)_{k,l}$ satisfying $(k,l)\notin\mathcal{K}_2$ and $p_{k,l}^*>0$ are the length-$(N-|\mathcal{K}_2|)$ GWBE sequences \cite{Viswanath : optimal0} such that
\begin{equation}\label{eq: GWBE}
\sum_{(k,l)\notin\mathcal{K}_2,p_{k,l}^*>0}{p_{k,l}^*}{\tilde{\underline{s}}_{k,l}^*}{\tilde{\underline{s}}_{k,l}^{*T}}
=\frac{\sum_{(k,l)\notin\mathcal{K}_2}{p_{k,l}^*}}{N-|\mathcal{K}_2|}I_{N-|\mathcal{K}_2|},
\end{equation}
where $|\mathcal{K}_2|$ denotes the cardinality of the set $\mathcal{K}_2$ and the identity matrix $I_{N-|\mathcal{K}_2|}$ is of size $(N-|\mathcal{K}_2|)$.
\end{theorem}

\begin{IEEEproof}
Given an optimal power matrix $P^*$, the inner optimization problem in (\ref{eq: inner}) becomes the sequence design problem for the single-code CDMA system considered in \cite{Viswanath : optimal0}, now having the number of active users equal to $\sum_{k=1}^K \sum_{l=1}^{\bar{n}_k}\text{sgn}(p_{k,l}^*)$.
If this number is not greater than the processing gain, i.e., $\sum_{k=1}^K \sum_{l=1}^{\bar{n}_k}\text{sgn}(p_{k,l}^*)\leq N$, then, as shown in \cite{Viswanath : optimal0}, a complete orthogonalization of the signature sequences is possible for this non-overloaded system with virtual single-code users.
Since the corresponding restricted FDMA system with the specified parameters has the sum of the bandwidth upper limits of the active users no greater than the total bandwidth, i.e., $\sum_{k=1}^K \sum_{l=1}^{\bar{n}_k}\text{sgn}(p_{k,l}^*)(1/2N)\leq 1/2$, every active virtual single-code user is non-undersized.
Thus, $(k,l)\in \mathcal{K}_2$ for all $p_{k,l}^*>0$, which makes the code design in (\ref{eq: opt seq}) lead to orthogonal signature sequences, which coincides with the result in \cite{Viswanath : optimal0} just mentioned.
On the other hand, if $\sum_{k=1}^K \sum_{l=1}^{\bar{n}_k}\text{sgn}(p_{k,l}^*)> N$, then the inner optimization problem of {Problem~\ref{problem: CDMA outer inner}} becomes the sequence design problem for the overloaded single-code CDMA system.
Again, the optimal signature sequences for the active virtual single-code users can be identified and constructed by using the results in \cite{Viswanath : optimal0}, which can be rewritten as (\ref{eq: opt seq}) and (\ref{eq: GWBE}) for the multi-code system except that $\mathcal{K}_2$ is replaced by the set $\mathcal{K}_1\subset \mathcal{K}_2$ of all oversized virtual single-code user indexes.
To show that $\mathcal{K}_2$ can be used instead of $\mathcal{K}_1$, let $(k',l')$ be the user index of a critically-sized virtual single-code user and choose any orthogonal matrix $U$, of which the first column is proportional to the GWBE signature sequence of the $(k',l')$th user constructed as described in (\ref{eq: opt seq}) and (\ref{eq: GWBE}) with $\mathcal{K}_2$ being replaced by $\mathcal{K}_1$.
By pre- and post-multiplying $U^T$ and $U$, respectively, to (\ref{eq: GWBE}), we have
\begin{IEEEeqnarray}{rCl}
&&U^T\left(\sum_{(k,l)\notin\mathcal{K}_1,p_{k,l}^*>0}{p_{k,l}^*}\tilde{\underline{s}}_{k,l}^*\tilde{\underline{s}}_{k,l}^{*T}\right)U \nonumber\\%
&=&U^T\left(\frac{\sum_{(k,l)\notin\mathcal{K}_1,p_{k,l}^*>0}{p_{k,l}^*}}{N-|\mathcal{K}_1|}I_{N-|\mathcal{K}_1|}\right)U \IEEEyessubnumber\\%
\!\!\!\!\!\!&=&\!\left[
\begin{array}{c|c}
p_{k',l'}^* & \underline{0}_{N-|\mathcal{K}_1|-1}^{T}\\%
\hline\\%
\!\!\!\underline{0}_{N-|\mathcal{K}_1|-1}\!\!\! & \displaystyle\frac{ \sum_{
\scriptsize
\begin{array}{c}
\!\!\!\!(k,l)\notin\mathcal{K}_1\cup\{(k',l')\}\\
p_{k,l}^*>0
\end{array}
\!\!\!\!\!
}{p_{k,l}^*}}{N-|\mathcal{K}_1|-1}I_{N-|\mathcal{K}_1|-1}\!\!\!\!
\end{array}
\right],\IEEEyessubnumber\label{eq: critic}\IEEEeqnarraynumspace
\end{IEEEeqnarray}
where (\ref{eq: critic}) comes from {Remark~\ref{remark: 4a}}-(a) and the method of componendo and dividendo that imply $p_{k',l'}^*=\sum_{\;(k,l)\notin\mathcal{K}_1,\;p_{k,l}^*>0}$

\noindent ${p_{k,l}^*}/({N-|\mathcal{K}_1|})=
\sum_{(k,l)\notin\mathcal{K}_1\cup\{(k',l')\},p_{k,l}^*>0}{p_{k,l}^*}/(N-|\mathcal{K}_1|$ $-1)$.
Thus, the signature sequence of the $(k',l')$th user is orthogonal to the subspace spanned by those of the remaining non-oversized virtual single-code users.
Moreover, the $(2,2)$th entry of the block matrix in (\ref{eq: critic}) shows that GWBE sequences are optimal as the signature sequences of the remaining virtual single-code users.
Obviously, this procedure can be repeated until every critically-sized virtual single-code user is allocated an orthogonal sequence.
Therefore, the conclusion follows.
\end{IEEEproof}

Here follows a simple consequence of this theorem.

\begin{remark}\label{remark: ortho}
If the $k$th user is a non-undersized multi-code user, i.e., $k\leq {K_2}$, then it is always allocated $\bar{n}_k$ orthogonal sequences.
\end{remark}

\begin{IEEEproof}
By (\ref{eq: wkl*K2}) and (\ref{eq: nk*K2}) in the proof of {Theorem~\ref{theorem: opt power}}, the $k$th user for $k\leq K_2$, has $\bar{n}_k$ virtual single-code users with $w_{k,l}^*=1/(2N),\forall l$.
Then, the equivalence of the restricted FDMA system with equal bandwidth upper limit $1/(2N)$ to the CDMA system with virtual single-code users implies that $(k,l)\in \mathcal{K}_2, \forall k\leq {K_2}, \forall l$.
Therefore, by {Theorem~\ref{theorem: opt seq}}, the conclusion follows.
\end{IEEEproof}

Note that each of the virtual single-code users of the oversized and the critically-sized multi-code users has the equal optimal FDMA-equivalent bandwidth of $1/(2N)$, which results in the information rate in (\ref{eq: CDMAmaxsumrate2}) of such a virtual single-code user given by
$1/(2N)\log(1+ Np_{k,l}^*)/(\sigma^2)=1/(2N)\log(1+ Np_{k})/(\sigma^2\bar{n}_k)$, where we used $p_{k,l}^*=p_k/\bar{n}_k, \forall k\leq K_2$ in (\ref{eq: opt power}).
In other words, a virtual single-code user of a critically-sized multi-code user occupies the same amount of signal dimension as that of an oversized multi-code user does.
This explains why the critically-sized virtual single-code users are also allocated orthogonal sequences as the oversized virtual single-code users.

A caution needs to be paid to the information rate of an undersized virtual single-code user.
As pointed out in \cite{Guess3}, the individual information rate of a single-code user that is allocated a GWBE sequence may not be uniquely determined.
This is because, though the maximum sum rate is uniquely determined, the sum rate of such users can be maximized by various combinations of time sharing or rate splitting combined with superposition encoding and successive interference cancelation decoding \cite{NIT}, \cite{Urbanke: RS}, which results in the different rate profiles of the users.
For the same reason, though the FDMA-equivalent bandwidth of a non-undersized virtual single-code user is the signal dimension actually allocated to the user, that of an undersized virtual single-code user is not.

{Remark~\ref{remark: ortho}} says that every virtual single-code user of an non-undersized multi-code user is allocated an orthogonal sequence but not the truth of the converse.
The next result shows that even some virtual single-code users of an undersized multi-code user can be allocated orthogonal sequences.

Fourth, we find the maximum number of orthogonal signature sequences to satisfy the constraint on $n_k$ for all $k$ and, at the same time, retain the maximum sum rate.
Recall that, as mentioned in Section~\ref{sec: sig sys}, this maximization may contribute to reducing the system complexity.
According to \cite{Viswanath : optimal0}, the GWBE sequences are allocated to the non-oversized single-code users, where it is not pointed out that a set of GWBE sequences may include an orthogonal sequence.
Now, {Remark~\ref{remark: ortho}} says that, in the single-code CDMA system, the use of non-orthogonal GWBE sequences is actually limited only to the undersized users because the GWBE sequences for the critically-sized single-code users are all orthogonal sequences.
Of course, {Remark~\ref{remark: ortho}} does not imply that, in the multi-code CDMA system, every signature sequence of an undersized multi-code user is a non-orthogonal GWBE sequence.
Rather, the use of non-orthogonal GWBE sequences can be further limited.
In other words, as the following result shows, the number of orthogonal sequences can be further increased.

\begin{theorem}\label{theorem: max ortho}
The maximum of the number $n_k^{\perp}$ of active orthogonal signature sequences of the $k$th multi-code user to still retain the maximum sum rate is given by
\begin{IEEEeqnarray}{C}\label{eq: nkmax}
\max n_k^{\perp}=\left\{
\begin{array}{l}
\bar{n}_k, \;\;\;\;\;\;\qquad\qquad\qquad \text{ for } 1\leq k\leq {K_2},  \\%
%\lfloor 2N {w_k^*} \rfloor, & \text{ for } {K_2} < k \leq K,
\Bigg\lfloor
\Bigg(N-\!\!\displaystyle\sum_{k'=1}^{{K_2}}\bar{n}_{k'}\Bigg)\!\displaystyle\frac{ p_{k}}{\sum_{k'={K_2}+1}^{K}p_{k'}}\Bigg\rfloor,
\end{array}
\right.\IEEEeqnarraynumspace\\%
\;\;\;\;\;\;\qquad\qquad\qquad\qquad\qquad\qquad \text{ for } {K_2} < k \leq K,\nonumber
\end{IEEEeqnarray}
where $\lfloor \cdot \rfloor$ denotes the flooring operation.
\end{theorem}

\begin{IEEEproof}
If the $k$th multi-code user is oversized or critically-sized, then its all $n_k^*=\bar{n}_k$ data streams are transmitted by using orthogonal sequences, as shown in {Remark~\ref{remark: ortho}}.
Thus, $\max n_k^{\perp}=\bar{n}_k, \forall k\leq {K_2}$.
If the $k$th multi-code user is undersized, then the equivalence of the CDMA system with virtual single-code users to the restricted FDMA system with equal bandwidth upper limit $1/(2N)$ implies that its $l$th data stream is transmitted by using an orthogonal signature sequence if  ${w_{k,l}^*}=1/(2N)$.
Thus, $\max n_k^{\perp}=\lfloor 2N {w_k^*}\rfloor, \forall k> {K_2}$.
Since the optimal FDMA-equivalent bandwidth ${w_k^*}$ of the $k$th multi-code user is given by (\ref{eq: FDMA equi sol2}), we have the conclusion.
\end{IEEEproof}

In other words, {Theorem~\ref{theorem: max ortho}} says that all the virtual single-code users of the non-undersized multi-code users are allocated orthogonal signature sequences, while some virtual single-code users of the undersized multi-code users can be allocated orthogonal signature sequences.
This result may be alternatively explained by using the properties of the PSDs of users in the equivalent restricted FDMA system.
By {Remark~\ref{remark: 4a}}-(a), all the critically-sized and the undersized multi-code users have the optimal PSDs equal among the corresponding users in the equivalent restricted FDMA system.
In addition, all the virtual single-code users of a critically-sized multi-code user that are allocated orthogonal sequences have the FDMA-equivalent bandwidth $1/(2N)$.
Suppose that there is a virtual single-code user of an undersized multi-code user having the FDMA-equivalent bandwidth $1/(2N)$.
Since it is not distinguishable in terms of the PSD and the FDMA-equivalent bandwidth from a virtual single-code user of a critically-sized multi-code user, it is no wonder why it is also allocated an orthogonal signature sequence.

Note that the equal-power distribution of {Remark~\ref{remark: equal power}} minimizes the number of orthogonal sequences.
This is because it makes the $k$th user for $K_2<k\leq K$, have $\bar{n}_k$ data streams satisfying $0< w_{k,l}^*=w_k^*/\bar{n}_k<1/(2N)$, for $l=1,2,\cdots,\bar{n}_k$, which leads to no orthogonal sequences for undersized multi-code users.

Fifth, we find the minimum number of active signature sequences for each multi-code user to still retain the maximum sum rate.
Though simple, the equal-power distribution of {Remark~\ref{remark: equal power}} is definitely not a desirable solution as far as the complexity of the transmitters and the receiver is concerned, because it maximizes the number of active signature sequences.
In addition, even the maximization of the number of orthogonal sequences in {Theorem~\ref{theorem: max ortho}} may still leave freedom in choosing the number of virtual single-code users for undersized multi-code users.
This is because, even after $\lfloor 2N {w_k^*}\rfloor$ virtual single-code users are each  allocated the FDMA-equivalent bandwidth of $1/(2N)$, the remaining FDMA-equivalent bandwidth $(2Nw_k^*-\lfloor 2N {w_k^*}\rfloor)/(2N)<1/(2N)$ can be arbitrarily distributed to the rest $\bar{n}_k-\lfloor 2N {w_k^*}\rfloor$ virtual single-code users.
Thus, it becomes of interest to investigate the problem of minimizing the number of active signature sequences.

\begin{theorem}\label{theorem: min number}
The minimum number $\min n_k^*$ of active signature sequences of the $k$th multi-code user to still retain the maximum sum rate is given by
%\begin{equation}\label{eq: nkmin}
%\min n_k^*=\left\{
%\begin{array}{ll}
%\bar{n}_k, & \text{ for } 1\leq k\leq {K_2},\\%
%\lceil 2N {w_k^*} \rceil, & \text{ for } {K_2} < k \leq K,
%\end{array}
%\right.
%\end{equation}
\begin{IEEEeqnarray}{C}\label{eq: nkmin}
\min n_k^*=\left\{
\begin{array}{l}
\bar{n}_k, \;\;\;\;\;\;\qquad\qquad\qquad \text{ for } 1\leq k\leq {K_2},  \\%
%\lfloor 2N {w_k^*} \rfloor, & \text{ for } {K_2} < k \leq K,
\Bigg\lceil
\Bigg(N-\!\!\displaystyle\sum_{k'=1}^{{K_2}}\bar{n}_{k'}\Bigg)\!\displaystyle\frac{ p_{k}}{\sum_{k'={K_2}+1}^{K}p_{k'}}\Bigg\rceil,
\end{array}
\right.\IEEEeqnarraynumspace\\%
\;\;\;\;\qquad\qquad\qquad\qquad\qquad\qquad \text{ for } {K_2} < k \leq K,\nonumber
\end{IEEEeqnarray}
where $\lceil \cdot \rceil$ denotes the ceiling operation.
\end{theorem}

\begin{IEEEproof}
For the oversized and the critically-sized multi-code users, we always have $n_k^*=\bar{n}_k$, as shown in {Theorem~\ref{theorem: opt power}}.
Thus, $\min n_k^*=\bar{n}_k, \forall k\leq {K_2}$.
For the undersized users, we always have $w_k^*< \bar{n}_k/(2N)$, as shown in {Remark~\ref{remark: 3a}}-(b).
By the condition (\ref{eq: conditions for wkl}), the minimum number of nonzero $({w_{k,l}^*})_l$ of the $k$th user of the equivalent FDMA system is given by $\lceil 2N {w_k^*} \rceil$.
Due to the equivalence in {Theorem~\ref{theorem: CDMA max sum rate}}, this number is the same as the minimum number of the active signature sequences of the $k$th user of the multi-code CDMA system.
Since the optimal FDMA-equivalent bandwidth ${w_k^*}$ of the $k$th multi-code user is given by (\ref{eq: FDMA equi sol2}), we have the conclusion.
\end{IEEEproof}

By using any $(w_{k,l}^*)_{k,l}$ that satisfies both the optimality condition in (\ref{eq: conditions for wkl}) and the condition $\sum_{l=1}^{\bar{n}_k}\text{sgn}({w_{k,l}^*})=\min n_k^*, \forall k$, to have the minimum numbers of multi-codes of users, the optimal power distribution can be obtained again as (\ref{eq: opt power}) in {Theorem~\ref{theorem: opt power}}.
Recall that this freedom in choosing ${w_{k,l}^*}$ applies only to the undersized multi-code users, because the non-undersized multi-code users have the unique optimal power distribution that results in ${w_{k,l}^*}=1/(2N),\forall k\leq {K_2}, \forall l$, and $\bar{n}_k=\max n_k^{\perp}=\min n_k^*$ orthogonal signature sequences.

Note that the maximum number $\max n_k^{\perp}$ of orthogonal signature sequences and the minimum number $\min n_k^*$ of signature sequences can be achieved simultaneously by allocating $\lfloor 2N {w_k^*} \rfloor$ orthogonal codes to each undersized multi-code user.
In this case, the number of non-orthogonal GWBE sequences allocated to an undersized multi-code user becomes $\min n_k^*-\max n_k^{\perp}=\lceil 2N {w_k^*} \rceil - \lfloor 2N {w_k^*} \rfloor =0$ or $1$.
This implies that the total minimum number of non-orthogonal GWBE sequences is upper bounded by $K-K_2$.
This also implies that, interestingly, an undersized multi-code user may not have any non-orthogonal GWBE sequence but only orthogonal sequences if $\min n_k^*-\max n_k^{\perp}=0$.
Consequently, the multi-code CDMA system with undersized multi-code users may only have orthogonal sequences if $\min n_k^*-\max n_k^{\perp}=0, \forall k>K_2$.

Finally, unlike the minimization in {Theorem~\ref{theorem: min number}} of the number of signature sequences given the power and the upper-limit profile of users, we may consider the minimization of the upper limits on the number of multi-codes, subject to the achievement of the sum capacity of the MAC.

\begin{remark}\label{remark: 12}
The minimal\footnote{For the definition of minimality, see \cite{Boyd}.} of the upper-limit profiles $\underline{\bar{n}}$ on the numbers of multi-codes of users to achieve the sum capacity of the MAC has the $k$th entry $\min \bar{n}_k$ given by
\begin{equation}\label{eq: min nk2}
\min \bar{n}_k= \left\lceil  N\cdot\left(\frac{p_k}{\sum_{k'=1}^{K}p_{k'}}\right) \right\rceil,
\end{equation}
for $k=1,2,\cdots,K$.
\end{remark}

\begin{IEEEproof}
Given any pair of the processing gain and the power profile of users,
we can always find $\underline{\bar{n}}$ with finite entries such that
\begin{equation}\label{eq: min nk2 cond}
\frac{p_k}{\bar{n}_k}\leq \frac{\sum_{k'=1}^{K}p_{k'}}{N}, \forall k.
\end{equation}
Note that ${p_1}/{\bar{n}_1}= \max {p_k}/{\bar{n}_k}\leq (\sum_{k'=1}^{K}p_{k'})/N$ if $\underline{\bar{n}}$ satisfies (\ref{eq: min nk2 cond}), while ${p_1}/{\bar{n}_1}> (\sum_{k'=1}^{K}p_{k'})/N$ otherwise.
Thus, for any such $\underline{\bar{n}}$, there is no oversized user by {Definition~\ref{def: over}}, i.e., ${K_1}=0$, and the FDMA-equivalent bandwidth of the optimal multi-code system is given by $w_k^*=p_k/(2\sum_{k'=1}^{K}p_{k'}), \forall k$, by (\ref{eq: FDMA equi sol}).
On the other hand, by (\ref{eq: FDMA equi sol2}) and (\ref{eq: nkmin}), we have $\min n_k^*=\bar{n}_k= 2Nw_k^*, \forall k\leq{K_2}$, and $\min n_k^*= \lceil 2Nw_k^* \rceil, \forall k>{K_2}$.
Thus, we can rewrite (\ref{eq: nkmin}) simply as $\min n_k^*=\lceil 2N {w_k^*} \rceil, \forall k$, in this case.
Therefore, the conclusion follows.
\end{IEEEproof}

By (\ref{eq: min nk2}), the total minimum number of multi-codes required to achieve the sum capacity of the MAC can always be upper bounded as
\begin{IEEEeqnarray}{rCl}\label{eq: upper}
&&\sum_{k=1}^K \min \bar{n}_k = \sum_{k=1}^K N\cdot\left(\frac{p_k}{\sum_{k'=1}^{K}p_{k'}}\right) \nonumber\\%
&&\qquad\;+ \sum_{k=1}^K\left\{ \left\lceil  N\cdot\left(\frac{p_k}{\sum_{k'=1}^{K}p_{k'}}\right) \right\rceil
-N\cdot\left(\frac{p_k}{\sum_{k'=1}^{K}p_{k'}}\right)\right\}\nonumber\\%
&&\qquad\;\leq N+(K-1),
\end{IEEEeqnarray}
regardless of the power profile of users, where we used $0\leq \lceil x \rceil - x < 1, \forall x$.
Note that this upper bound coincides with that in \cite[Lemma~5]{Guess3}, where it is shown that the upper bound on the total minimum number of multi-codes is $N+(K-1)$ to have no oversized users regardless of the rate profile of users in minimizing the sum power.
Unlike the result in \cite[Lemma~5]{Guess3}, we have derived this upper bound without assuming the equal-power distribution to the virtual single-code users of a multi-code user.

Recall that it is necessary for a single-code CDMA system to be critically-loaded or overloaded in order to achieve the sum capacity of the MAC.
Otherwise, some system resource is wasted because only $K$ orthogonal sequences are allocated for the $N>K$ dimensional signal space.
%For example, given $K$ users, (\ref{eq: upper}) says that there exists an optimal single-code CDMA system that achieves the sum capacity of the MAC regardless of the power profile, only
%when $N = 1$.
On the contrary, it is not necessary for a multi-code CDMA system to be critically-loaded or overloaded in order to achieve the sum capacity of the MAC.
This is because there exists an upper-limit profile that satisfies (\ref{eq: min nk2 cond}) as far as total of at least $N+(K-1)$ multi-codes are allowed, even when the system is underloaded, which may be a more interesting system loading condition in some applications.
% regardless of the power profile and the system loading of the multi-code system.
%
%, regardless of the power profile,
%
%a multi-code CDMA system can achieve the sum capacity of the MAC regardless of the power profile,
%
%given $K$ users, (\ref{eq: upper}) says that
%
%Even in those loading conditions that do not waste any signal dimension, it is not guaranteed that there exists a system that achieves the sum capacity of the MAC.
%
%For example, given $K$ users, (\ref{eq: upper}) says that a single-code CDMA system can achieve the sum capacity of the MAC regardless of the power profile, only when $N=1$.
%On the contrary, given $K$ users, (\ref{eq: upper}) says that a multi-code CDMA system can achieve the sum capacity of the MAC regardless of the power profile,
%
%there exists a multi-code system with total $N+(K-1)$ multi-codes that achieves the sum capacity of the MAC, even when the system is underloaded, which may be a more interesting system loading condition in some applications.
Thus, this finiteness of the upper bound shows that the sub-optimality of the CDMA as a multiple-access scheme is due only to the excessive restriction on the numbers of multi-codes and that it can always be overcome by a bounded system complexity.

%In other words, given $K$ users, the sum capacity of the MAC can be achieved whether the system is underloaded, critically-loaded, or overloaded, as far as at least the total of $N+(K-1)$ multi-codes are allowed in the system.

%미니멈 코드수를 정한후에 코드별 파워를 equal-power로 하는것도 가능
%=-=-=-=-=-=-=-=-=-=-=-=-=-=-=-=-=-=-=-=-=-=-=-=-=-=-=-=-=-=-=-=-=-=-=-=-=-=-=-=-=-=-
\subsection{Extension to Symbol-Asynchronous but Chip-Synchronous Multi-Code System}\label{sec: CDMA-asynch}
%=-=-=-=-=-=-=-=-=-=-=-=-=-=-=-=-=-=-=-=-=-=-=-=-=-=-=-=-=-=-=-=-=-=-=-=-=-=-=-=-=-=-
So far, only symbol-synchronous multi-code system is considered.
We now consider in this subsection the symbol-asynchronous but chip-synchronous multi-code CDMA system.

In \cite{Ulukus4}, the maximum sum rate is completely characterized for the symbol-asynchronous but chip-synchronous single-code CDMA system as an extension of the work in \cite{Viswanath : optimal0} for the symbol-synchronous single-code system.
The optimal signature sequences are derived and shown to be orthogonal sequences for oversized single-code users, while generalized asynchronous WBE (GAWBE) sequences for non-oversized users.
It is also shown that, regardless of the delay profile, the maximum sum rate of the asynchronous single-code system is the same as that of the synchronous single-code system as long as the two systems have the processing gain and the power profile of users in common.

In this subsection, we briefly discuss how to joint optimally distribute the power and allocate signature sequences to maximize the sum rate of the symbol-asynchronous but chip-synchronous multi-code system as an extension of the work in the previous subsections for the symbol-synchronous multi-code system.
To proceed, define
\begin{IEEEeqnarray}{rCl}
\underline{\tau}_k&\triangleq& [\tau_{k,1},\tau_{k,2},\cdots,\tau_{k,\bar{n}_k}],\IEEEyessubnumber\label{eq: tau_k}\\%
\noalign{\noindent{\text{and}}\vspace{\jot}}%
\underline{\tau}&\triangleq& [\underline{\tau}_1,\underline{\tau}_2,\cdots,\underline{\tau}_K],\IEEEyessubnumber\label{eq: tau}
\end{IEEEeqnarray}
as the relative delay profile of the $k$th multi-code user for $k=1,2,\cdots,K$, and the relative delay profile of all $\sum_{k=1}^{K}\bar{n}_k$ data streams of the multi-code system, respectively.
Since it is reasonable to model that all $\bar{n}_k$ data streams of the $k$th user have the same delay, we assume that there exists $\tau_k\in\{0,1,2,\cdots,N-1\}$ such that
\begin{equation}
\tau_k= \tau_{k,l},\forall l,
\end{equation}
for each $k=1,2,\cdots,K$.

By the same reason as {Lemma}~\ref{lemma: diagonal}, the data symbols are assumed independent without loss of generality, not only when they are from different multi-code users but also when they are from the same multi-code user.
This enables us to view the asynchronous multi-code system as an asynchronous system with virtual single-code users, which leads to the following theorem.

\begin{theorem}\label{theorem: asynch}
Given a power profile $\underline{p}$, an upper-limit profile $\underline{\bar{n}}$, a processing gain $N$, and the variance $\sigma^2$ per dimension of the AWGN,
the maximum sum rate $\mathcal{C}_{\text{CDMA}}^{\text{async}}(\underline{p},\underline{\tau},\underline{\bar{n}},N,\sigma^2)$ of the symbol-asynchronous but chip-synchronous multi-code CDMA system with delay profile $\underline{\tau}$ is the same as that of the symbol-synchronous multi-code CDMA system, i.e.,
%
%The maximum sum rate $\mathcal{C}_{\text{CDMA}}^{\text{Async}}(\underline{p},\underline{\tau},\underline{\bar{n}},N,\sigma^2)$ of the symbol-asynchronous but chip-synchronous multi-code CDMA system with power profile $\underline{p}$, delay profile $\underline{\tau}$, upper-limit profile $\underline{\bar{n}}$, processing gain $N$, and variance $\sigma^2$ per dimension of the AWGN is equal to that of the symbol-synchronous multi-code CDMA system with the same $\underline{p}$, $\underline{\bar{n}}$, $N$, and $\sigma^2$, i.e.,
\begin{equation}
\mathcal{C}_{\text{CDMA}}^{\text{async}}(\underline{p},\underline{\tau},\underline{\bar{n}},N,\sigma^2)=\mathcal{C}_{\text{CDMA}}(\underline{p},\underline{\bar{n}},N,\sigma^2),
\end{equation}
where $\mathcal{C}_{\text{CDMA}}(\underline{p},\underline{\bar{n}},N,\sigma^2)$ is given by (\ref{eq: CDMAmaxsumrate}).
\end{theorem}

\begin{IEEEproof}
A sketch of the proof is as follows.
If we view the system as an asynchronous system with virtual single-code users, then we can formulate a double maximization problem to find the maximum sum rate by taking the same procedures as those taken to obtain {Problem~\ref{problem: CDMA outer inner}} for the synchronous multi-code system.
The outer optimization of this double maximization problem is again to be solved over the diagonal power matrix $P$ under the same constraints (\ref{eq: sumnk3}) and (\ref{eq: sumnk33}) as that of {Problem~\ref{problem: CDMA outer inner}} is, while the inner optimization is now to be solved over the signature sequences of the asynchronous virtual single-code users.
As already mentioned, the asynchronous single-code system has the same maximum sum rate as the synchronous system does \cite{Ulukus4}.
So, given a feasible $P$, the inner optimization over the signature sequences returns the same maximum sum rate as the corresponding synchronous system with virtual single-code users does.
Moreover, by {Lemma~\ref{lemma: replace}}, the synchronous system with virtual single-code users has the same maximum sum rate as the equivalent restricted FDMA system does.
Thus, we again reach {Problem~\ref{problem: replaced}} by replacing the inner optimization problem with the sum-rate maximization problem for the equivalent restricted FDMA system.
The next procedures are exactly the same as those used to obtain {Theorem~\ref{theorem: CDMA max sum rate}}.
Therefore, we have the conclusion.
\end{IEEEproof}

Since the sum-rate maximization problems for the synchronous and the asynchronous multi-code systems share {Problem~\ref{problem: replaced}}, it follows that the necessary and sufficient condition for the asynchronous multi-code system to achieve the sum capacity of the MAC is the same as the condition given in  {Theorem~\ref{theorem: nece suffi}}, and that the optimal power distribution is the same as the distribution given in {Theorem~\ref{theorem: opt power}}.
Once an optimal power distribution $(p_{k,l}^*)_{k,l}$ is found, the optimal signature sequences for the asynchronous system with virtual single-code users can be found by directly using the results in \cite{Ulukus4}.
Other results in the previous sections can also be similarly extended.

%=-=-=-=-=-=-=-=-=-=-=-=-=-=-=-=-=-=-=-=-=-=-=-=-=-=-=-=-=-=-=-=-=-=-=-=-=-=-=-=-=-=-
\section{Numerical Results and Discussions}\label{sec: discussion}
%=-=-=-=-=-=-=-=-=-=-=-=-=-=-=-=-=-=-=-=-=-=-=-=-=-=-=-=-=-=-=-=-=-=-=-=-=-=-=-=-=-=-
In this section, we provide numerical results and discussions, which include the comparison of the sum-rate optimal multi-code CDMA system with the multi-code CDMA system having random signature sequences.
%=-=-=-=-=-=-=-=-=-=-=-=-=-=-=-=-=-=-=-=-=-=-=-=-=-=-=-=-=-=-=-=-=-=-=-=-=-=-=-=-=-=-
\subsection{Sum-Rate Optimal Multi-Code CDMA System}\label{sec: tradeoff}
%=-=-=-=-=-=-=-=-=-=-=-=-=-=-=-=-=-=-=-=-=-=-=-=-=-=-=-=-=-=-=-=-=-=-=-=-=-=-=-=-=-=-

\begin{figure}[tbp]\centering%
%\vspace{-0.02in}
\includegraphics[width=6in]{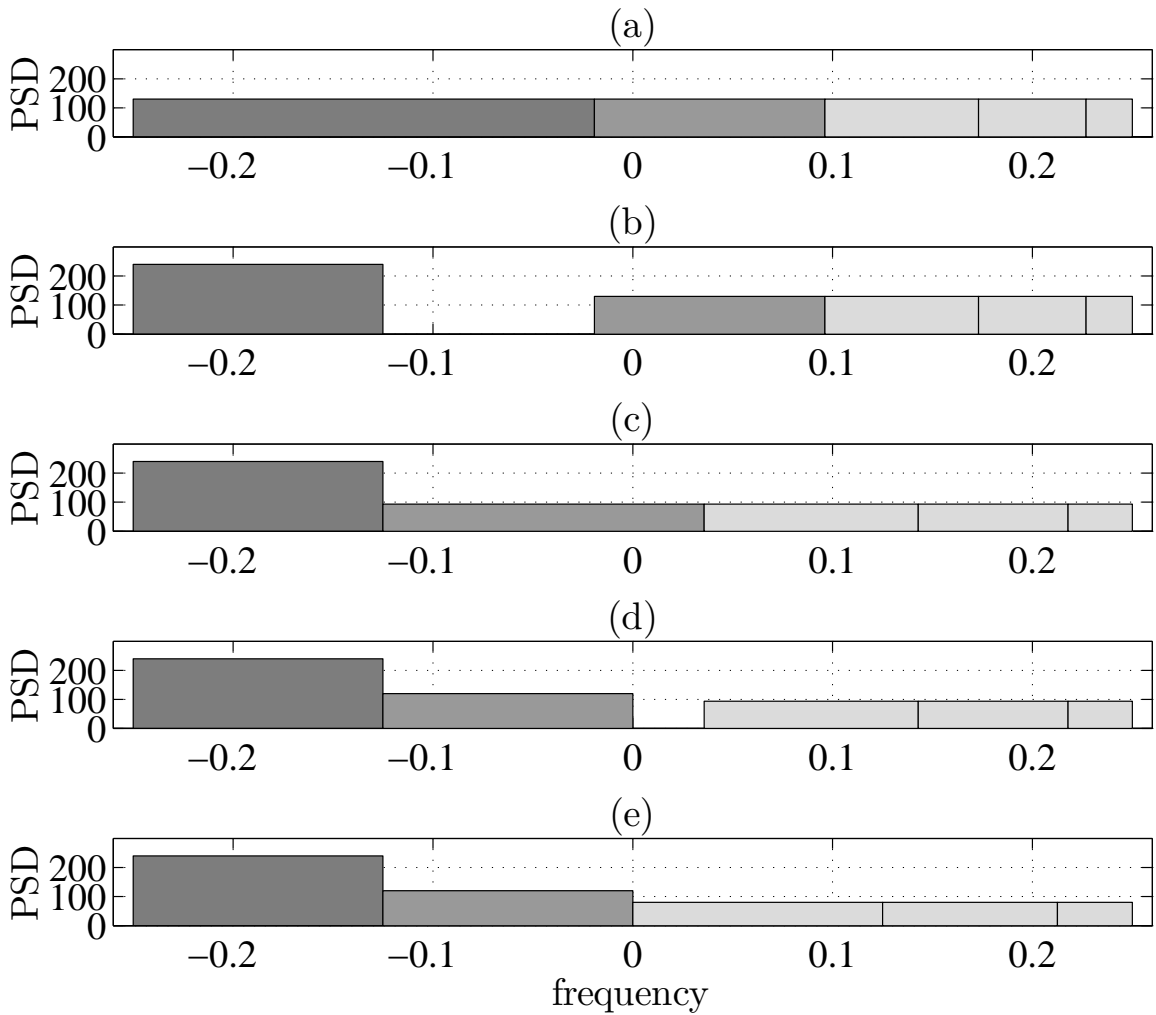} %,height=2.2in%4.5in
\caption{Allocation of the FDMA-equivalent bandwidths by the iterative algorithm in TABLE~\ref{table: 1}. (a) after line $3$ of the first iteration, (b) after the first iteration, (c) after line $3$ of the second iteration, (d) after the second iteration, and (e) after the third iteration.}\label{fig: non-over}
\end{figure}

% The use of the user PSDs also helps us visually understand the above remark that shows how the imposition of the upper limits affects the bandwidth allocation.
The first numerical result is to illustrate how the bandwidths and the PSDs of users in the equivalent FDMA system are updated as the algorithm in Table~\ref{table: 1} iterates.
In this example, there are $5$ users with processing gain $8$, power profile $\underline{p}=[30,15,10,7,3]$, and upper-limit profile $\underline{\bar{n}}=[2,2,2,2,2]$, which results in ${K_1}=2$, ${K_2}=3$, and the profile of the optimal FDMA-equivalent bandwidths $\underline{w}^*=[1,1,1,0.7,0.3]/8\approx[0.125,0.125,0.125,0.088,0.038]$.
Fig.~\ref{fig: non-over}-(a) shows the tentative allocation in the line $3$ of the first iteration, where the bandwidths are the due shares given by $[30/65,15/65,10/65,7/65,3/65]/2\approx[0.231,0.115,0.077,0.054,0.023]$.
Fig.~\ref{fig: non-over}-(b) shows the optimal bandwidth of the first user and the tentative allocation to the other users after the first iteration, where the first user is allocated less bandwidth than the due share because the FDMA-equivalent bandwidth-constraint profile is $\underline{\bar{w}}=[1,1,1,1,1]/8$.
Fig.~\ref{fig: non-over}-(c) shows the optimal bandwidth of the first user and the tentative allocation computed in the line $3$ of the second iteration.
Note that a non-oversized user has the tentatively allocated bandwidth greater than that in the first iteration and, consequently, its user PSD has become smaller.
Fig.~\ref{fig: non-over}-(d) shows the optimal bandwidths of the first two users and the tentative allocation to the other users after the second iteration.
Fig.~\ref{fig: non-over}-(e) shows the optimal bandwidths allocated after the third, which is final in this example, iteration.
Note that a non-oversized user has the allocated bandwidth greater than that in the second iteration and, consequently, its user PSD has become smaller.
We can see from Fig.~\ref{fig: non-over}-(e) that the critically-sized user with index $k=3$ is allocated the bandwidth upper limit, and that the undersized-user with index $k=4$ or $5$ is allocated a less bandwidth than the bandwidth upper limit, as shown in Remark~\ref{remark: 3a}.
We can also see that the non-oversized users with indexes $k=3,4,$ and $5$ have the same PSDs, that the optimal PSDs of users are non-increasing, and that the oversized user with index $k=1$ or $2$ has a greater PSD than that of the non-oversized user with index $k=3,4,$ or $5$, as shown in Remark~\ref{remark: 4a}.

%TTTTTTTTTTTTTTTTTTTTTTTTTTTTTTTTTTTTTTTTTTTTTTTTTTTTTTTTT
%TTTTTTTTTTTTTTTTTTTTTTTTTTTTTTTTTTTTTTTTTTTTTTTTTTTTTTTTT

\begin{figure}[tbp]\centering%
\includegraphics[width=6in]{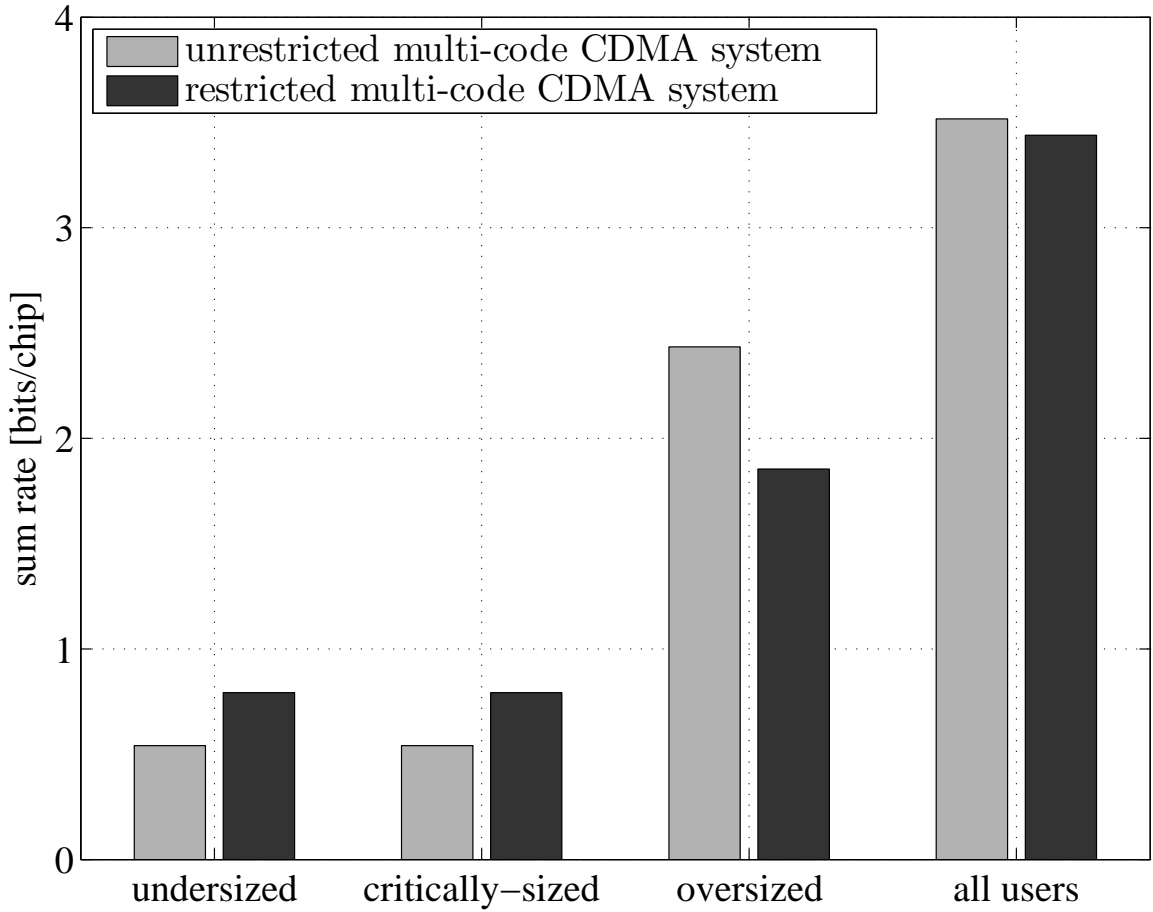} %,height=2.2in%4.5in
\caption{Comparison of the sum rates of the users of the unrestricted system of Fig.~\ref{fig: non-over}-(a) and those in the restricted multi-code CDMA system of Fig.~\ref{fig: non-over}-(e).
%The restricted system increases the sum rate of the non-oversized users at the cost of the reduced sum rate of the oversized users.
}\label{fig: sum rate}
\end{figure}

The second numerical result compares the total sum rate and the sum rates of the oversized, the critically-sized, and the undersized users of the restricted multi-code CDMA system in Fig.~\ref{fig: non-over}-(e) against those of the unrestricted CDMA system in Fig.~\ref{fig: non-over}-(a).
The sum rates of the oversized, the critically-sized, and the undersized users of the unrestricted multi-code CDMA system are, respectively, the summations of the rates of the corresponding users of the restricted system.
Note that, in Fig.~\ref{fig: non-over}-(e), the second user that is oversized has the optimal FDMA-equivalent bandwidth $1/8=0.125$.
However, in Fig.~\ref{fig: non-over}-(a), the second user has the due share $15/130\approx 0.115$, which is smaller than its optimal FDMA-equivalent bandwidth and leads to the second user's due rate less than its optimal rate.
This is the counter example mentioned in Remark~\ref{remark: 5.6}-(d).
We can see from Fig.~\ref{fig: sum rate} that the imposition of the equal upper limits mitigates the unfairness inherent in the sum-rate maximization in the sense that that the sum rate of the undersized users is also increased, the rate of the critically-sized user is increased, but that the sum rate of the oversized users is decreased, as shown in Remark~\ref{remark: 5.6}.
We can also see that a decrease in the total sum rate is unavoidable, as shown in Theorem~\ref{theorem: nece suffi} and, equivalently, in Remark~\ref{remark: 5a}, in order to improve the fairness.

%TTTTTTTTTTTTTTTTTTTTTTTTTTTTTTTTTTTTTTTTTTTTTTTTTTTTTTTTT
%TTTTTTTTTTTTTTTTTTTTTTTTTTTTTTTTTTTTTTTTTTTTTTTTTTTTTTTTT
%There are $5$ users with processing gain $N=4$, power profile $\underline{p}=[30\;15\;10\;7\;3]$, and upper limit profile $\underline{\bar{n}}=[1\;1\;1\;1\;1]$.

\begin{figure}[tbp]\centering%
\includegraphics[width=6in]{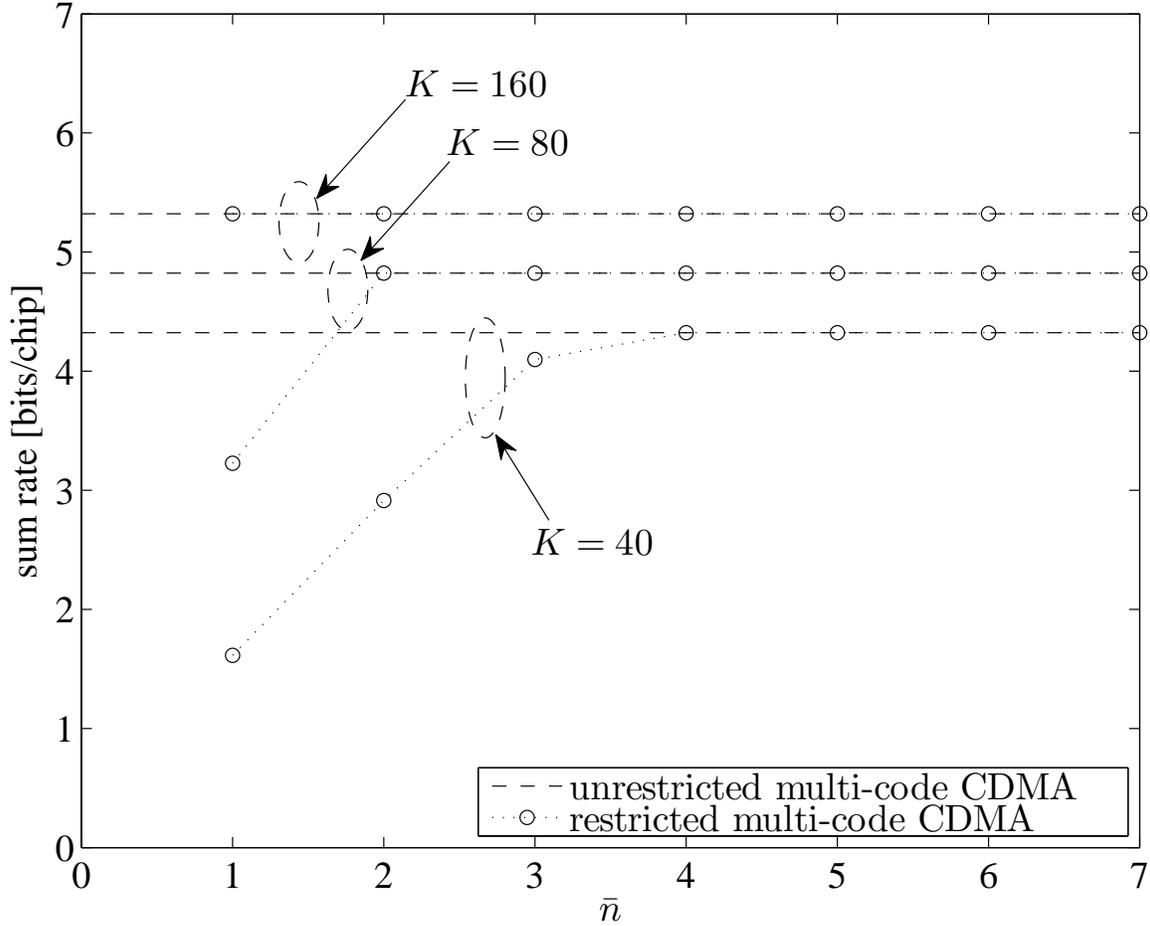} %,height=2.2in%4.5in
\caption{Effect of the equal upper limit $\bar{n}$ on $n_k, \forall k$, on the sum rate of the restricted multi-code CDMA system with $K=40, 80,$ and $160$, $N=128$, and $p_k/\sigma^2=10$ [dB], $\forall k$.}\label{fig: equal power}
\end{figure}

The third numerical result shows the sum rate of the restricted multi-code CDMA system as a function of the upper limit on the number of multi-codes.
For simplicity, we consider three cases of system loading with $K=40, 80$, and $160$, and $N=128$, where the $k$th user has the upper limit $\bar{n}$ on the number $n_k$ of multi-codes and the power that leads to $p_k/\sigma^2=10$ [dB] for all $k$.
Fig.~\ref{fig: equal power} shows both the sum rates of the three restricted multi-code CDMA systems that are evaluated by using the formula in Theorem~\ref{theorem: CDMA max sum rate} and the sum rates of the three corresponding unrestricted multi-code CDMA systems that are evaluated by using the formula in (\ref{eq: MAC}).
In \cite{Rupf}, it is shown that a single-code CDMA system with symmetric-power users achieves the sum capacity of the MAC if and only if the system is overloaded or critically loaded.
Thus, for low $K/N < 1$ such as state-of-the-art CDMA \cite{Verdu : LSA}, the maximum sum rate of a single-code CDMA system falls short of the sum capacity, which we can see from Fig.~\ref{fig: equal power} with $K/N=40/128$ and $80/128$.
However, even for low $K/N < 1$, the sum rate of the multi-code systems increase as $\bar{n}$ increases and eventually becomes equal to the sum capacity.
This is because, for $\bar{n}\leq N/K$, each user has the FDMA-equivalent bandwidth $\bar{n}/(2N)$ by (\ref{eq: FDMA equi sol}) in Proposition~\ref{prop: eFDMA}, which makes the total allocated FDMA-equivalent bandwidth less than the total available system bandwidth, i.e., $\bar{n}K/(2N)<1/2$, but makes more and more FDMA-equivalent bandwidth utilized by the system as $\bar{n}$ increases.
For $\bar{n}\geq N/K$, the necessary and sufficient condition to achieve the sum capacity of the MAC in Theorem~\ref{theorem: nece suffi} is met by the multi-code CDMA system with symmetric-power users.
Note that the system with a heavier loading achieves a higher sum capacity $(1/(2N))\log(1+(Kp_k)/(N_0/2))$ because the more the symmetric-power users the larger the total system power.
%우리는 Fig.~\ref{fig: equal power}을 통해 이 영역에서 multi-code CDMA system이 single-code CDMA에 비하여 sum rate가 얼마나 향상되는지를 볼 수 있다.
%asymmetric power의 경우는 $\bar{n}K/N$값이 1이 된다고 해서 반드시 sum capacity of MAC을 achieve하는 것은 아니다.

%TTTTTTTTTTTTTTTTTTTTTTTTTTTTTTTTTTTTTTTTTTTTTTTTTTTTTTTTT
%TTTTTTTTTTTTTTTTTTTTTTTTTTTTTTTTTTTTTTTTTTTTTTTTTTTTTTTTT

The above arguments are alternatively justified with the help of Remark~\ref{remark: equal power}, which shows that the equal distribution of each user's  power to its multiple data streams is optimal.
Thus, when the upper limits $\bar{n}_k$ are equal to $\bar{n}$ for all $k$, a multi-code CDMA system with $K$ symmetric-power users can be replaced by a CDMA system with $\bar{n}K$ symmetric-power virtual single-code users as far as the maximum sum rate is concerned.
Since the CDMA system with symmetric-power virtual single-code users has the system loading $\bar{n}K/N$, the sum rate increases for $\bar{n}K/N\leq 1$ by allocating orthogonal sequences and becomes equal to the sum capacity of the MAC for $\bar{n}K/N> 1$ by allocating WBE sequences.
Although this approach much straightforwardly explains the behavior of the optimal sum rate as a function of $\bar{n}$, the use of $\bar{n}K$ WBE sequences is not the only optimal way to design a sum-rate optimal system.
As mentioned immediately after Theorem~\ref{theorem: max ortho}, even an undersized multi-code user may have orthogonal sequences, which comes from an optimal unequal distribution of each user's  power to its multiple data streams.
This complexity issue is discussed in the next result.

\begin{figure}[tbp]\centering%
\includegraphics[width=6in]{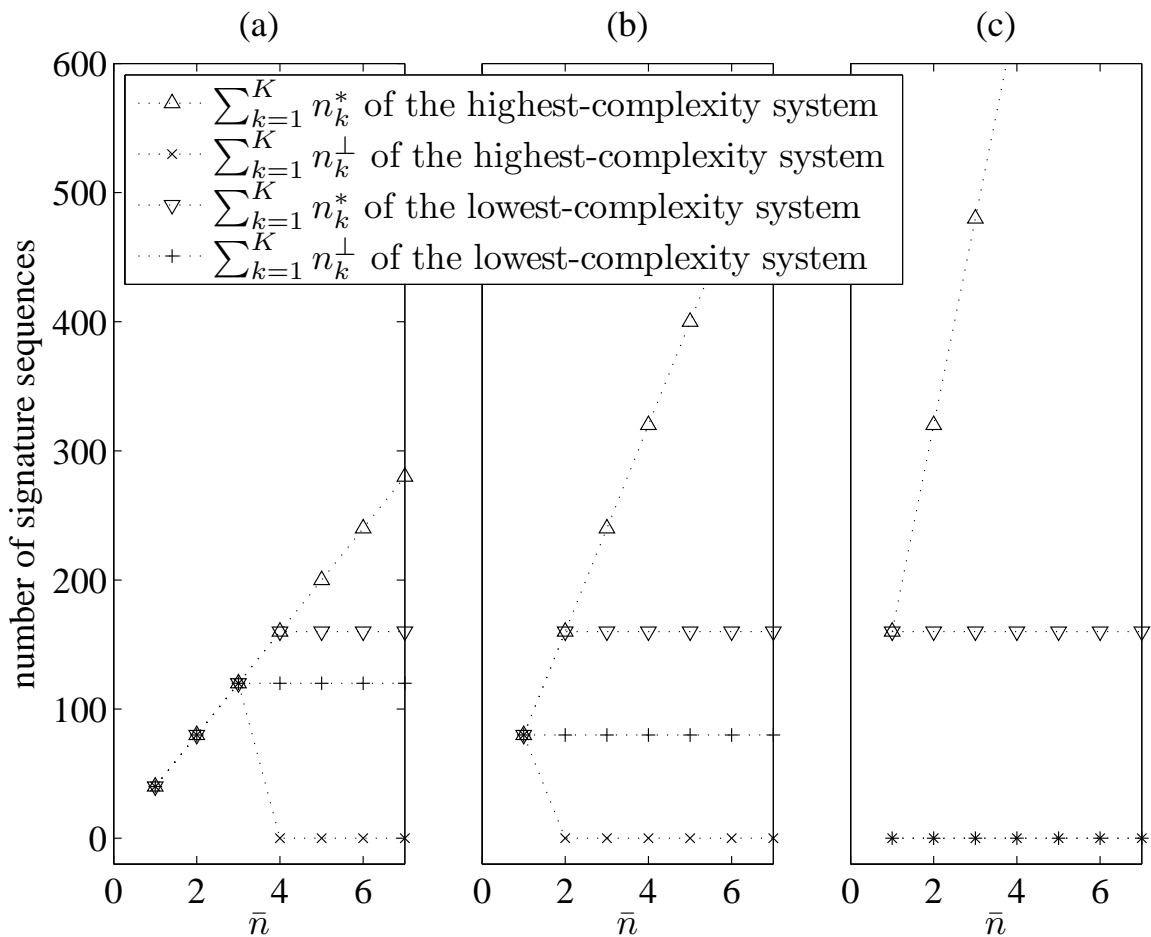} %,height=2.2in%4.5in
\caption{Numbers of total and orthogonal signature sequences of the highest- and the lowest-complexity restricted multi-code CDMA systems having the parameters in common with Fig.~\ref{fig: equal power}: (a) $K=40$, (b) $K=80$, and (c) $K=160$.}\label{fig: unequal power2}
\end{figure}

The fourth numerical result shows the total number of signature sequences and the number of orthogonal signature sequences of the highest- and the lowest-complexity restricted multi-code CDMA systems having the parameters in common with Fig.~\ref{fig: equal power}.
The highest-complexity system is the system considered in Remark~\ref{remark: equal power}, where each user has the same  power $p_k, \forall k$, and distributes it equally to all available $\bar{n}$ data streams of each user.
This system has the highest complexity in the sense that the total number of signature sequences is maximized as $n_k^*=\bar{n}, \forall k,$ but the number of orthogonal sequences is minimized as $n_k^\perp=0$ for $\bar{n}K/N>1$, among all optimal multi-code CDMA systems.
The lowest-complexity system is the system considered in Theorems~\ref{theorem: max ortho} and~\ref{theorem: min number}, which maximizes the number of orthogonal codes and at the same time minimizes the total number of multi-codes, among all optimal multi-code CDMA systems.
We can see from Fig.~\ref{fig: unequal power2} that, for $\bar{n}K/N\leq 1$, the highest- and the lowest-complexity systems have the same total numbers of signature sequences and the same numbers of orthogonal sequences.
This is because all the multi-code users of an optimal multi-code CDMA system are non-undersized in this case by Definition~\ref{def: over}, which makes all the optimal systems have $n_k^*=n_k^\perp=\bar{n}, \forall k$, by Remark~\ref{remark: ortho}.
We can also see that the portion of the complexity reduction obtained from a large number of orthogonal sequences vanishes only for the system loading $K/N\geq 1$ because the lowest-complexity system with the minimum number of signature sequences $n_k^*=\bar{n}, \forall k,$ also has $n_k^\perp=0, \forall k$.

%TTTTTTTTTTTTTTTTTTTTTTTTTTTTTTTTTTTTTTTTTTTTTTTTTTTTTTTTT
%TTTTTTTTTTTTTTTTTTTTTTTTTTTTTTTTTTTTTTTTTTTTTTTTTTTTTTTTT

%=-=-=-=-=-=-=-=-=-=-=-=-=-=-=-=-=-=-=-=-=-=-=-=-=-=-=-=-=-=-=-=-=-=-=-=-=-=-=-=-=-=-
\subsection{Comparison to Multi-Code CDMA System Having Random Signature Sequences}\label{sec: compare to random sequence}
%=-=-=-=-=-=-=-=-=-=-=-=-=-=-=-=-=-=-=-=-=-=-=-=-=-=-=-=-=-=-=-=-=-=-=-=-=-=-=-=-=-=-

In this subsection, we discuss the spectral efficiency of the multi-code CDMA systems having jointly sum-rate optimal power distribution and signature sequences.
In particular, the spectral efficiency is compared to that of the CDMA systems in \cite{Honig2}--\cite{Verdu : fading} having random signature sequences.
In \cite{Verdu : LSA}, the spectral efficiency of the single-code CDMA system with symmetric-power users is derived when random signature sequences and the optimal encoder-decoder pair are employed.
Since it is difficult to derive a closed-form spectral efficiency as a function of $N$, $K$, and the bit energy per noise density $E_b/N_0$, an asymptotic analysis later named a large system analysis is performed by letting $K$ and $N$ tend to infinity with $K/N$ being fixed.
In \cite{Verdu : fading}, this result is extended to single-code CDMA systems with independent and identically distributed (i.i.d.) flat-fading channel for each user.
In \cite{Honig2}, these results are extended to multi-code CDMA systems with i.i.d. flat-fading or frequency-selective fading channel for each multi-code user.
However, due to the difficulty in computing the spectral efficiency with a large number of multi-code users, the cases only with a small number of multi-code users or only with a large number of single-code users are considered in \cite{Honig2}.

In the previous section, we derived the sum rate of the multi-code CDMA system with asymmetric-power users, when the upper limits are imposed on the numbers of multi-codes of users and the jointly optimal signature sequences and encoder-decoder pair are employed.
To compare with the results in \cite{Verdu : LSA}, we only consider the cases with symmetric-power users, equal upper limits, and no channel fading.

\begin{remark}\label{remark: 13}
When the equal upper limits are imposed, the spectral efficiency in [bits/chip] of the sum-rate optimal multi-code CDMA system with symmetric-power users is given by
\begin{equation}
\textsf{C}_\text{CDMA}\left(\frac{\bar{n}K}{N},\frac{E_b}{N_0}\right)\!\triangleq\!\left\{
\begin{array}{ll}
\displaystyle \!\!\!\!\left(\frac{\bar{n}K}{N}\right)\textsf{C}_\text{SU}\left(\frac{E_b}{N_0}\right),&\!\!\!\!\text{for }\displaystyle\frac{\bar{n}K}{N}\leq1,\vspace{0.1in}\\
\displaystyle\!\!\textsf{C}_\text{SU}\left(\frac{E_b}{N_0}\right),&\!\!\!\!\text{elsewhere,}
\end{array}
\right.\label{eq: SE vs KN equal2}
\end{equation}
where $\bar{n}$ is the equal upper limit and $\textsf{C}_\text{SU}(E_b/N_0)$ satisfying
\begin{IEEEeqnarray}{lCl}\label{eq: AWGN}
\textsf{C}_\text{SU}\left(\frac{E_b}{N_0}\right)=\frac{1}{2}\log\left(1+2\textsf{C}_\text{SU}\left(\frac{E_b}{N_0}\right)\frac{E_b}{N_0}\right)
\end{IEEEeqnarray}
is the spectral efficiency in [bits/dimension] of the optimal single-user system operating over AWGN.
\end{remark}

\begin{IEEEproof}
Let $p_\text{tot}$ be the total signal power and $\bm{1}_K\triangleq [1,1,\cdots,1]^T$ be the length-$K$ all-one vector.
Then, from Theorem~\ref{theorem: CDMA max sum rate}, we have the maximum sum rate given by
\begin{IEEEeqnarray}{lCl}
\mathcal{C}\triangleq\mathcal{C}_\text{CDMA}\bigg(\frac{p_\text{tot}}{K}\bm{1}_K,\bar{n}\bm{1}_K,N,\sigma^2\bigg)\nonumber\\
=\left\{
\begin{array}{ll}
\displaystyle\!\!\!\frac{1}{2}\left(\frac{\bar{n}K}{N}\right)\log\left(1+\bigg(\frac{p_\text{tot}}{\sigma^2}\bigg)\big/\bigg(\frac{\bar{n}K}{N}\bigg)\right),&\!\!\!\text{for } \displaystyle\frac{\bar{n}K}{N}\leq 1,\vspace{0.1in}\\
\displaystyle\!\!\!\frac{1}{2}\log\left(1+\frac{p_\text{tot}}{\sigma^2}\right),&\!\!\!\text{elsewhere.}
\end{array}
\right.\nonumber\vspace{-4pt}\\
\label{eq: SE vs KN equal}\vspace{-7pt}
\end{IEEEeqnarray}
Since $p_\text{tot}/\sigma^2$, $\mathcal{C}$, and $E_b/N_0$ are related as \cite[Eq. (3)]{Verdu : LSA}
\begin{equation}
\frac{p_\text{tot}}{\sigma^2}=2\frac{E_b}{N_0}\mathcal{C},
\end{equation}
$\mathcal{C}$ in (\ref{eq: SE vs KN equal}) can be rewritten as $\mathcal{C}=(1/2)({\bar{n}K}/{N})\log(1+2\mathcal{C}(E_b/N_0)/({\bar{n}K}/{N}))$, for ${\bar{n}K}/{N}\leq 1$, and $\mathcal{C}=(1/2)$$\log(1+2\mathcal{C}({E_b}/{N_0}))$, elsewhere.
Therefore, by using (\ref{eq: AWGN}), we have (\ref{eq: SE vs KN equal2}).
\end{IEEEproof}

Note that, given $E_b/N_0$, the spectral efficiency in (\ref{eq: SE vs KN equal2}) of the sum-rate optimal multi-code CDMA system with symmetric-power users is a function of $\bar{n}$, $K$, and $N$ only through $\bar{n}K/N$.
Thus, the systems with different $(\bar{n},K,N)$ triples have the same performance as far as their effective system loadings $\bar{n}K/N$ are the same, which makes it needless to perform an asymptotic analysis that is done in \cite{Verdu : LSA}.

\begin{figure}[tbp]\centering%
\includegraphics[width=6in]{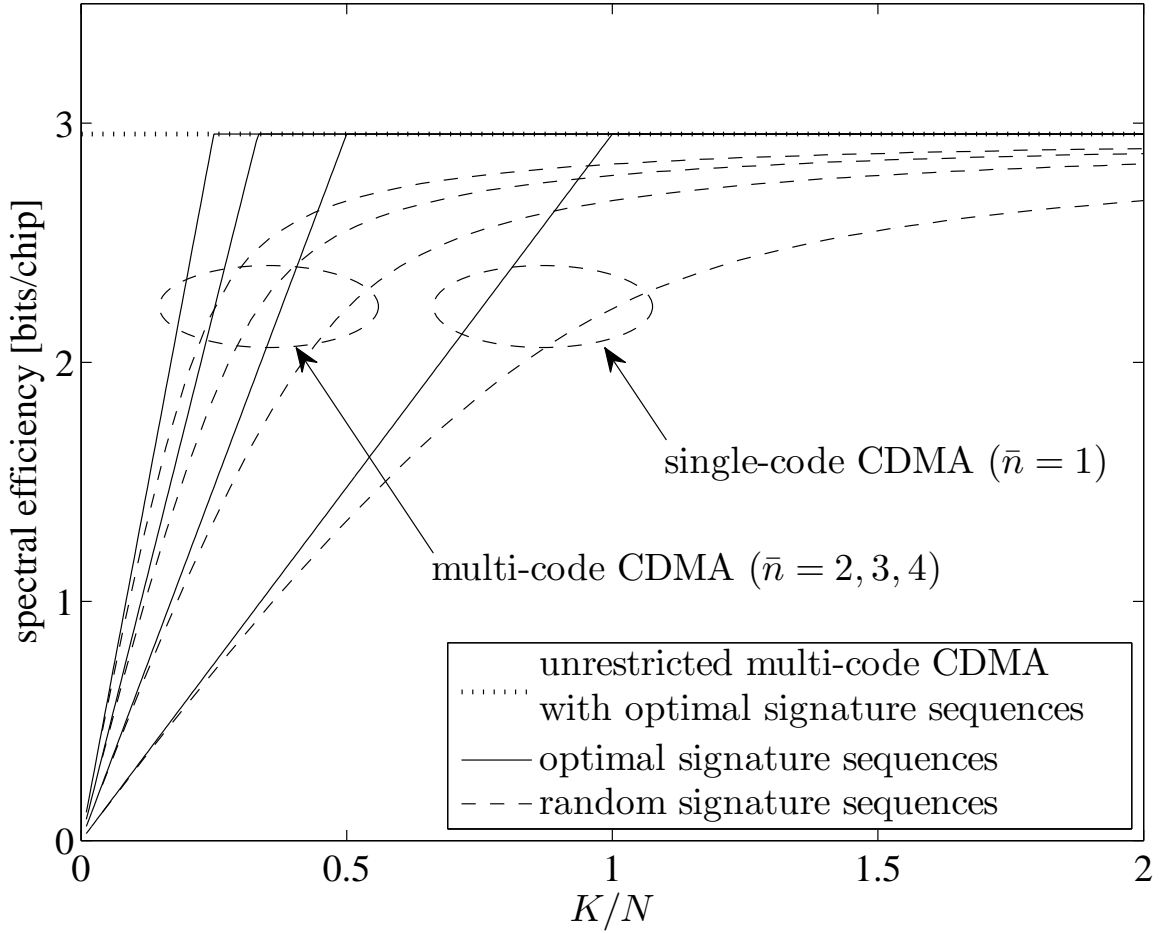} %,height=2.2in%4.5in
\caption{Comparison of spectral efficiency of the CDMA systems with sum-rate optimal sequences and random sequences, when $E_b/N_0=10$ [dB] and each symmetric-power user has the equal upper limit $\bar{n}$ on the number of multi-codes.}\label{fig: SE ver KN equalpower}
\end{figure}

Fig.~\ref{fig: SE ver KN equalpower} compares the spectral efficiency in (\ref{eq: SE vs KN equal2}) with the spectral efficiency of the single- and multi-code CDMA systems having random signature sequences, when $E_b/N_0=10$ [dB].
The unrestricted multi-code CDMA system having no upper limits always achieves the spectral efficiency $\textsf{C}_\text{SU}({E_b}/{N_0})$ of the optimal single-user system, while the single- and multi-code systems having sum-rate optimal sequences have the spectral efficiency linearly increasing for $K/N\leq 1/\bar{n}$ by using orthogonal sequences, while achieving $\textsf{C}_\text{SU}({E_b}/{N_0})$ elsewhere by using orthogonal, WBE, and/or GWBE sequences depending on the power distribution.
Note that, as $\bar{n}$ increases, the range of the system loading $K/N$ increases on which the system optimally utilizes the spectrum.
The spectral efficiency of a single-code CDMA system having random sequences is obtained by using the result in \cite{Verdu : LSA}, while that of a multi-code CDMA system having random sequences is obtained by scaling that of the single-code system in the horizontal axis under the assumption that every data stream has the same power.
Obviously, given $\bar{n}$ and $K/N$, the sum-rate optimal CDMA system always outperforms the CDMA system having random sequences.
Particularly, the performance gap is maximized at $K/N=1/\bar{n}$, i.e., when the effective system loading is critical, which extends the result in \cite{Verdu : LSA} obtained for the single-code systems.

%TTTTTTTTTTTTTTTTTTTTTTTTTTTTTTTTTTTTTTTTTTTTTTTTTTTTTTTTT
%TTTTTTTTTTTTTTTTTTTTTTTTTTTTTTTTTTTTTTTTTTTTTTTTTTTTTTTTT

\begin{figure}[tbp]\centering%
\includegraphics[width=6in]{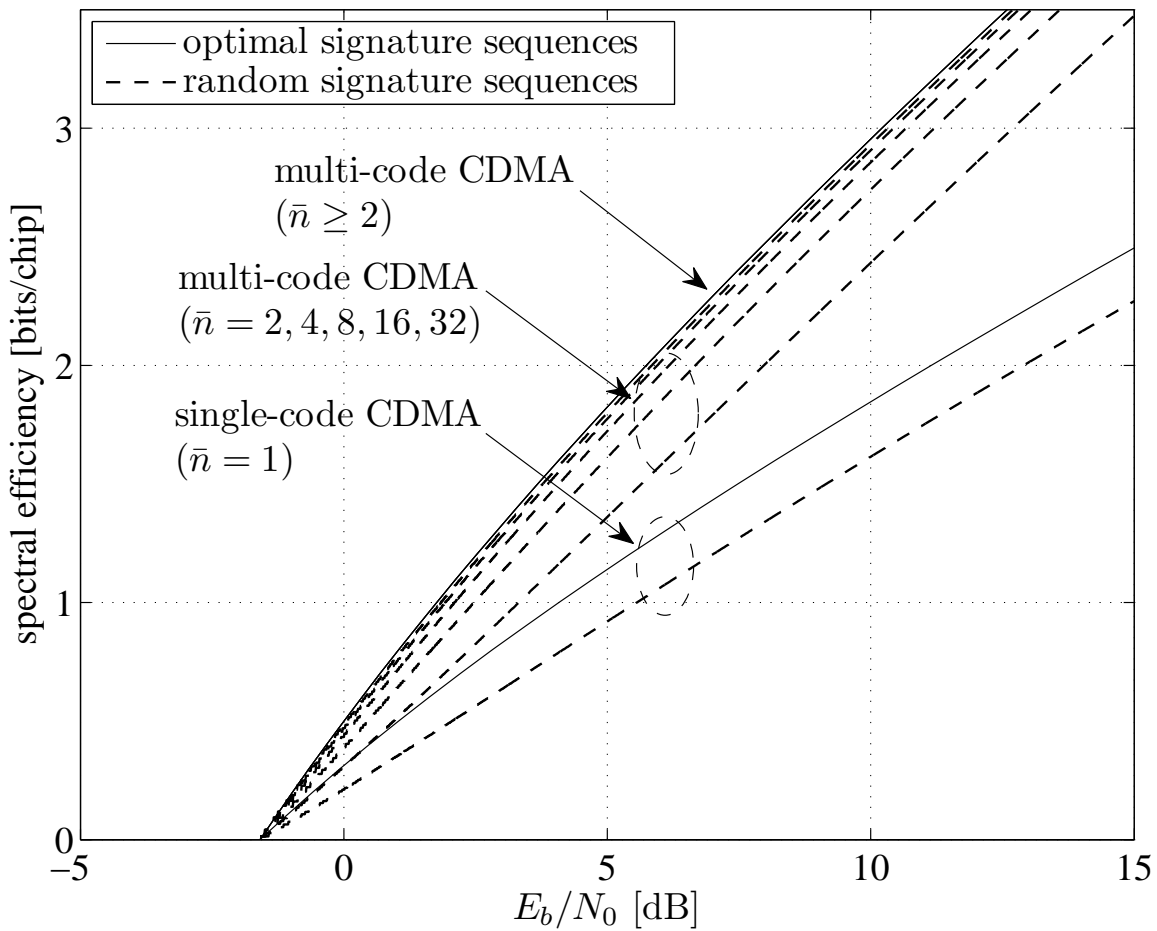} %,height=2.2in%4.5in
\caption{Comparison of spectral efficiency of the CDMA systems with sum-rate optimal sequences and random sequences, when $K/N=80/128=0.625$ and $\bar{n}=1,2,4,8,16,32$.
The uppermost line coincides with the spectral efficiency of the optimal single-user system.}\label{fig: SE ver EbN0 equalpower}
\end{figure}

Fig.~\ref{fig: SE ver EbN0 equalpower} performs the same comparison as Fig.~\ref{fig: SE ver KN equalpower} does.
Now, the spectral efficiency is plotted as a function of $E_b/N_0$ for a fixed system loading of $K/N=80/128=0.625$ and more values of $\bar{n}=1,2,4,8,16,$ and $32$.
In \cite{Verdu : LSA}, it is pointed out that the spectral efficiency of the single-code system having random sequences approaches $\textsf{C}_\text{SU}({E_b}/{N_0})$ as $K/N$ tends to infinity.
Recall that increasing $\bar{n}$ of the multi-code CDMA system having random sequences corresponds to increasing the system loading of the virtual single-code user CDMA system having random sequences.
Thus, as we can see from Fig.~\ref{fig: SE ver EbN0 equalpower}, the spectral efficiency of the multi-code system having random sequences approaches the spectral efficiency $\textsf{C}_\text{SU}({E_b}/{N_0})$ of the optimal single-user system, as the number $\bar{n}$ of multi-codes per user tends to infinity.
On the contrary, the sum-rate optimal multi-code system achieves $\textsf{C}_\text{SU}({E_b}/{N_0})$ for $\bar{n}\geq 2$ and, moreover, this can be done by using only two multi-codes per user as exemplified in Fig.~\ref{fig: unequal power2}-(b) with $K=80$ and $N=128$.
Similar results can be obtained for various system loading.

%TTTTTTTTTTTTTTTTTTTTTTTTTTTTTTTTTTTTTTTTTTTTTTTTTTTTTTTTT
%TTTTTTTTTTTTTTTTTTTTTTTTTTTTTTTTTTTTTTTTTTTTTTTTTTTTTTTTT

%TTTTTTTTTTTTTTTTTTTTTTTTTTTTTTTTTTTTTTTTTTTTTTTTTTTTTTTTT
\begin{figure}[tbp]\centering%
\includegraphics[width=6in]{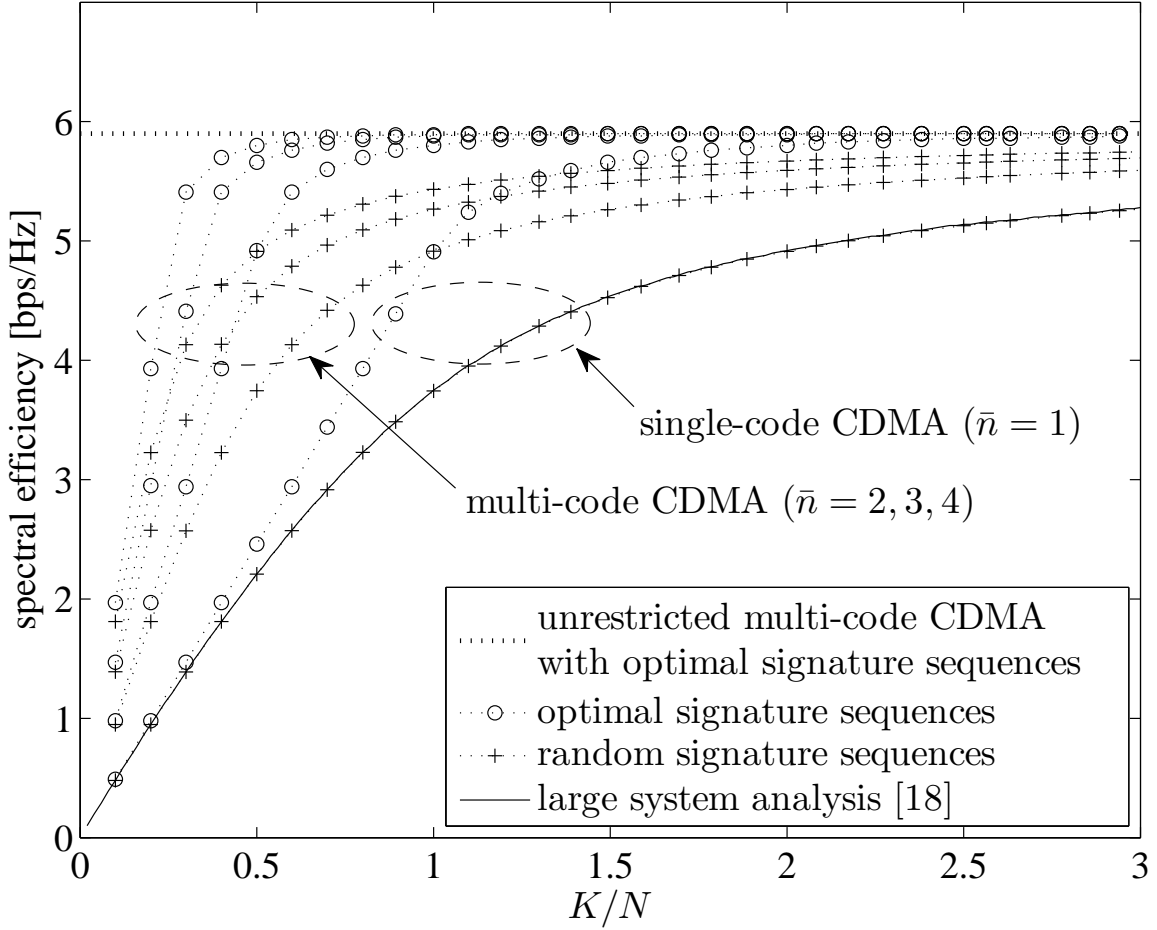} %,height=2.2in%4.5in
\caption{Comparison of spectral efficiency of the CDMA systems with sum-rate optimal sequences and random sequences, when the asymmetric-power users have i.i.d. Rayleigh flat-fading channels and  $\bar{E}_b/N_0=10$ [dB].}\label{fig: SE ver KN Raylpower}
\end{figure}
%TTTTTTTTTTTTTTTTTTTTTTTTTTTTTTTTTTTTTTTTTTTTTTTTTTTTTTTTT

Fig.~\ref{fig: SE ver KN Raylpower} performs the same comparison as Fig.~\ref{fig: SE ver KN equalpower} does except that there are asymmetric-power users.
In \cite{Verdu : fading}, the spectral efficiency of the single-code CDMA system having random sequences is derived by using a large system analysis, when each single-code user has an i.i.d.~flat-fading channel.
We adopt this result now with Rayleigh flat-fading channel for the single-code CDMA system.
However, this result cannot be used to obtain the spectral efficiency of the multi-code CDMA system having random sequences because the multiple data streams of each multi-code user do not suffer from i.i.d.~fading channels but from the same fading channel.
In \cite{Honig2}, the spectral efficiency of the multi-code CDMA system having random sequences or so-called isometric sequences\footnote{The multiple data streams of a  multi-code user are orthogonalized by using the columns from a Haar-distributed unitary random matrix. For details, see \cite{Debbah : isometric}.} is derived again by using a large system analysis, when each multi-code user has an i.i.d.~flat-fading or frequency-selective fading channel.
However, this large system analysis is only for the case where $\bar{n}$ and $N$ tend to infinity with $\bar{n}K/N$ being kept as a constant and is not for the systems with a large number of multi-code users.
Thus, instead of performing a large system analysis, the spectral efficiency of the multi-code CDMA system having random sequences in Fig.~\ref{fig: SE ver KN Raylpower} is obtained by generating $K= 100$ i.i.d.~exponential random numbers per each Monte-Carlo simulation run as the powers of the multi-code users that suffer Rayleigh flat fading.
To obtain the spectral efficiency of the sum-rate optimal restricted multi-code CDMA system, the sum-rate formula (\ref{eq: CDMAmaxsumrate}) is converted for the bandpass signaling or, equivalently, for the complex-baseband signaling, and the same method is used for each Monte-Carlo simulation run.
The performance of the unrestricted multi-code CDMA system is computed through numerical integration by using the well-known probability density functions \cite{Leon} of the sum of i.i.d. exponential random variables.
Fig.~\ref{fig: SE ver KN Raylpower} shows the spectral efficiency in [bps/Hz] obtained at the average bit energy per noise density $\bar{E}_b/N_0$ of $10$ [dB], where each marker is the average of $10^5$ Monte-Carlo simulation runs.
Similar to Fig.~\ref{fig: SE ver KN equalpower}, the spectral efficiency improves as $\bar{n}$ increases, and the sum-rate optimal CDMA system outperforms the CDMA system having random sequences.
Although not shown in Fig.~\ref{fig: SE ver KN Raylpower}, this is still the case when the isometric sequences are used.
In contrast to Fig.~\ref{fig: SE ver KN equalpower}, for $K/N\geq 1/\bar{n}$, the spectral efficiency of the sum-rate optimal restricted multi-code CDMA system is less than the spectral efficiency of the  unrestricted multi-code CDMA system.
This is because the probability of event that there exists at least one oversized multi-code user is not zero even if the effective system loading $\bar{n}K/N$ is greater than or equal to the unity.
\section{Conclusions and Future Work}\label{sec: Con}
%=-=-=-=-=-=-=-=-=-=-=-=-=-=-=-=-=-=-=-=-=-=-=-=-=-=-=-=-=-=-=-=-=-=-=-=-=-=-=-=-=-=-

In this paper, we have completely characterized the maximum sum rate of a multi-code CDMA system.
The maximum sum rate is derived as a function of the processing gain, power profile, and the upper-limit profile of the users on the number of multi-codes.
It turns out that the maximum sum rate is the same as those of the equivalent restricted FDMA and TDMA systems with upper limits on the bandwidths and the duty cycles of users, respectively.
This equivalence greatly enhances our understanding of the sum-rate optimal restricted multi-code systems by using, e.g., the optimal bandwidth allocation and the corresponding PSDs of users in the equivalent restricted FDMA systems.

The optimal distribution scheme of each user's  power to its multiple data streams is also derived that maximize the sum rate jointly with the optimal signature sequences.
Interestingly, it is shown that the same maximum sum rate can be achieved in general by various combinations of the total number of signature sequences and that of orthogonal signature sequences.
It is also shown that the restricted multi-code CDMA is a multiple-access scheme that trades off the maximum sum rate with the fairness among the users by imposing different upper limit profile on the number of multi-codes of users.
It turns out that the CDMA system is one of the optimal multiple-access schemes that achieve the sum capacity of the MAC, as far as enough number of multi-codes are assigned to the multi-code users.

In \cite{Massey: spread spectrum}, an optimal coding-spreading tradeoff problem for CDMA systems is posed to investigate the effects on the spectral efficiency of different allocations of the system bandwidth to coding and spreading.
Some partial answers can be found in \cite{Verdu : LSA} and \cite{Veeravalli: coding spreading} for  single-code CDMA systems.
A more complete answer to this open problem in terms of the Shannon and the Fourier bandwidths \cite{Massey: spread spectrum} remains as a future work, which may be obtained by extending the results in this paper to a continuous-time bandpass multi-code CDMA system.
\section*{Appendix}
%=-=-=-=-=-=-=-=-=-=-=-=-=-=-=-=-=-=-=-=-=-=-=-=-=-=-=-=-=-=-=-=-=-=-=-=-=
%-----------------------------------------------------
\subsection{Proof of {Lemma~\ref{lemma: diagonal}}}
%-----------------------------------------------------
%\begin{IEEEproof}
Let $(P,S)$ be a feasible solution to {Problem~\ref{prob: CDMA 1}}.
If we find an orthogonal matrix $U_k$ and a diagonal matrix $\tilde{P}_k$ by using the orthogonal eigen-decomposition of $P_k$ such that
\begin{equation}
P_k = U_k\tilde{P}_k U_k^{T},
\end{equation}
then the signal correlation matrix $\mathbb{E}\{\underline{\bm{x}}_k\underline{\bm{x}}_k^T\}$ of the $k$th multi-code user can be rewritten as
\begin{equation}
S_k P_k S_k^T=(S_k U_k) \tilde{P}_k (S_k U_k)^T.
\end{equation}
To make every column of $\tilde{S}_k\triangleq S_k U_k = [\underline{\tilde{s}}_{k,1},\underline{\tilde{s}}_{k,2},...,$ $\underline{\tilde{s}}_{k,\bar{n}_k}]$ have norm $N$, we define $\underline{\hat{s}}_{k,l}$ and $\hat{p}_{k,l}$ as $\underline{\hat{s}}_{k,l} \triangleq \sqrt{N}\underline{\tilde{s}}_{k,l}/ {\|\underline{\tilde{s}}_{k,l}\|}$ and $\hat{p}_{k,l}\triangleq{\tilde{p}_{k,l}}{\|\underline{\tilde{s}}_{k,l}\|^2}/N$ with $\tilde{p}_{k,l}$ being the $l$th diagonal entry of $\tilde{P}_k$.
Then, $S_k P_k S_k^T=\tilde{S}_k \tilde{P}_k \tilde{S}_k^T$ can be rewritten as
\begin{IEEEeqnarray}{rCl}
\tilde{S}_k \tilde{P}_k \tilde{S}_k^T&=&\sum_{l=1}^{\bar{n}_k}\tilde{p}_{k,l}\underline{\tilde{s}}_{k,l}\underline{\tilde{s}}_{k,l}^T\IEEEyessubnumber\\
&=&\sum_{l=1}^{\bar{n}_k}\hat{p}_{k,l}\underline{\hat{s}}_{k,l}\underline{\hat{s}}_{k,l}^T=\hat{S}_k\hat{P}_k\hat{S}_k^T,\IEEEyessubnumber
\end{IEEEeqnarray}
where $\hat{S}_{k}\triangleq [\underline{\hat{{s}}}_{k,1},\underline{\hat{{s}}}_{k,2},...,\underline{\hat{{s}}}_{k,\bar{n}_k}]$ and
$\hat{P}_k\triangleq\text{diag}(\hat{p}_{k,1},$ $\hat{p}_{k,2},...,\hat{p}_{k,\bar{n}_k})$.
Thus, if we define $\hat{S}\triangleq[\hat{S}_1,\hat{S}_2,...,\hat{S}_K]$ and $\hat{P}\triangleq\text{diag}(\hat{P}_1,\hat{P}_2,...,\hat{P}_K)$, then the correlation matrix $S P S^{T}=\sum_{k=1}^{K}S_k P_k S_k^{T}$ of the signal component in the objective function (\ref{eq: obj}) can be rewritten as
\begin{equation}
SPS^{T}=\hat{S}\hat{P}\hat{S}^T,
\end{equation}
where $\hat{S}$ now consists of column vectors of norm $N$ and $\hat{P}$ is diagonal with non-negative entries.
Therefore, the conclusion follows.
%\end{IEEEproof}
%
%-----------------------------------------------------
\subsection{Proof of {Lemma~\ref{lemma: kappa}}}
%-----------------------------------------------------

 %
First, we show that $\hat{w}_k$ defined in (\ref{eq: classification rule}) satisfies
\begin{IEEEeqnarray}{rCl}
\hat{w}_k > \bar{w}_k&\Longrightarrow&  \hat{w}_l > \bar{w}_l, \forall l\leq k,\text{ and}\IEEEyessubnumber\label{pass}\\%
\hat{w}_k \leq \bar{w}_k&\Longrightarrow& \hat{w}_l \leq \bar{w}_l, \forall l\geq k,\IEEEyessubnumber\label{fail}
\end{IEEEeqnarray}
i.e., if the $k$th user is oversized then every user having a smaller user index is also oversized, while if the $k$th user is non-oversized then every user having a bigger index is non-oversized, too.
To see this, assume $\hat{w}_k > \bar{w}_k$.
For $l\in \{1,2,...,k\}$, we have
\begin{IEEEeqnarray}{rCl}
\;&&p_k\bigg(w_\text{tot}-\sum_{k'=1}^{k-1}\bar{w}_{k'}\bigg)>\bar{w}_k\sum_{k'=k}^{K}p_{k'}\IEEEyessubnumber\label{eq: 48a} \\
\Rightarrow\;&&p_{l}\bigg(w_\text{tot}-\sum_{k'=1}^{k-1}\bar{w}_{k'}\bigg)>\bar{w}_{l}\sum_{k'=k}^{K}p_{k'}\IEEEyessubnumber\label{eq1}\\
\Rightarrow\;&&p_{l}\bigg(w_\text{tot}-\sum_{k'=1}^{k-1}\bar{w}_{k'}\bigg)+p_{l}\sum_{k'=l}^{k-1}\bar{w}_{k'}>\bar{w}_{l}\sum_{k'=k}^{K}p_{k'}+\bar{w}_{l}\sum_{k'=l}^{k-1}p_{k'}\IEEEeqnarraynumspace\IEEEyessubnumber\label{eq2}\\
\Leftrightarrow\;&&p_{l}\bigg(w_\text{tot}-\sum_{k'=1}^{l-1}\bar{w}_{k'}\bigg)>\bar{w}_{l}\sum_{k'=l}^{K}p_{k'},\IEEEyessubnumber
\end{IEEEeqnarray}
where (\ref{eq: 48a}) is by {Definition~\ref{def: classification}}, and (\ref{eq1}) and (\ref{eq2}) come from (\ref{powerordered}) and the assumption $\hat{w}_k > \bar{w}_k$.
Thus, $\hat{w}_l>\bar{w}_l, \forall l\leq k$, holds.
Now, assume $\hat{w}_k \leq \bar{w}_k$.
For $l\in \{k+1,k+2,...,K\}$, we have
\begin{IEEEeqnarray}{rCl}
\;&&p_k\bigg(w_\text{tot}-\sum_{k'=1}^{k-1}\bar{w}_{k'}\bigg)\leq\bar{w}_k\sum_{k'=k}^{K}p_{k'}\IEEEyessubnumber \label{eq: 49a}\\
\Rightarrow\;&&p_{l}\bigg(w_\text{tot}-\sum_{k'=1}^{k-1}\bar{w}_{k'}\bigg)\leq\bar{w}_{l}\sum_{k'=k}^{K}p_{k'}\IEEEyessubnumber\label{eq3}\\
\Rightarrow\;&&p_{l}\bigg(w_\text{tot}-\sum_{k'=1}^{k-1}\bar{w}_{k'}\bigg)-p_{l}\sum_{k'=k}^{l-1}\bar{w}_{k'}\leq\bar{w}_{l}\sum_{k'=k}^{K}p_{k'}-\bar{w}_{l}\sum_{k'=k}^{l-1}p_{k'}\IEEEeqnarraynumspace\IEEEyessubnumber\label{eq4}\\
\Leftrightarrow\;&&p_{l}\bigg(w_\text{tot}-\sum_{k'=1}^{l-1}\bar{w}_{k'}\bigg)\leq\bar{w}_{l}\sum_{k'=l}^{K}p_{k'},\IEEEyessubnumber
\end{IEEEeqnarray}
where (\ref{eq: 49a}) is by {Definition~\ref{def: classification}}, and (\ref{eq3}) and (\ref{eq4}) come from (\ref{powerordered}) and the assumption $\hat{w}_k \leq \bar{w}_k$.
Thus, $\hat{w}_l\leq\bar{w}_l,\forall l\geq k$, holds.

Then, we evaluate $\hat{w}_k$ from $k=1$ to $k=K$ to find the largest, thus unique, index ${K_1}$ satisfying $\hat{w}_{K_1} > \bar{w}_{K_1}$.
If $\hat{w}_1\leq\bar{w}_1$, then all the users are non-oversized by (\ref{fail}).
Thus, ${K_1}=0$ and (\ref{eq: kapp}) holds.
If $\hat{w}_1>\bar{w}_1$, then ${K_1}\in \{1,2,...,K\}$ exists and (\ref{eq: kapp}) holds by (\ref{pass}) and (\ref{fail}).
Therefore, the conclusion follows.

%-----------------------------------------------------
\subsection{Proof of {Proposition~\ref{proposition: FDMAsol}}}
%-----------------------------------------------------

%\begin{IEEEproof}
Define the Lagrangian function as
\begin{IEEEeqnarray}{l}
\mathcal{L}(w_k,\mu_k,\tilde{\mu}_k,\mu)\triangleq\sum_{k=1}^{K}w_k\ln\left(1+\frac{p_k}{N_0w_k}\right)\qquad\qquad\qquad\;\;\;\nonumber\\
\;-\sum_{k=1}^{K}\mu_k(w_k-\bar{w}_k)+\sum_{k=1}^{K}\tilde{\mu}_kw_k-\mu\left(\sum_{k=1}^{K}w_k-w_\text{tot}\right),\IEEEeqnarraynumspace
\end{IEEEeqnarray}
where $(\mu_k)_k,(\tilde{\mu}_k)_k,$ and $\mu$ are the dual variables.
Note that we use the natural logarithmic function $\ln(\cdot)$ just for convenience because any positive scaling of the objective function does not affect the solution.
Then, the Karush-Kuhn-Tucker (KKT) conditions \cite{Boyd} for {Problem~\ref{prob: FDMA}} are given by
\begin{IEEEeqnarray}{ll}\label{eq: KKT}
&\ln\left(1+\frac{p_k}{N_0{w_k^*}}\right)-\frac{p_k/{w_k^*}}{N_0+p_k/{w_k^*}}\qquad\qquad\qquad\nonumber\\
&\qquad\qquad\qquad\qquad\qquad-\mu_k+\tilde{\mu}_k-\mu=0,\forall k,\IEEEyessubnumber\label{eq: c1}\\
&{w_k^*}-\bar{w}_k\leq 0,\forall k,\IEEEyessubnumber\label{eq: c2}\\
&{w_k^*}\geq0,\forall k,\IEEEyessubnumber\label{eq: c3}\\
&\sum_{k=1}^{K}{w_k^*}-w_\text{tot}\leq 0,\IEEEyessubnumber\label{eq: c4}\\
&\mu_k\geq 0,\forall k,\IEEEyessubnumber\label{eq: c5}\\
&\mu_k({w_k^*}-\bar{w}_k)=0,\forall k,\IEEEyessubnumber\label{eq: c6}\\
&\tilde{\mu}_k\geq 0,\forall k,\IEEEyessubnumber\label{eq: c7}\\
&\tilde{\mu}_k {w_k^*}=0,\forall k,\IEEEyessubnumber\label{eq: c8}\\
&\mu\geq 0,\text{ and}\IEEEyessubnumber\label{eq: c9}\\
&\mu\left(\sum_{k=1}^{K}{w_k^*}-w_{\text{tot}}\right)=0,\IEEEyessubnumber\label{eq: c10}
\end{IEEEeqnarray}
where (\ref{eq: c1}) is the stationarity condition; (\ref{eq: c2})-(\ref{eq: c4}) are the primal feasibility conditions; (\ref{eq: c5}), (\ref{eq: c7}), and (\ref{eq: c9}) are the dual feasibility conditions; and (\ref{eq: c6}), (\ref{eq: c8}), and (\ref{eq: c10}) are the complementary slackness conditions.
Let $s\geq0$ be defined as
\begin{equation}
s\triangleq \frac{\sum_{k'={K_1}+1}^{K}p_{k'}}{w_\text{tot}-\sum_{k'=1}^{{K_1}}\bar{w}_{k'}}.\label{eq: s}
\end{equation}
The positivity of the denominator in (\ref{eq: s}) will be shown soon.
Now, the optimal solution (\ref{eq: optband}) can be rewritten as ${w_k^*}=\bar{w}_k,\forall k\leq{K_1}$ and ${w_k^*}=p_k/s,\forall k>{K_1}$.
Thus, $s$ can be interpreted as the common PSD of non-oversized users, if exist.
Then, the claim is that $(\mu_k)_k,(\tilde{\mu}_k)_k,$ and $\mu$ given by
\begin{IEEEeqnarray}{lCl}\label{eq: KKT multi}
\!\!\!\!\!\!\!\!\!\!\!\mu_k&=&\left\{\!
\begin{array}{ll}
\ln\left(1+\frac{p_k}{N_0\bar{w}_{k}}\right)-\frac{p_k/\bar{w}_{k}}{N_0+p_k/\bar{w}_{k}}\\
\,-\left(\ln\left(1+\frac{s}{N_0}\right)-\frac{s}{N_0+s}\right),&\!\!\!\!\text{for } 1\leq k \leq{K_1},\\
0,&\!\!\!\!\text{for } {K_1} < k \leq K,
\end{array}
\right.\IEEEyessubnumber\label{eq: muk}\\
\!\!\!\!\!\!\!\!\!\!\!\tilde{\mu}_k&=&0,\forall k,\text{ and}\IEEEyessubnumber\label{eq: mutildek}\\
\!\!\!\!\!\!\!\!\!\!\!\mu&=&
\begin{array}{ll}
\ln\left(1+ \frac{s}{N_0}\right)-\frac{s}{N_0+s},\IEEEyessubnumber\label{eq: mu}
\end{array}
\end{IEEEeqnarray}
together with (\ref{eq: optband}) satisfy all the KKT conditions in (\ref{eq: KKT}).

Substituting (\ref{eq: optband}), (\ref{eq: muk}), (\ref{eq: mutildek}), and (\ref{eq: mu}) into (\ref{eq: c1}), we can see that (\ref{eq: c1}) is satisfied for all $k$.
Next, from (\ref{eq: optband}), note that ${w_k^*}-\bar{w}_k=0,\forall k\leq{K_1}$.
In addition, if ${K_1}<K$, since $w_{{K_1}+1}^*=\hat{w}_{{K_1}+1}\leq\bar{w}_{{K_1}+1}$ by (\ref{eq: kappa2}), $w_{{K_1}+1}^*=p_{{K_1}+1}/s\leq\bar{w}_{{K_1}+1}$ implies ${w_k^*}=p_{k}/s\leq\bar{w}_{k},\forall k>{K_1}$, which comes from (\ref{powerordered}).
Thus, the condition (\ref{eq: c2}) is satisfied.

Next, if ${K_1}=0$ then the condition (\ref{eq: c3}) is satisfied for all $k$ by (\ref{eq: optband}).
If ${K_1}\ne 0$ then, since ${w_k^*}=\bar{w}_k \text{ for } 1\leq k\leq{K_1}$ by (\ref{eq: optband}) and $\bar{w}_k>0,\forall k$, by assumption, the condition (\ref{eq: c3}) is satisfied for $k\leq {K_1}$.
On the other hand, if (\ref{eq: kappa}) is evaluated at $k={K_1}>0$, then $\hat{w}_{K_1}>\bar{w}_{K_1}$, which leads to ${p_{K_1}}(w_\text{tot}-\sum_{k'=1}^{{K_1}-1}\bar{w}_{k'})>\bar{w}_{K_1}{\sum_{k'={K_1}}^{K}p_{k'}}>\bar{w}_{K_1} p_{K_1}$.
So, we have
\begin{equation}\label{eq: 57}
w_\text{tot}-\sum_{k'=1}^{{K_1}}\bar{w}_{k'}>0
\end{equation}
due to the assumption $p_k>0,\forall k$.
Thus, the condition (\ref{eq: c3}) is satisfied for $k> {K_1}$, because ${w_k^*}>0$ for $k>{K_1}$.
By the way, (\ref{eq: 57}) verifies that the denominator of $s$ is positive when ${K_1}\ne 0$.
Note that it is obviously the case with $K_1= 0$.

Next, if ${K_1}=K$, then the condition (\ref{eq: c4}) is satisfied by (\ref{eq: 57}).
If ${K_1}<K$, then the condition (\ref{eq: c4}) is satisfied again as
\begin{equation}\label{eq: 58}
\sum_{k=1}^{K}{w_k^*}=
\displaystyle\sum_{k=1}^{{K_1}}\bar{w}_k+\sum_{k={K_1}+1}^{K}\frac{p_k}{s} = w_\text{tot},
\end{equation}
by (\ref{eq: optband}) and (\ref{eq: s}).

Next, if ${K_1}=0$, then the condition (\ref{eq: c5}) is satisfied for all $k$ by (\ref{eq: muk}).
If (\ref{eq: kappa}) is evaluated again at $k={K_1}>0$, then this time $p_{K_1}(w_\text{tot}-\sum_{k=1}^{{K_1}-1}\bar{w}_k)>\bar{w}_{K_1}\sum_{k={K_1}}^{K}p_k$ implies ${p_{K_1}}/{\bar{w}_{K_1}}>s$, which leads to $p_k/\bar{w}_k>s,\forall k\leq{K_1}$, by (\ref{powerordered}).
So, we have $\mu_k>0,\forall k\leq{K_1}$, because $\ln(1+x/N_0)-x/(N_0+x)$ is a  monotone increasing function of $x>0$.
In addition, $\mu_k=0,\forall k>{K_1}$, from (\ref{eq: muk}).
Thus, the condition (\ref{eq: c5}) is satisfied.

Next, substituting (\ref{eq: muk}) and (\ref{eq: optband}) into (\ref{eq: c6}), we can see that the condition (\ref{eq: c6}) is satisfied.
By (\ref{eq: mutildek}), the conditions (\ref{eq: c7}) and (\ref{eq: c8}) are satisfied immediately.
Note that $\ln(1+x/N_0)-x/(N_0+x)$ is a  monotone increasing function of $x$ and that $\ln(1+0/N_0)-0/(N_0+0)=0$.
Thus, by the non-negativity of $s$, the condition (\ref{eq: c9}) is satisfied.

Finally, if ${K_1}=K$, then the condition (\ref{eq: c10}) is satisfied because $s=0$ by (\ref{eq: s}), which leads to $\mu=0$.
If ${K_1}<K$, then $\sum_{k=1}^{K}{w_k^*}=w_\text{tot}$ as already shown in (\ref{eq: 58}).
Thus, the condition (\ref{eq: c10}) is satisfied.

It can be easily verified that the constraints (\ref{bandub}) and (\ref{tbandconst}) satisfy Slater's condition \cite{Boyd}.
Thus, these KKT conditions become necessary and sufficient for a feasible solution to be optimal.
Moreover, by evaluating the objective function with the optimal bandwidths in (\ref{eq: optband}), we obtain the maximum sum rate as (\ref{eq: FDMAmaxsumrate}).
Therefore, the conclusion follows.

%=-=-=-=-=-=-=-=-=-=-=-=-=-=-=-=-=-=-=-=-=-=-=-=-=-=-=-=-=-=-=-=-=-=-=-=-=-=-

\enlargethispage{-0.0in}
%=-=-=-=-=-=-=-=-=-=-=-=-=-=-=-=-=-=-=-=-=-=-=-=-=-=-=-=-=-=-=-=-=-=-=-=-=-=-
\end{document}